\documentclass [10pt,twocolumn]{IEEEtran}
\usepackage{algorithm,algorithmic,amsbsy,amsmath,amssymb,epsfig,bbm,mathrsfs,multirow,amsthm}
\usepackage[T1]{fontenc}
\usepackage[utf8]{inputenc}
\usepackage{amsmath}
\usepackage{algorithm}
\usepackage{array,multirow,graphicx}
\usepackage{color}
\usepackage{cite}
\usepackage{float}
\usepackage{makecell}
\usepackage[x11names,dvipsnames,table]{xcolor} %for use in color links
\usepackage{colortbl}
\usepackage{multirow}
\usepackage{lipsum}
\usepackage{graphicx} % 导言区引入
\usepackage{booktabs}
\usepackage{tabularx} % 自动列宽
\usepackage[colorlinks=true,linkcolor=black,citecolor=colorlinks=true,linkcolor=black,citecolor=blue,urlcolor=blue,urlcolor=blue]{hyperref}

\definecolor{mygray}{gray}{0.6}
\definecolor{myblue}{rgb}{0.8,0.85,1} 
\usepackage{array}
\newcolumntype{L}[1]{>{\raggedright\let\newline\\\arraybackslash\hspace{0pt}}m{#1}}
\newcolumntype{C}[1]{>{\centering\let\newline\\\arraybackslash\hspace{0pt}}m{#1}}
\newcolumntype{R}[1]{>{\raggedleft\let\newline\\\arraybackslash\hspace{0pt}}m{#1}}

\usepackage{array, tabularx, boldline}
\usepackage{eurosym}
\usepackage{amstext} % for \text
\DeclareRobustCommand{\officialeuro}{%
  \ifmmode\expandafter\text\fi
  {\fontencoding{U}\fontfamily{eurosym}\selectfont e}}

\usepackage{array, tabularx, boldline}
\usepackage{graphicx}
\usepackage{cellspace}
\begin{document}
%\[\title{Pricing Models in Internet of Things: A Survey}
\title{Incentive Mechanism Design for Resource Management in Satellite Networks: A Comprehensive Survey}

\author{Nguyen Cong Luong, Zeping Sui, \textit{Member, IEEE}, Duc Van Le, \textit{Senior Member, IEEE}, Jie Cao, \textit{Member, IEEE}, Bo Ma, \textit{Member, IEEE}, Nguyen Duc Hai, \textit{Member, IEEE}, Ruichen Zhang, Vu Van Quang, Dusit Niyato, \textit{Fellow, IEEE}, and Shaohan Feng, \textit{Member, IEEE}.
\thanks{Nguyen Cong Luong, Nguyen Duc Hai and Vu Van Quang are with the Phenikaa School of Computing, Phenikaa University, Hanoi 12116, Vietnam. E-mails: luong.nguyencong@phenikaa-uni.edu.vn,  21010560@st.phenikaa-uni.edu.vn, and quang.vuvan@phenikaa-uni.edu.vn.}
\thanks{Zeping Sui is with the School
of Computer Science and Electronics Engineering, University of Essex, Colchester CO4 3SQ, U.K. E-mail: zepingsui@outlook.com.}

\thanks{Duc Van Le is with the School of Electrical Engineering and Telecommunications, University of New South Wales, NSW 2052, Australia. E-mail: duc.le1@unsw.edu.au.}

\thanks{Jie Cao is with the Department of Information Science and Technology,
Harbin Institute of Technology (Shenzhen), Shenzhen 518055, China. E-mail:
caojhitsz@ieee.org.}

\thanks{Bo Ma and Shaohan Feng are with the School of Information and Electronic Engineering Sussex Artificial Intelligence Institute, Zhejiang Gongshang University, Hangzhou 310018,
China. E-mail: mabo@mail.zjgsu.edu.cn and feng\_shaohan@mail.zjgsu.edu.cn.}

\thanks{Ruichen Zhang and Dusit Niyato are with the College of Computing and Data Science, Nanyang
Technological University, Singapore 639798. E-mails: ruichen.zhang@ntu.edu.sg and dniyato@ntu.edu.sg.}
 
}

\maketitle
%====================================================================
\begin{abstract}
Resource management is one of the challenges in satellite networks due to their high mobility, wide coverage, long propagation distances, and stringent constraints on energy, communication, and computation resources. Traditional resource allocation approaches rely only on hard and rigid system performance metrics. Meanwhile, incentive mechanisms, which are based on game theory and auction theory, investigate systems from the "economic" perspective in addition to the "system" perspective. Particularly, incentive mechanisms are able to take into account rationality and other behavior of human users into account, which guarantees benefits/utility of all system entities, thereby improving the scalability, adaptability, and fairness in resource allocation. This paper presents a comprehensive survey of incentive mechanism design for resource management in satellite networks. The paper covers key issues in the satellite networks, such as communication resource allocation, computation offloading, privacy and security, and coordination. We conclude with future research directions including learning-based mechanism design for satellite networks. 
\end{abstract}


%main

%=====================================================================
\section{Introduction}
Satellite networks are rapidly transforming the global landscape of communication, sensing and navigation. Satellite networks comprise spaceborne platforms which include Geosynchronous Earth Orbit (GEO), Medium Earth Orbit (MEO) and Low Earth Orbit (LEO) satellites \cite{10679152}. With the increasing utilization of satellite communications, modern satellite networks are no longer isolated infrastructures with trivial applications. Instead, they are gradually becoming integral components of next-generation communication systems, complementing terrestrial networks to provide reliable and ubiquitous services \cite{kodheli2020satellite, giordani2020non,sui2025multi}. Satellite networks have found various applications, notably satellite-based Internet-of-Things (SIoT) for navigation systems, agriculture, tracking and healthcare \cite{routray2019satellite}.

One of the primary challenges in satellite networks is resource management \cite{peng2018review, zhou2021machine}, which includes access control, spectrum allocation, power allocation, computation offloading and security. First, it is important to consider the heterogeneous nature of satellite networks, which usually contain different types of entities such as satellites, dynamic topologies, and service requirements \cite{10670196}. Second, the satellites usually possess high mobility and wide coverage to serve multiple ground entities \cite{11185315,10183832,10250854,10217007,10129061}, while their resources including power, bandwidth and computing resources are limited. Therefore, the allocation of resources must be completed in a way that ensures Quality-of-Service (QoS) while being adaptive enough to handle the frequent handovers and dynamic link availability of satellite communications. Moreover, with the emergence of mega-constellations such as StarLink, resource management must possess scalability with minimal coordination overhead. Finally, long distances between satellites and ground users can impose wireless security issues within the network. 

Against such challenges of modern satellite networks, traditional resource allocation techniques are becoming increasingly insufficient. They typically require a centralized controller and require complete network information. Additionally, most traditional resource management algorithms do not take into account dynamic topologies and time-varying resource demands of satellites \cite{zhou2021machine}. Moreover, satellite networks can be of high mobility, wide coverage and large scale, making traditional algorithms suffer from complexity and scalability problems \cite{zhang2022joint}. 

% Recently, incentive mechanisms have emerged as a powerful alternative. Already a powerful paradigm in federated learning \cite{zhan2021survey}, crowdsensing \cite{wu2024research} and peer-to-peer (P2P) systems \cite{haddi2015survey}, incentive mechanisms aim to leverage principles from economics and game theory to align the objectives of diverse entities within satellite networks. Compared to traditional algorithms, incentive mechanisms possess several advantages as follows. First, incentive mechanisms such as Stackelberg game  or cooperative game  enable decentralized decision-making with low coordination, as well as fairness and truthfulness in resource trading. They also boast superior scalability and robustness against selfish or malicious factors within the systems. Second, computation offloading schemes based on incentive mechanisms  allow decentralized task allocation across satellites, edge servers and users. For instance, the non-cooperative game allows fully selfish decision-making with stable equilibria, improving system efficiency and lowering costs. Additionally, the auction game allows price negotiation between buyers and sellers, ensuring scalability and economic motivation. Third, incentive mechanisms enhance security approaches with adaptive anti-jamming and secure spectrum allocation , as well as truthful, privacy-preserving allocation for  . They also efficiently address trade-offs between privacy and performance in cooperative localization  and incentivize cooperative defense across satellites, UAVs and IoT devices . 

In contrast, incentive mechanisms, which are essentially based on economics, game theory and auction theory, have emerged as powerful alternatives. They enable decentralized decision-making \cite{lin2023leo, peng2024cloud, wang2019game, tang2024digital}, which helps reduce coordination overhead, and ensure fairness, truthfulness, and robustness against selfish or malicious behaviors. Importantly, incentive mechanisms investigate systems from the economic perspective in addition to the systematic perspective. The latter relies only on hard, inflexible metrics to measure system performance, while the former can take into account rationality and other behavior of human users \cite{luong2016data, luong2017resource}. For example, Stackelberg \cite{zhang2022joint, zhu2020two} and cooperative games \cite{wang2022qos, du2019coalitional} promote stable resource sharing, while non-cooperative games \cite{wang2019game} support efficient equilibria even under fully selfish actions. Stackelberg games are also suitable for hierarchical optimization of systems, in which the entities act symmetrically \cite{zhang2022joint, xu2023adaptive}. Auction-based schemes \cite{jin2020double, huang2024profit} allow price negotiation between buyers and sellers, enhancing scalability and economic motivation. Regarding security applications, incentive mechanisms also mitigate security issues by supporting adaptive anti-jamming \cite{li2022secure}, secure spectrum allocation \cite{han2020spatial}, and privacy-preserving resource sharing for cooperative relaying and crowdsensing \cite{zhao2022tensor, ma2022blockchain}. They further address trade-offs between privacy and performance in cooperative defense \cite{shi2021clap} and incentivize cooperative defense across satellites, UAVs, and IoT devices \cite{huang2025consolidated}. Other key research issues such as coordination for satellite networks \cite{du2022game, deng2020ultra} have also been addressed with various models of incentive mechanisms. 

There have been some relevant surveys regarding incentive mechanisms and satellite networks. However, none of them provides a comprehensive view into the utilization of incentive mechanisms for satellite networks' resource allocation problems. Particularly, the survey in \cite{luong2018applications} discusses the applications of economic and pricing models for resource allocation problems in 5G wireless networks such as user association, spectrum allocation, and interference and power management. {\color{black}Meanwhile, the survey in \cite{luong2017resource} presents a comprehensive literature review on the integration of economics and pricing models into resource management in cloud networking, extending to resource sharing in edge computing and cloud-based software-defined wireless networking issues. A survey specifically addressing auction designs for edge computing and its three common paradigms, i.e., cloudlet, fog computing and mobile edge computing (MEC) can be found in \cite{qiu2022applications}. A work in \cite{luong2016data} provides a comprehensive survey on utilizing economic analysis and pricing models for data collection and wireless communication in IoT. Recently, the work in \cite{jiang2024game} targets game-theory based solutions in satellite communication networks, addressing various problems such as satellite/user grouping, relay selection, virtual network embedding, task offloading and access control. Other notable surveys regarding satellite communications can be found in \cite{kodheli2020satellite, wang2018high, xiaogang2016survey, fontanesi2025artificial, chen2024survey} where the vital aspects of satellite networks such as communications, convergence, computation offloading and routing are studied. Particularly, the work in \cite{fontanesi2025artificial} provides a comprehensive survey on applications of machine learning for physical layer design and resource allocation in satellite networks. However, these aforementioned works all lack the specialized focus on the applications of incentive mechanisms for resource management in satellite networks.} In contrast to cellular terrestrial networks such as 5G/6G, satellite networks exhibit unique characteristics that require distinct approaches to incentive mechanism design. First, satellite networks span wide geographical areas and serve diverse types of ground users (e.g., mobile devices, ships, and aircraft). Therefore, incentive mechanisms must ensure differentiated QoS requirements while accounting for varying service priorities. Second, satellites are inherently mobile, incentive mechanisms must take into account limited connection durations between satellites and ground users, as well as potential handover challenges. Third, the considerable distance between satellites and ground users imposes constraints on communication, making the design of incentive mechanisms more challenging, i.e., for obtaining a solution. Finally, satellite networks involve significant deployment expenses. Hence, incentive mechanisms should significantly gain the network service providers’ revenues to sustain their motivation for resource provisioning. {\color{black} Table~\ref{tab:compa} show the major differences between our survey and related surveys. Our survey provides a comprehensive survey on applications of incentive mechanisms for satellite networks.}

\begin{table*}[h!]
\caption{Summary of the Related Surveys and Papers.}
\centering
\renewcommand{\arraystretch}{1.7} % reduced line height for compact appearance
\setlength{\tabcolsep}{2pt}
\begin{tabularx}{\linewidth}{|>{\centering\arraybackslash}X|
>{\centering\arraybackslash}X|
>{\centering\arraybackslash}X|
>{\centering\arraybackslash}X|
>{\centering\arraybackslash}X|
>{\centering\arraybackslash}X|
>{\centering\arraybackslash}X|
>{\centering\arraybackslash}X|
>{\centering\arraybackslash}X|}
\hline
\rowcolor[HTML]{C0C0C0}
\makecell{\bfseries Existing \\ \bfseries surveys} &
\makecell{\bfseries Incentive \\ \bfseries mechanism} &
\makecell{\bfseries Satellite \\ \bfseries networks} &
\makecell{\bfseries Resource \\ \bfseries allocation} &
\makecell{\bfseries Computation \\ \bfseries offloading} &
\makecell{\bfseries Privacy \\ \bfseries and security} &
\makecell{\bfseries Routing} &
\makecell{\bfseries Relay \\ \bfseries selection} &
\makecell{\bfseries Caching} \\
\hline
\cite{luong2018applications} & $\checkmark$ &  & $\checkmark$ &  &  &  &  &  \\
\hline
\cite{luong2017resource} & $\checkmark$ &  &  & $\checkmark$ &  &  &  &  \\
\hline
\cite{xiaogang2016survey} &  & $\checkmark$ &  &  &  & $\checkmark$ &  &  \\
\hline
\cite{jiang2024game} & $\checkmark$ & $\checkmark$ &  & $\checkmark$ &  &  & $\checkmark$ &  \\
\hline
\cite{chen2024survey} &  & $\checkmark$ &  & $\checkmark$ & $\checkmark$ &  &  &  \\
\hline
\cite{fontanesi2025artificial} &  & $\checkmark$ & $\checkmark$ &  & $\checkmark$ &  &  &  \\
\hline
\textbf{This paper} & $\checkmark$ & $\checkmark$ & $\checkmark$ & $\checkmark$ & $\checkmark$ & $\checkmark$ & $\checkmark$ & $\checkmark$ \\
\hline
\end{tabularx}
\label{tab:compa}
\end{table*}

\begin{figure}
    \centering
\includegraphics[width=0.9\linewidth]{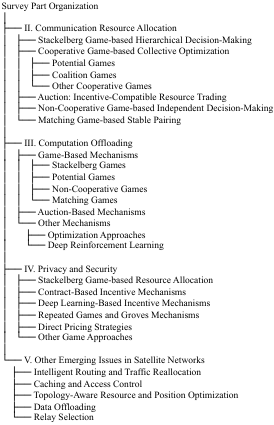}
    \caption{\color{black}Taxonomy of incentive mechanism designs for resource management in satellite networks.}
    \label{fig:taxonomy}
    \vspace{-2em}
\end{figure}

This motivates us to provide an extensive review of how incentive mechanisms are applied to address resource allocation issues of satellite networks. The contributions of this survey are as follows:
\begin{itemize}
    \item We review and discuss various game-theoretic incentive approaches to solve communication resource allocation problems for satellite networks, namely access control, spectrum allocation, power allocation and offloading issues. The discussions aim to highlight their ability to reduce latency, ensure fairness and QoS, as well as scaling in mega-constellations. 
    \item We survey and analyze innovative game-based, auction-based and learning-based mechanisms for computation offloading for satellite networks. Specifically, we focus on the problems such as offloading strategy, resource pricing and allocation; showing the effectiveness of incentive mechanisms in energy saving, fairness and scalability.  
    \item We present related works about incentive-compatible security mechanisms for satellite networks under jamming, spoofing and privacy attacks. Specifically, we provide insights into problems such as anti-jamming, routing optimization, covert communication,  cooperative defense and diversified access; demonstrating how economic incentives enable adaptive defenses beyond traditional centralized schemes. 
    \item We investigate recent advances of incentive mechanisms for offloading and coordination in satellite networks. We demonstrate their effectiveness in addressing problems such as traffic scheduling, multipath routing, caching, clustering and scalable offloading. 
    \item Finally, we discuss promising research directions related to the integration of incentive mechanisms into resource allocation for satellite networks. 
\end{itemize}
The remainder of the paper is organized as follows. Section \ref{section:comm_resource_alloc} reviews the applications of incentive mechanisms for communication resource allocation for satellite networks. Section \ref{section:com_offload} discusses incentive-based approaches for computation offloading for satellite networks. Section \ref{Section 4} provides reviews of incentive mechanisms for security issues in satellite networks. Section \ref{section:mixed} discusses the integration of game-theoretic incentives into solving coordination problems in satellite networks. Section \ref{section:concl_future} highlights key challenges and future research directions, and finally concludes this paper. {\color{black}Fig.~\ref{fig:taxonomy} provides a taxonomy of the survey part of this work.}~{\color{black}Abbreviations frequently used in this work is listed in Table~\ref{tab:notation-table}.}

\begin{table}[h]
\caption{List of Abbreviations Frequently Used in This Paper.}
\centering
\renewcommand{\arraystretch}{1.3}
\resizebox{\columnwidth}{!}{%
\begin{tabular}{|l|l|}
\hline
\rowcolor[HTML]{C0C0C0}
\textbf{Abbreviation} & \textbf{Description} \\ \hline
AUV & Autonomous Underwater Vehicles \\ \hline
CoIV & Cost of Information Value \\ \hline
EPG & Exact Potential Game \\ \hline
GEO & Geosynchronous Earth Orbit \\ \hline
HAP & High-Altitude Platform \\ \hline
IST-IOV & Satellite Terrestrial Internet of Vehicles \\ \hline
LEO & Low Earth Orbit \\ \hline
MDP & Markov Decision Process \\ \hline
MEC & Mobile Edge Computing \\ \hline
NE & Nash Equilibrium \\ \hline
PE Game & Pursuit-evasion Differential Game \\ \hline
SAGINs & Space-Air-Ground Integrated Networks \\ \hline
SEC & Satellite Edge Computing \\ \hline
UAV & Unmanned Aerial Vehicle \\ \hline
VCG & Vickrey–Clarke–Groves \\ \hline
\end{tabular}%
}
\label{tab:notation-table}
\end{table}

\section{Satellite Communication Resource Allocation}
\label{section:comm_resource_alloc}
%=============================================================================

Efficiently distributing limited resources among users and services is a primary challenge in satellite networks\cite{jiang2024game}. Key problems include access control, spectrum allocation, power allocation, computation offloading, and security\cite{smith2021ran}. Managing these resources faces significant challenges, as satellite networks are highly heterogeneous, involving diverse entities, dynamic topologies, and varied service requirements. Resource allocation must ensure QoS while adapting to frequent handovers and dynamic link conditions. Moreover, secure and incentive-compatible mechanisms are essential in decentralized networks. These prevent malicious or selfish resource usage from degrading overall system performance.
In this section, game-theoretical approaches for communication resource allocation in satellite networks are reviewed.

\begin{table*}[ht]
\caption{Summary of Incentive Mechanisms for Satellite Communication Resource Allocation.}
\centering
\renewcommand{\arraystretch}{1.2}
\begin{tabular}{|l |p{3.4cm}| p{4.8cm}| p{2.8cm}| p{4.3cm}|}
\hline
\textbf{Ref.} & \textbf{Scenario} & \textbf{Problem focus} & \textbf{Methodology} & \textbf{Players}\\
\hline
\cite{zhang2022joint} & Ultra-dense LEO integrated satellite-terrestrial networks	 & High mobility and large scale complicating optimal communication mode selection & Stackelberg game + evolutionary game (user mode selection) & Leader: Network operator; Followers: End users\\
\hline
\cite{yan2022electromagnetic} & Satellite network electromagnetic confrontation & Antagonistic spectrum access and interference minimization under confrontation & Stackelberg game & Leader: Blue Team; Followers: Red Team \\
\hline
\cite{wen2024hierarchical}	 & Cognitive satellite-terrestrial networks (CSTNs) & Lack of fairness in resource trading between satellite and terrestrial networks & Two-layer Stackelberg game & Leader: Satellite; Followers: Terrestrial base stations (BSs, via coalition collaboration, adversarial network)\\
\hline
\cite{xu2023adaptive}	 & GEO relay systems	 & Spectrum waste caused by equipment-oriented spectrum management & Task-oriented Stackelberg game & Leader: GEO satellite; Followers: End users\\
\hline
\cite{xu2022hierarchical} & Space-air-ground integrated networks (SAGINs)	 & Multi-tier bandwidth pricing and resource allocation optimization for multicast services & Four-stage Stackelberg game & Leader: LEO satellites; Intermediate followers: UAVs; Final followers: Social communities\\
\hline
\cite{xu2022ubiquitous}	 & Space-air-ocean integrated networks (SAOINs)	 & Ubiquitous transmission and resource pricing for maritime scenarios & Modified three-stage Stackelberg game & Leader: LEO satellites; Intermediate followers: UAVs; Final followers: Unmanned Surface Vessels (USVs)\\
\hline
\cite{liu2024demand} & Multilayer heterogeneous satellite networks (LEO/MEO)	 & Distributed inter-satellite link allocation and resource matching & Stackelberg game & Leader: LEO satellites; Followers: MEO satellites\\
\hline
\cite{zhu2020two} & Cloud-based integrated terrestrial-satellite networks	 & Resource allocation and user service selection optimization under multi-QoS tiers & Stackelberg game & Leader: Network operator; Followers: End users\\
\hline
\end{tabular}
\label{tab:offloading1}
\end{table*}

%==========================	
\subsection{Stackelberg Game-based Hierarchical Decision-Making}
%==========================

The Stackelberg game is applicable to hierarchical decision-making and precisely models the inherent information asymmetry and resource control characteristics within satellite networks.
Therefore, several works have employed Stackelberg game frameworks to address hierarchical resource management in space computing networks\cite{liu2021research}.
The general idea is that a leader (e.g., satellite operator and LEO satellite) sets rules or parameters, and followers (e.g., end users and MEO satellites) maximize their utility accordingly. 

The authors in \cite{zhang2022joint} investigated ultra-dense LEO integrated satellite-terrestrial networks, where high mobility and large scale complicate optimal communication mode selection. To resolve this challenge, the interaction among network entities is modeled as a Stackelberg game, where the network operator acts as the leader to determine access prices for terrestrial base stations (BSs) and LEO satellites. End users serve as followers and adopt an evolutionary game to select communication modes that maximize their data rate. It is proven that backward induction yields the Stackelberg Equilibrium (SE), where the operator’s revenue is maximized and users converge to stable mode selections. Similarly, \cite{zhu2020two} studies cloud-based integrated terrestrial-satellite networks, where the operator offers two QoS/price tiers and allocates resources, while users select services via an evolutionary game. 
In the context of electromagnetic confrontation, \cite{yan2022electromagnetic} proposes a Stackelberg game between a blue team and red team in satellite networks.  The game decomposes into two exact potential sub-games, and the distributed hierarchical confrontation channel selection algorithm finds the SE. However, neither \cite{zhang2022joint} nor \cite{yan2022electromagnetic} addresses fairness in resource trading.
For cognitive satellite-terrestrial networks (CSTNs) with fairness concerns, \cite{wen2024hierarchical} designs a two-layer Stackelberg framework. In the considered framework, the satellite sets resource trading terms, while terrestrial BSs form coalitions via a distributed merge-and-split algorithm to negotiate resource demands.  This framework ensures a “win-win” outcome for both the satellite and BSs.

\begin{figure}
    \centering
    \includegraphics[width=0.8\linewidth]{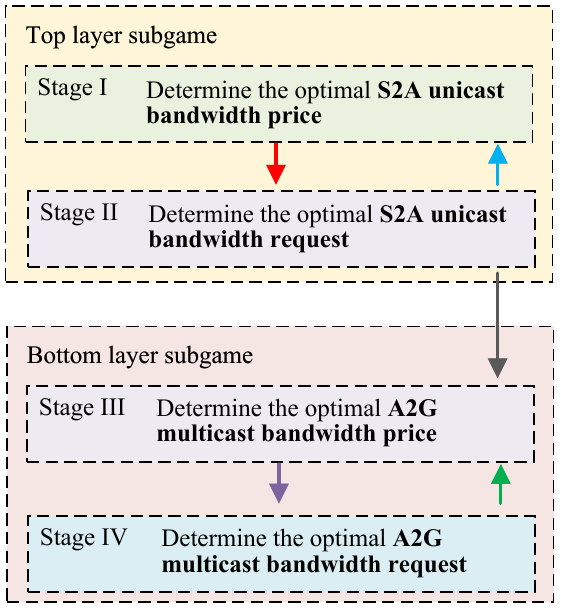}
    \caption{Diagram of four-stage Stackelberg game-based bandwidth allocation framework in \cite{xu2022hierarchical}.}
    \label{fig:Four-stage-Stackelberg}
\end{figure}
In GEO relay systems, \cite{xu2023adaptive} notes that equipment-oriented spectrum management causes a waste of spectrum resources. Thus, a task-oriented Stackelberg game is proposed, with the GEO satellite using two pricing modes, including U-prior (maximizing user utility) and S-prior (maximizing spectrum utilization). Then, end users adjust resource consumption based on pricing. While a Nash Equilibrium (NE) exists when users prioritize individual utility, the SE optimizes the dual-mode strategy, outperforming baselines in throughput and latency with minimal revenue loss.
\textcolor{black}{Building on the hierarchical logic of Stackelberg games and extending to space-air-ground integrated networks, \cite{xu2022hierarchical} models a four-stage Stackelberg game, as shown in Fig.~\ref{fig:Four-stage-Stackelberg}. It can be decomposed into two typical two-stage Stackelberg subgames. The top subgame has the LEO satellite set the space-to-air (S2A) unicast bandwidth price to allocate bandwidth to Unmanned Aerial Vehicles (UAVs). The bottom subgame features UAVs as leaders and communities as followers, negotiating air-to-ground (A2G) multicast bandwidth pricing and requests. The proposed method accelerates decision convergence via multi-stage decentralized decision-making, including fast pricing in LEO, local optimization of UAVs and instant bandwidth selection within the community. It also reduces mid-transmission mode switches through incentive alignment, thereby lowering congestion, decision, and handover latency.} 
Similarly, \cite{xu2022ubiquitous} addresses ubiquitous transmission in space-air-ocean integrated networks with a modified three-stage Stackelberg game. Specifically, LEO satellites set LEO-to-UAV data rate prices, and UAVs set UAV-to-unmanned surface vessel (USV) prices. Then, USVs select data rates under budget constraints. 
Through non-linear programming methods, optimal strategies for USVs and UAVs were derived under varying data rate budgets. Also, LEOs use accelerated conjugate gradient descent for pricing, significantly improving the utility of USV.
To solve distributed link allocation in multilayer heterogeneous satellite networks, \cite{liu2024demand} models LEO satellites as leaders, \textcolor{black}{while MEO satellites act as followers.  Stackelberg’s hierarchical structure enables LEO leaders to generate targeted link requests by leveraging their global state awareness, thus eliminating request conflicts at the source to reduce reconfiguration latency. Meanwhile, MEO followers perform distributed time-slot allocation based on local data to cut down inter-satellite link  communication rounds, and the stability of SE locks in long-term links to avoid ping-pong handovers, consequently reducing latency.}

%\cite{wang2022qos}\\
%\cite{liu2021research} \\
%\cite{fei2023research} \\
%\cite{zhang2022joint} \\ (Access control/Handover)\\
%\cite{chen2020correlated} \\ (Access control)\\
%===============================
\subsection{Cooperative Game-based Collective Optimization}
%===============================	
Cooperative games are able to resolve satellite-specific bottlenecks, such as high interference, limited resources at individual nodes, and coverage blind spots. These issues cannot be resolved by any single participant acting alone.
In satellite networks, cooperative games focus on collaboration among entities to maximize collective utility, with subcategories including exact potential games\cite{wang2022qos,zhang2021potential,wang2020distributed} (where a global potential function aligns individual and group gains) and coalition formation games\cite{du2019coalitional,gao2023sum,gao2021files} (where entities form stable groups).  Such game strategies are particularly suited to handover coordination, interference mitigation, and resource pooling in satellite networks.

\subsubsection{Exact Potential Games}

The authors in \cite{wang2022qos} address QoS-centric handover for civil aviation in ultra-dense LEO networks, where high aircraft/satellite mobility causes frequent handovers. Multiple aircraft cooperate with interference-related peers to select target satellites, and the game is proven an exact potential game. Distributed best response and collaborative mixed handover strategy are proposed for guaranteeing civil aviation QoS while enhancing network utility.  
For uplink multibeam Satellite IoT (S-IoT) networks, \cite{zhang2021potential} transforms the sum capacity maximization problem into an interference avoidance problem, which is modeled as an exact potential game. S-IoT users collaboratively schedule access and allocate power, and the game has multiple NEs. An iterative algorithm converges to an NE, improving sum capacity and user fairness compared to conventional methods.
In cognitive satellite networks, \cite{wang2020distributed} first considers coupled channel access and power optimization under a distributed framework. Terrestrial cognitive users adjust access and power strategies, and the game is proven an exact potential game with at least one pure NE. Joint-Strategy Iteration is proposed to solve discrete power control and address optimal NE in two-dimensional strategy space. 

\subsubsection{Coalition Formation Games}
 
To improve air-to-ground spectrum utilization, \cite{du2019coalitional} proposes a coalitional graph game for air-to-air links. The proposed distributed algorithm approaches the performance of optimal exhaustive search with low complexity, demonstrating that air-to-air communication is an effective complement to air-to-ground systems.
For LEO-UAV integrated multi-tier computing networks, \cite{gao2023sum} splits optimization into two coalition formation games. The first one is that data nodes and UAVs form coalitions via Max-Satisfaction coalition formation games to minimize UAV data load. Also, UAVs and LEOs form groups via a many-to-one matching game.  
In LEO satellite-terrestrial integrated networks, \cite{gao2021files} optimizes file delivery via non-orthgonal multiple access (NOMA)-based coalition formation game. The coalition formation game uses a preference order and a best-response algorithm finds stable coalitions, minimizing total delivery and sharing costs while ensuring successful NOMA transmission.

\subsubsection{Other Cooperative Games}

\cite{kim2020cognitive} introduces solidarity values into cognitive satellite spectrum management, where satellite and terrestrial agents dynamically access idle bands. A two-step interactive game leverages synergy to balance marginalism and egalitarianism, outperforming existing protocols in spectrum efficiency and fairness.
To resolve resource mismatch in cascaded satellite-wide-area mobile base station downlinks, \cite{li2024resource} establishes a resource collaborative scheduling mechanism for cascaded downlinks based on cooperative game theory. The utility function of the Nash product is converted into a max-min problem, and the non-convexity issue is resolved through a convex transformation. The proposed scheme improves resource utilization and cascaded transmission rates by adapting to distinct scheduling systems and QoS constraints.
The study in \cite{kim2021bargaining} addresses spectrum allocation for high-speed railway communications in integrated terrestrial-satellite networks. The interaction between satellite and terrestrial networks is modeled using a weighted proportional gain and loss bargaining solution. The goal was to adaptively share limited spectrum resources by exploring mutual benefits between the two networks. Unlike rigid resource allocation schemes, this bargaining paradigm reaches reciprocal consensus under dynamically changing communication conditions.
For satellite communication systems with terrestrial agents, The study in \cite{li2021agent} proposes a dynamically optimal cooperation scheme based on stochastic processes and the optimal contract principle. Terrestrial agents act as spectrum sales agents, reselling bulk-allocated satellite spectrum to end users.  This work fills a gap in agent-based spectrum management by linking cooperative incentives to real-world market dynamics.
In multi-beam satellite communication systems for mobile terminals, \cite{hu2020multi} develops a cooperative multi-agent deep reinforcement learning (MADRL) framework to address bandwidth allocation challenges. Each beam is treated as a cooperative player, aiming to satisfy traffic demands with flexible payloads. The cooperative game leverages inter-agent synergy to enhance transmission efficiency while maintaining low complexity. 
For satellite transponder power control, \cite{wan2021study} establishes a discrete cooperative game model to coordinate competing user terminals. The cooperative game decomposes the power allocation problem into a nonlinear optimization problem, with its Karush-Kuhn-Tucker conditions analyzed and solved both analytically and numerically. The resulting optimal power control scheme allocates power based on channel gain and power-noise ratio, minimizing interference and maximizing transponder throughput. 

%\cite{liu2020spectrum} \\%10
%\cite{chamberlain2024spectrum}\\%11
%\cite{wang2020novel} \\%12
%\cite{wang2021game}\\%13
%\cite{yan2022electromagnetic}\\%14
%\cite{zhang2020auction} \\%15
%\cite{chamberlain2024facilitating} \\%16
%\cite{wen2024hierarchical} \\%17
%\cite{kim2020cognitive} \\%18
%\cite{li2024game} \\%19
%\cite{liu2022spectrum} \\%20
%\cite{yuan2023game} \\%21
%\cite{du2019coalitional} \\%22
%\cite{kim2021bargaining} \\%23   
%\cite{xu2023adaptive} \\%24
%\cite{xu2022hierarchical} \\%25
%\cite{xu2022ubiquitous} \\%26
%\cite{li2019spectral} \\%27
%\cite{cheng2023satellite} \\%28
%\cite{smith2021ran} \\%29
%\cite{zhang2020vickrey} \\%30
%\cite{yang2023lightweight} \\%31
%\cite{li2021agent} \\%32
%\cite{li2019spectrum} \\%33
%\cite{tun2024joint} \\%34
%\cite{han2023anti} \\%35
%\cite{qin2021joint} \\ (Jointly spectrum allocation, power control, horizontal coordinate)\\%36
%\cite{hu2020multi}\\%37
%\cite{gao2023sum} \\%38
%\cite{li2024resource}\\%39
%\cite{liu2024demand} \\ %time alllocation40
%\cite{jiang2024game} \\%41
%\cite{jia2020joint}\\(Access and data backhaul of remote area users)\\%42

%===============================
\subsection{Auction: Incentive-Compatible Resource Trading}
%===============================	

Auctions like Vickrey and Vickrey–Clarke–Groves (VCG) enforce truthful bidding, where the winner pays the second-highest bid or compensates others for opportunity costs, eliminating incentives to lie about resource value. Satellite networks rely on decentralized participants that act selfishly—without proper incentives, these players may submit false bids or underperform. Therefore, auction mechanisms are non-negotiable for satellite systems, where a single dishonest bid could waste scarce spectrum or disrupt primary services. Specifically, auction mechanisms ensure truthful bidding and efficient resource allocation in satellite communication networks, where an auctioneer  (e.g., primary satellite network) allocates spectrum/relays to bidders  (e.g., terrestrial IoT clusters, UAVs) in exchange for services (e.g., cooperative relaying).

The study in \cite{wang2021game} addresses bandwidth allocation for discontinuous frequency bands in cognitive satellite-terrestrial networks, which is formulated as a bandwidth auction game. Specifically, terrestrial cognitive users (bidders) compete for discontinuous idle spectrum bands, with the auctioneer (satellite/network controller) ensuring allocations meet individual user bandwidth requirements, user priorities, and total bandwidth limits. The game is proven to have at least one pure-strategy NE and an improved best response learning algorithm based on exploration and exploitation was proposed. 
The authors in \cite{zhang2020auction} design auctions for hybrid satellite-terrestrial IoT networks, in which the primary S-IoT network allocates spectrum to T-IoT cluster heads in exchange for cooperative relaying. 
For lightweight LEO downlink transmitters, \cite{cheng2023satellite} proposes distributed dynamic auctions, in which ground stations periodically bid for sub-channels, and LEO satellites allocate resources based on bids.
In cognitive hybrid satellite-terrestrial overlay networks, \cite{zhang2020vickrey} uses a one-shot distributed Vickrey auction for secondary relay selection, in which relays  compete for selection, using decode-and-forward or amplify-and-forward (AF) protocols, with NOMA in the second transmission phase. The auction outperforms conventional relay selection, improving total capacity.
To address security threats in SAGIN spectrum sharing, \cite{yang2023lightweight} proposes a blockchain-based sequential Vickrey auction, in which the primary satellite allocates spectrum to UAVs that relay satellite signals. The blockchain ensures privacy and trust, and the auction solves multi-relay selection while maximizing primary user throughput.

%=============================
\subsection{Non-Cooperative Game-based Independent Decision-Making}
%=============================
Non-Cooperative game-based entities act selfishly to maximize individual utility, which can be used for power control, spectrum sensing, anti-jamming, and admission control in satellite networks.

In downlink CSTNs, \cite{chen2020correlated} investigates distributed power control considering primary satellite link QoS, in which users adjust power to balance transmission rate and mutual interference. 
To resolve satellite-terrestrial spectrum conflict, \cite{liu2020spectrum} models satellite users as secondary users (SUs) and a terrestrial user as the primary user (PU). SUs cooperate to sense PU channels and determine sensing ratios via game theory, maximizing SU throughput without disrupting PU communication. 
\cite{chamberlain2024spectrum} focuses on spectrum sharing between Earth Exploration Satellite Service (EESS) users (passive microwave radiometers) and commercial 5G users. To evaluate commercial users’ incentive to utilize shared spectrum (despite EESS preemption), the authors propose a joint queuing and game-theoretic model. Commercial users (players) decide whether to access shared spectrum based on preemption risks and pay-as-you-go admission fees set by a provider. The results show that while increasing commercial user participation lowers incentives for new users, the socially optimal state aligns with the profit-maximizing fee. 
For maritime area detection in cloud-edge collaboration satellite networks, \cite{li2024game} leverages a non-cooperative game to optimize task offloading ratios and minimize total delay. 
For cognitive satellite networks, \cite{wang2020novel} splits optimization into multi-channel access and power control problems. Strategies differ for users inside/outside fixed-satellite station protection distance, and each game has at least one NE. The Multi-channel Access and Power Optimization algorithm converges to NEs, improving spectrum efficiency.
In space-air-ground integrated networks with passive sensor incumbents, \cite{chamberlain2024facilitating} models commercial users (choosing primary/post-paid or secondary/free tiers) and a provider (setting priority fees), where all users prefer the primary tier. 
For 6G IoT, \cite{yuan2023game} proposes an energy-efficient spectrum sharing game. Providers and demanders use distributed reinforcement learning to select strategies with smart contracts executing transactions. The game converges to NE, reducing energy consumption/latency and improving fairness.
\cite{li2019spectral} explores game-theoretic spectrum pricing and auctions for dynamic satellite spectrum allocation, addressing broadband satellite service demands (e.g., satellite IoT, cloud services). Specifically, interference pricing in multi-beam satellite systems, differential pricing for heterogeneous spectrum as well as congestion pricing for non-GEO satellites are considered. 
For satellite systems with heterogeneous user preferences, \cite{li2019spectrum} proposes a differential pricing scheme based on the Hotelling model (a non-cooperative game framework). Existing spectrum pricing methods often adopt terrestrial network models, ignoring satellite-specific characteristics (e.g., long, variable link distances). The adopted Hotelling game models user choices between satellite spectrum bands with different pricing and link quality, with the satellite setting differential prices to maximize spectrum efficiency. Numerical results show that the scheme outperforms uniform pricing in matching user preferences to spectrum bands and improving overall utilization, highlighting its adaptability to satellite link variability.
To mitigate satellite IoT (SIoT) jamming, \cite{han2023anti} modeled satellites and ground cells (players) selecting non-jammed satellites for coverage swaps and optimizing beam angles. The game maximizes satellite cluster transmission rate during jamming, maintaining high achievable rates.
In LEO/MEO two-layer networks, \cite{zhang2021game} uses the super-modular game theory to prove NE existence for dynamic power allocation. LEOs optimize power via Newton iteration, and MEOs control penalty factors. 
For satellite transponders, \cite{wan2023differential} establishes a differential game for user power control. 
In satellite-vehicular networks (SVNs), \cite{zhang2023rate} proposes a non-cooperative game to incentivize vehicles to report CSIT and participate in power allocation, minimizing outage probability and maximizing system weighted sum rate.
For NOMA satellite multi-beam networks, \cite{wang2020admission} models admission control and power allocation as a super-modular game. A two-stage dynamic access process and iterative power allocation algorithm maximize supported users while ensuring QoS.

%=============================
\subsection{Matching Game-based Stable Bilateral/Multilateral Pairing}
%=============================
 
Non-cooperative games and matching games both address distributed decision-making, yet they tackle diametrically opposed satellite challenges. Non-cooperative games are applicable to resource competition where nodes act independently\cite{li2024game,wang2020novel,chamberlain2024facilitating,yuan2023game,li2019spectral}, which is most suitable when satellite nodes vie for shared, indivisible resources (e.g., transmission power or a single spectrum band), and cooperation is impractical.
Matching games are suited to stable pairings that avoid frequent reallocations. This game mechanism excels when two independent sets (e.g., PUs and SUs, LEO beams and users) require long-term and low-overhead pairing\cite{liu2022spectrum,tun2024joint,qin2021joint,jia2020joint,li2025game}. This is crucial for satellite networks, as reallocations may waste time and spectrum resources.

In cognitive satellite networks, \cite{liu2022spectrum} proposes a stable matching algorithm for spectrum allocation, where PUs and SUs (players) negotiate independently to form matches. The algorithm outperforms greedy/random matching in utility, sum rate, throughput, and spectrum utilization, with low complexity for dynamic networks.
\cite{tun2024joint} addresses joint UAV deployment and resource allocation in THz-assisted MEC-enabled integrated space-air-ground networks. To solve the mixed-integer nonlinear programming problem, the authors decomposes it into four sub-problems, one of which uses a matching game for THz sub-band assignment. UAVs and THz sub-bands are matched based on channel quality and UAV task requirements, ensuring efficient use of THz bandwidth. The matching game’s stable solution avoids sub-band conflicts and maximizes UAV task offloading efficiency. 
In 6G heterogeneous IoT networks, \cite{qin2021joint} transforms the resource allocation sub-problem into a many-to-one spectrum matching game. UAVs prioritize subchannels based on channel gain and interference, and subchannels prioritize UAVs based on traffic demands. This matching game is coupled with power control to maximize the system’s uplink sum rate.  
For SAG networks in remote areas, \cite{jia2020joint} proposes two matching algorithms. The first one is a restricted three-sided matching algorithm for users, HAPs, and LEOs. The second one is a two-tier algorithm adapting to dynamic HAP-LEO connections.  
In order to balance convergence time and accuracy of the resource allocation in LEO/GEO mega hybrid constellations, \cite{li2025game} proposes a two-layer matching game. The upper layer matches LEO beams to LEO users via a distributed beam matching learning algorithm, and the lower layer optimizes LEO beam power via a game solved by a dynamic power allocation logarithmic learning algorithm.

\subsection{Lessons Learned}
Table~\ref{tab:offloading1} provides a summary of incentive mechanisms for satellite communication resource allocation in satellite networks. The existing studies extensively leverage game-theoretic mechanisms to optimize communication resource management in satellite networks, encompassing critical resources such as spectrum\cite{xu2023adaptive}, bandwidth\cite{xu2022hierarchical},  data rate allocations\cite{xu2022ubiquitous}  and inter-satellite links\cite{liu2024demand}. These mechanisms provide decentralized frameworks where diverse players (e.g., LEO/GEO satellites, terrestrial BSs, UAVs, end users, and even adversarial teams) make strategic decisions. The objective is to optimise utility metrics related to communication performance, including throughput, latency, fairness, interference resilience, and profitability. 

However, notable limitations persist in these approaches. First, many rely on iterative algorithms (e.g., backward induction for Stackelberg Equilibrium, gradient descent for pricing optimization, or coalition merge-and-split processes) that require multiple rounds of information exchange between players to reach equilibrium\cite{zhang2022joint,yan2022electromagnetic,wen2024hierarchical,xu2023adaptive,xu2022hierarchical,xu2022ubiquitous,liu2024demand}. This imposes high communication overheads, particularly over resource-constrained inter-satellite links or dynamic satellite-terrestrial channels, where frequent data transmission can degrade real-time service quality. Second, satellite networks exhibit inherent high dynamism. LEO satellites’ rapid orbital motion shifts constellation topology, while traffic demands fluctuate unpredictably. As a result, equilibrium points computed using outdated local state information often become suboptimal or unstable, failing to adapt to real-time network changes.

Promising solutions to address these gaps emerge from integrated architectures such as software-defined satellite networks or cloud-edge-satellite collaboration, which can aggregate global network state across constellations, reducing coordination rounds and minimizing inter-satellite communication overheads. Additionally, integrating edge computing enables localized, low-latency resource adjustment, allowing players to update strategies based on fresh, proximity-based state information rather than outdated global data. These architectures bridge the gap between the decentralized flexibility of game theory and the dynamic adaptability required for practical satellite network deployment.
%=============================================================================
\section{Computation Offloading}
\label{section:com_offload}
%=============================================================================

\begin{table*}[ht]
\caption{Summary of Incentive Mechanisms For Computation Offloading.}
\centering
\renewcommand{\arraystretch}{1.2}
\begin{tabular}{|l |p{3.4cm}| p{3.7cm}| p{4.6cm}| p{3.2cm}|}
\hline
\textbf{Ref.} & \textbf{Scenario} & \textbf{Problem focus} & \textbf{Methodology} & \textbf{Players}\\
\hline
\cite{lin2023leo} & Tactical satellite MEC & Offloading strategy and resource pricing & Stackelberg game & LEO satellites, UAVs, ground MEC servers, tactical users\\
\hline
\cite{xiang2024edge} & Satellite MEC & Offloading strategy & Stackelberg game & LEO satellites, UAVs, ground users\\
\hline
\cite{chai2024computation} & Satellite IoV & Resource allocation and pricing & Stackelberg game & Service provider, vehicles\\
\hline
\cite{kim2024hierarchical} & Aerial computing IoT & Offloading strategy and resource pricing & Stackelberg evolutionary game & HAPs, UAVs\\
\hline
\cite{wang2024two} & Space-sea integrated network & Offloading strategy and resource pricing & Stackelberg and Bargaining games & LEO satellites, MASSs, AUVs\\
\hline
\cite{peng2024cloud} & Air-ground integrated IoT & Offloading strategy and resource allocation & Potential game & Floating airships, UAVs, IoT devices\\
\hline
\cite{gao2023game} & Satellite MEC & Offloading strategy and resource allocation & Stochastic potential game & LEO satellites, ground users\\
\hline
\cite{zhang2024dogs} & SAGIN IoT & Offloading strategy &  Weighted potential game & IoT devices\\
\hline
\cite{sun2021game} & Multi-satellite system & Observation task assignment & Potential game & LEO satellites\\
\hline
\cite{wang2019game} & Satellite MEC & Offloading strategy & Non-cooperative game & Edge devices\\
\hline
\cite{qiao2024orbit} & Satellite MEC & Offloading strategy & Stochastic non-cooperative game & LEO satellites\\
\hline
\cite{li2024joint} & Hybrid satellite MEC & Offloading strategy and resource allocation & Coalition formation and non-cooperative games & LEO satellites, cloud server, users\\
\hline
\cite{wang2024dynamic} & Satellite MEC & Resource pricing & Non-cooperative game & LEO satellites, ground edge server, users\\
\hline
\cite{yu2024computation} & Satellite MEC & Offloading strategy and resource pricing & Many-to-one matching game & LEO satellites, ground users\\
\hline
\cite{fang2022matching} & Satellite MEC & Offloading strategy and resource allocation & Many-to-one matching game & LEO and GEO satellites\\
\hline
\cite{jin2020double} & LEO satellite system & Resource pricing & Double-auction model & LEO satellites, space station\\
\hline
\cite{huang2024profit} & Satellite MEC & Resource pricing & Reverse auction model & LEO satellites, ground BS\\
\hline
\cite{cheng2024energy} & Satellite MEC & Offloading strategy and resource allocation & Lyapunov and stochastic optimization & LEO satellites, ground users\\
\hline
\cite{zhang2024cost} & Satellite IoT & Offloading strategy and resource allocation & Successive convex approximation & LEO satellites, IoT devices\\
\hline
\cite{yang2024joint} & Satellite MEC & Offloading strategy & Deep DRL & LEO satellites, ground users\\
\hline
\end{tabular}
\label{tab:offloading}
\end{table*}

Computation offloading in the satellite networks often requires jointly deciding the offloading destinations 
(i.e., one-device, ground stations, edge servers, UAVs, or satellites, etc) and allocating 
the right amount of limited computing resource under appropriate pricing. 
However, the high dynamics of network conditions and the heterogeneity of servers and user devices 
render the fully centralized offloading optimization impractical due the scalability, latency, fairness, and high overhead communication concerns. The game-theoretic approaches address these issues by enabling the decentralized, strategic decision-making frameworks in which the users, edge servers, and satellites can cooperate or compete to make their optimal offloading decisions independently. For example, the non-cooperative game mechanisms allow the multiple users 
compete in selecting the right satellites for offloading their tasks such that their own utilities can be maximized. 
Moreover, the auction mechanisms can let the users and satellites to negotiate the resource computing price 
as the buyers and sellers, respectively. 
In this section, we will review existing game-theoretic satellite computation offloading approaches 
in the following three categories: game-based, auction-based, and other mechanisms. 
Table~\ref{tab:offloading} provides a summary of key studies reviewed in this section.

\subsection{Game-based Mechanisms}

\begin{figure}
    \centering
    \includegraphics[width=\linewidth]{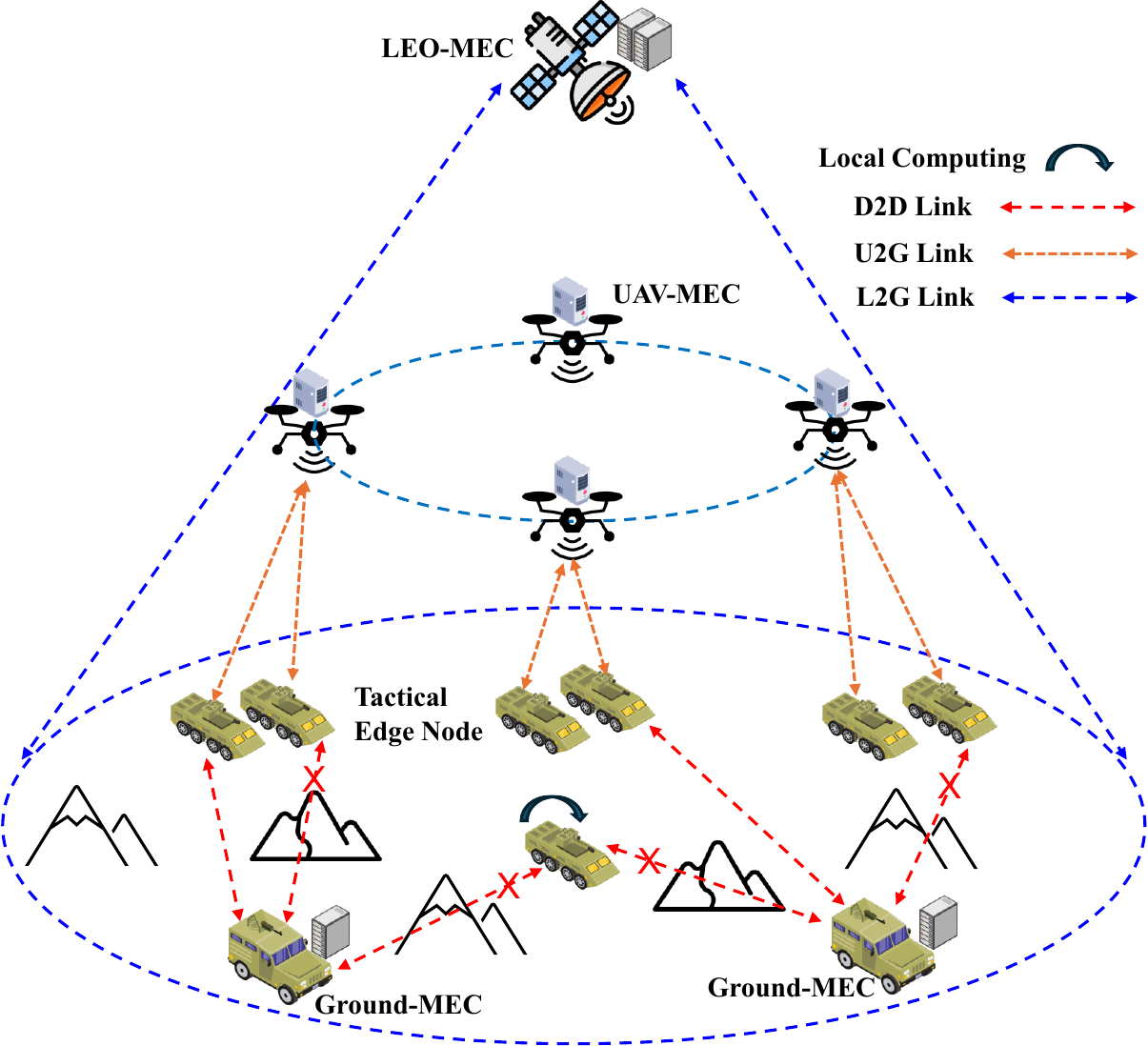}
    \caption{A tactical ad-hoc mobile edge computing (MEC) system~\cite{lin2023leo} allows the edge tactical nodes to offload their computation tasks to the ground-MEC, unmanned aerial vehicle (UAV)-MEC, low-Earth orbit (LEO)-MEC servers via the device-to-device (D2D), UAV-to-ground (U2G), and LEO-to-ground (L2G) links, respectively.}
    \label{fig:tactical-satellite}
\end{figure}

%%Stackelberg game %%%%%%%%%%

\subsubsection{Stackelberg Games}

The studies in~\cite{lin2023leo,xiang2024edge,chai2024computation,kim2024hierarchical,wang2024two,gao2024edge,fan2024graph} formulate various Stackelberg game models in which the mobile edge computing (MEC) servers 
(e.g., the satellites and ground edge servers) and the users play as the leaders and followers 
to optimize their resource allocation and offloading decisions. 
For instance, the authors in~\cite{lin2023leo} investigate a tactical ad-hoc MEC system as illustrated in Fig.~\ref{fig:tactical-satellite}, in which the ground-MEC, UAV-MEC and LEO-MEC satellite servers are incentivized to provide the tactical edge nodes with the heterogeneous computing resources using an SAGIN under the harsh communication conditions of the tactical environments. 
Specifically, a multi-leader and multi-follower Stackelberg game (MLMF-SG) is constructed to model the interactions between the edge nodes and MEC severs, in which the MEC servers act as the leaders to set the appropriate price for selling their computing resources with the objective of maximizing their total payment reward received from the edge nodes. Given the computing resource prices from the MEC servers, a follower subgame is formulated as a multi-mode computation task offloading game for the edge nodes to select their optimal offloading links and task partition ratios such that their task execution delay and energy consumption can be minimized. The existence of SE in the proposed MLMF-SG is theoretically proved. 
Moreover, the follower subgame is proved to be an exact potential game (EPG) with the NE 
(i.e., suboptimal offloading strategy). Finally, a hierarchical distributed iterative algorithm is proposed to find the MLMF-SG's SE solution via multiple iterations, in which the MEC servers' computing resource price and edge nodes' offloading strategies are updated in different time scales. The simulation results show that the proposed algorithm can help the edge tactical nodes reduce their total energy consumption by up to $5.1\times$ and improve their utility by up to $5.2\times$, 
compared with the one-device, full-ground offloading, and random-ratio offloading approaches. 
However, the effectiveness of the proposed approach is not well justified due to lack of performance comparison with 
strong competitors such as the game-based, optimization-based and learning-based approaches.

%In~\cite{xiang2024edge} investigates the offloading problem in a three-layer edge computing network 
%which consists of multiple ground user, UAVs and LEO satellites in the first, second and third layers, respectively.
%Each user can offload its computation tasks to the MEC servers deployed on the UAVs and satellites. 
In~\cite{xiang2024edge}, the MEC servers (i.e., the UAVs and LEO satellites) can forward a portion of the tasks 
received from the ground users to their neighboring servers. In the formulated Stackelberg game model, the ground users act as the leaders to adjust their offloading strategy to minimize their task execution delay and energy consumption. As the followers, 
the MEC servers adjust their forwarding percentage to minimize their energy consumption based the users' offloading strategies. The existence of NE in the formulated game is also proved. Then, an iterative offloading algorithm, namely NEIO-SG is proposed to the find the game's NE solution i.e., the optimal offloading and forwarding percentage strategies of the users and servers, respectively. The simulation results indicate that the proposed NEIO-SG can reduce the total system latency by 
about 13\% and the server's energy consumption by about 35\%, compared with the random offloading, centralized offloading, reinforcement learning (RL)-based offloading and genetic-based optimization approaches.

Moreover, the study in~\cite{chai2024computation} considers an integrated satellite-terrestrial Internet of Vehicles 
(IST-IOV) system managed by an edge computation service provider (ECSP) that is responsible for collecting the computation tasks from the vehicles and allocating the computing resources of the roadside units (RSUs) and LEO satellites for task processing.
The interaction between the ECSP and vehicles is modeled as a cooperative Stackelberg game in which as the leader, 
the ECSP aims at determining the computing resource prices of the RSU and satellites and power of the RSU for transmitting the tasks to the satellites such that the ECSP's utility (i.e., the difference between total payment reward and communication energy usage cost) can be maximized. Given the computing resource prices set by the ECSP, the vehicles participate the game as the followers to adjust to their task offloading ratio with the objective of maximizing their utility (i.e., the difference between their offloading satisfaction plus social interaction and their communication energy usage cost).     
Then, a dynamic cooperative game algorithm, called DCGT-IOV is proposed to find the NE of the formulated game via 
multiple iterations. The simulation results indicate that the proposed algorithm can reduce the required convergence time 
by up to about 1.4x, compared with the exact potential and congestion game frameworks. Moreover, it also increases the total utility of the ECSP and vehicles by up to about 73.3\%, compared with the five heuristic and game-based offloading and resource allocation approaches proposed in~\cite{zhao2020novel,lim2021dynamic,li2022joint}.

In stead of using the LEO satellites in~\cite{lin2023leo,xiang2024edge,chai2024computation}, the hierarchical aerial computing system proposed in~\cite{kim2024hierarchical} employs a high-altitude powerful platform (HAP) to process a potion of the computation tasks which the UAVs receive from the IoT devices in its coverage. 
%Then, the UAV can further forward a certain portion of its task to one high-altitude powerful platform (HAP) equipped with a %powerfull computing capability to minimize the task processing delay. 
%Specifically, the HAP charges the UAVs with a specific price per unit of data bits for its computation service.
Specifically, a Stackelberg evolutionary game (i.e., a combination of the classical Stackelberg and evolutionary games) is constructed to depict the interactions between the HAP and UAVs. As the leader, the HAP determines its computation price according to the evolutionary learning process. As the followers, the UAVs make their offloading decision (i.e., task offloading ratio) according to the multi-objective bargaining solution which adaptively responses to the price set by the HAP. 
Then, an interactive learning algorithm is proposed for the HAP and UAVs to determine their optimal pricing and offloading strategies which provide well-balanced system performance between contradictory offloading service requirements. 
The evaluation results show the proposed algorithm can improve the system throughput and resource utilization up to about 10 \%, and 15 \%, respectively, compared with the four existing hierarchical aerial computing algorithms.

In addition, the study in~\cite{wang2024two} considers a space-sea integrated network in which multiple autonomous underwater vehicles (AUVs) are deployed on the seabed to monitor the marine environment and then offload a portion of their collected data (e.g., marine biological images and intelligence reconnaissance videos) to the maritime autonomous surface ships (MASSs) for processing. Upon receiving the workloads from the AUVs, the MASSs process a portion and upload the remaining workloads to the LEO satellites. Two game models are formulated to efficiently find the optimal task offloading strategies and resource pricing policies for the UAVs, MASSs and satellites in a distributed way. Specifically, the interactions between a pair of AUV and MASS is modeled as a Stackelberg game in which the MASS acts as the leader to determine its pricing strategies for task processing and reward acquisition to maximum its revenue. Subsequently, as the follower, the AUV determines its offloading ratio to maximum its utility function (i.e., the difference between its level of satisfaction in performing computation offloading and the total cost paid to buy the MASS's computation resource and energy usage of its local computation processing). On the other hand, the interaction between the MASSs and satellites is modeled as a Bargaining game where the MASS is the buyer aiming purchasing the computation services for task offloading from the satellites as the seller. The MASSs and satellites aim to determine their offloading ratios and pricing strategies, respectively, with the objective of maximizing their utility (i.e., difference between their received reward and energy usage cost). Then, a two-tier task offloading (TTA) algorithm is proposed to obtain the equilibrium solutions of two formulated games based on the pricing-based incentive mechanisms. 
The simulation results show that the proposed algorithm can improve the utility of the AUV, MASS, and satellite by up to about 
$1.7\times$, $3.1\times$, and $1.4\times$, respectively, compared with the eight offloading and pricing algorithms proposed by the previous studies.

%In the scenarios with existence of three layers, i.e., LEO satellites, UAVs, and ground users, a three-layer Stackelberg game can %be proposed as shown in \cite{gao2024edge}. Therein, the LEO satellite with its strong computing capability and wide coverage %plays a leader, while the UAVs are the sub-leaders and the users are the followers. However, how to find and prove the existence %and uniqueness of the Nash equilibrium in such a game may be challenging. 

%The satellites are limited in computation and storage resources, and they may need to cooperatively execute tasks as presented in %\cite{fan2024graph}. However, this is feasible as the satellites are within in an closed-enough region. Importantly, the rewards %obtained from the task execution are proportional to the contributions of the satellites. To address this issue, the Shapley %value \cite{roth1988shapley}, a well-known concept in cooperative game theory, can be used. 

\subsubsection{Potential Games}
In stead of the Stackelberg game-based approaches in~\cite{lin2023leo,xiang2024edge,chai2024computation,kim2024hierarchical,wang2024two,gao2024edge,fan2024graph}, 
the studies in~\cite{sun2021game,peng2024cloud,gao2023game,gao2023game,chen2025game,gao2023multi} propose the potential
game-based approaches in which the participating incentives of the game players (e.g., the users, UAVs, satellites) are aligned through a shared potential function (e.g., the average task execution delay, weighted sum of users' utilities). 
For example, the authors in~\cite{peng2024cloud} consider an air-ground integrated IoT network in which the ground IoT devices can offload their computation tasks to the UAVs and floating airships equipped with the MEC servers via the device-to-deice (D2D) or cellular links. The joint offloading and resource allocation problem is formulated as a mixed-integer nonlinear programming (MINLP) formulation to find the optimal task offloading decision and communication channel selection for the IoT devices as well as the optimal resource allocation for the UAVs and airships. The objective function aims at minimizing the task processing latency and device energy consumption, subject to the constraints on the computing resource capability of the UAVs and airships. Solving the formulated MINLP is challenging due to its nonlinear and non-convex nature. 
Thus, it is decomposed into the separate offloading and resource allocation subproblems. 
Then, a distributed computing offloading and resource allocation (DCORA) algorithm is proposed 
to jointly solve these two subproblems. 
Specifically, the offloading subproblem is modeled as a completely potential game with the proved existence of the NE solution. Thus, the IoT devices can exchange the information with each other to learn their optimal offloading decision and channel selection via multiple iterations. The resource allocation subproblem is formulated as an unconstrained convex optimization 
which can be solved using a Karush-Kuhn-Tucker (KKT) condition-based algorithm. The simulation results show 
that the proposed DCORA can reduce the required convergence time and average task processing delay by about 13.3\% and 10\%, respectively, compared with the four popular offloading and resource allocation approaches.

Similar to~\cite{peng2024cloud}, the authors in~\cite{gao2023game} also formulate a joint offloading and resource allocation problem as an optimization problem to find the optimal offloading strategy, communication channel selection, and power control for the users to offload their tasks to the satellites. The objective is to minimize the users' computation overhead which is defined as a weighted sum of the task execution latency and energy usage. Due to its NP-hard nature, solving the formulated optimization is challenging. Thus, it is further decomposed into two subproblems for power control and offloading strategy. 
To solve these two subproblems, the interactions between the users are modeled as a stochastic potential game 
in which the user independently adjusts its offloading strategy and power level to minimize its computation overhead via multiple stages. Then, a game-based approach consisting of the synchronous log-linear learning-based power control algorithm and joint offloading strategy and power optimization algorithm based on stochastic learning automate is proposed for the users to reach the NE of the constructed game. The simulation results show that the proposed approach can reduce the response latency, network overhead, and energy consumption by up to about 2.1x, 8.2x, 1.6x, respectively, compared with the five offloading and power control baseline approaches.

The study in~\cite{zhang2024dogs} introduces a generic computing offloading framework for 
the IoT devices to make the optimal offloading decisions in a SAGIN framework. 
Specifically, each IoT device can offload its computation tasks to an UAV or a LEO satellite. 
The offloading problem is formulated as an optimization problem with the objective 
of minimizing the system cost as a weighted sum of the task execution delay and device's energy consumption, 
while satisfying the maximum execution latency requirements. 
A stochastic game is constructed to model the interactions between the devices for jointly determining 
their optimal offloading decisions.
Specifically, the constructed game is proved to be a weighted potential game with at least one NE. 
Then, a multiagent entropy-enhanced stochastic learning (MESL) algorithm is proposed to allow the devices reach the game's NE in a fully distributed manner without the need of information exchange between the devices.
Moreover, the proposed MESL algorithm adopts the entropy of decision probability for each device 
to accelerate the convergence of learning their optimal offloading decisions. The experimental results indicate that the proposed MESL can reduce the average cost of the system and energy consumption by up to approximately $5.1\times$ and $1.8\times$, respectively, compared to the five heuristic-based and deep reinforcement learning (DRL)-based offloading algorithms.

Differently, the study in~\cite{sun2021game} investigates a multi-satellite system in which each satellite selects to execute a set of clients' observation tasks located within its field of regard (FOR). The multi-satellite task assignment is formulated as an unconstrained optimization problem to minimize the total observing and swaying cost of all satellites while maximizing the total number of executed tasks. However, solving the formulated problem in a centralized manner suffers from the concerns of scalability and robustness. Therefore, a fully distributed algorithm is proposed to achieve the optimal task assignment via the interactions between the satellites in a potential game consist multiple autonomous and rational players. 
Specifically, each satellite interacts with its neighbors to select its task set which maximizes its utility via 
multiple iterations. The proposed algorithm is proved to be converged with probability 1 to the NE which corresponds to near-optimal task assignment solution. The simulation results justify that the proposed algorithm can achieve the best solution efficiency, compared with the four existing distributed learning algorithms. 
However, it requires a higher computation time for convergence.

%\cite{wondmagen2025review}: Do not review because it is a review paper, not a technical research paper.

Moreover, the study in~\cite{chen2025game} investigates the offloading problem in an SAGIN in which the ground users aim at making the optimal task offloading decisions (i.e., local computing or offloading to the LEO satellites or base stations) to minimize their energy consumption. 
%This is an NP-hard problem that is challenging to be solved with centralized algorithm due to the high competition among the users on the limited resources and the scalability issue. 
The interaction between the users is modeled as a game in which each user makes its local offloading decision to minimize its energy cost. The formulated game is then proved as an exact potential game with an NE where no user can decrease its overhead 
by changing its decision independently while other users' decisions do not change. 
Simulation results show that the proposed game approach can reduce the total offloading cost by up to $85.2$\%, 
compared with the case the users locally execute the tasks. The results further show the scalability of the proposed game approach when demonstrating that it can work up to $90$ users. However, the time-varying channels and the task execution latency are not considered.

Similar to \cite{chen2025game}, the authors in~\cite{gao2023multi} also formulate a potential game to model the interactions between the ground users. The goal of the users to make the optimal offloading decisions (i.e., local computing or offloading to the LEO satellites) such that their task execution latency and energy consumption can be minimized. 
The dynamic channels between the users and the satellite are considered, which make this system model more realistic. Nevertheless, with the introduction of the dynamics of the channels, it is more challenging to find the NE of the game. 
To address this, the random learning~\cite{sastry2002decentralized} is used. 
Simulation results demonstrate that the use of the random learning enables the proposed game to converge in a small number of iterations. Moreover, it is able to reduce the user cost up to 30\%, compared to the potential game in which the Best Response~\cite{monderer1996potential} is used to find the NE. However, it can be better if the work accounts for the fairness among the users. For example, the Jain's fairness index can be used to quantify the fairness among the users as proposed in~\cite{wei2025network}.

\subsubsection{Non-Cooperative Games}

Unlike the above studies in~\cite{lin2023leo,xiang2024edge,chai2024computation,kim2024hierarchical,wang2024two,gao2024edge,fan2024graph,sun2021game,peng2024cloud,gao2023game,gao2023game,chen2025game,gao2023multi} which align the game players (e.g., ground users, UAVs, AUVs, and satellites) through the hierarchical decision structures in in the Stackelberg games or the shared functions in the potential games, the studies in~\cite{wang2019game,qiao2024orbit,wang2024dynamic} adopt non-cooperative game approaches, allowing players to make independent decisions to maximize their own utility without forming binding agreements.
For instance, the study in~\cite{wang2019game} considers a satellite edge computing system in which an edge device cannot offload its tasks a satellite at any time due to the intermittent terrestrial-satellite communication caused by satellite orbiting. 
In particular, the computation offloading problem is formulated as a non-cooperative game in which 
each device selfishly selects an optimal offloading strategy which determines the percentage of its tasks 
to the satellite such that its cost function including the average task response time and energy consumption 
can be minimized. The existence of the Nash equilibrium of the formulated game is theoretically proved. 
Then, an iterative algorithm is proposed to find the optimal computation offloading strategies for all devices 
via multiple iterations. Specifically, each device iteratively searches its best offloading strategy and computes 
its minimum cost function using the Lagrange multiplier approach. 
The simulation results show that the proposed iterative algorithm can reduce the average device cost by up to $36.6\times$, 
compared with the even offloading, random offloading, and only-local computing approaches.

In~\cite{qiao2024orbit}, an overloaded ordinary satellite can further offload a portion of its 
tasks received from the ground users to another idle satellite. A repeated stochastic non-cooperative game is formulated to model the interactions between the satellites, in which a head satellite gives the task offloading strategies to each ordinary with the objective of maximizing the overall energy utility of the entire satellite system based on its observation on the dynamic network conditions. The ordinary satellite only accepts an offloading strategy if it maximizes its own energy utility. Specifically, the energy utility of a satellite is defined as the remaining energy after accounting for the energy consumed for the data communication and processing. Then, a Lyapuno-based online data balancing algorithm is proposed to find the coarse correlated 
equilibrium for the formulated game. The simulation results show that the proposed algorithm can reduce the satellite energy  consumption by up to $61.49$\%, compared with the multi-hop offloading, only-local computing, and random offloading approaches.

Similar to~\cite{qiao2024orbit}, the study in~\cite{li2024joint,li2022game} considers a hybrid MEC-based LEO satellite network, the LEO satellite can also forward their received tasks to a ground cloud server, in stead of the neighboring satellites.  
%Each user needs to pay a certain monetary cost for purchasing the computation resource of the satellite for task processing. 
%An additional cost is charged for the communication energy usage if the satellite forwards its tasks to the ground cloud. 
An NP-hard optimization problem is formulated to jointly optimize the offloading decision-making, 
communication channel selection and transmit power control for the users, as well as the resource allocation for the satellites with the objective of minimizing the overall user cost, subject to the constraints on the user's maximum transit power and satellite's computation capability. Then, a two-level hierarchical game-based approach is proposed to efficiently find the optimal solutions. Specifically, the proposed approach decomposes the formulated problem into the offloading and resource allocation subproblems, in which the offloading subproblem is formulated as a hedonic coalition formation game for the users to determine their offloading decisions by jointly considering their own utility and the coalition utility in the upper level. 
In the lower level, under the user task offloading decisions, a non-cooperative game is established to determine the joint power control and channel selection for the users in each coalition. The existence of the NE of the formulated noncooperative game 
is proved. As such, the joint offloading and resource allocation problem can be efficiently solved. 
The simulation results show that the proposed game-based approach can reduce the average user cost, 
task execution delay, and task delay violation probability by up to about $7.1\times$, $3.7\times$, $34$\%, respectively, 
compared with the four offloading and resource allocation schemes~\cite{tang2021computation}.

Moreover, the authors in~\cite{wang2024dynamic} introduce a two-stage offloading model in which the ground user pays a certain resource purchasing price to offload  a part of its tasks to the satellite in the first stage. In the second stage, each satellite executes a portion of its received tasks and forwards the remaining tasks to a ground MEC server which employs a specific pricing strategy to sell its computation resource. A two-stage task pricing mechanism is proposed to model the offloading and pricing decision-making processes of the users and the ground server based on the differential game theory. 
Specifically, the proposed mechanism adopts Pontryagin's maximum principle~\cite{onori2015pontryagin} to construct the Hamiltonian functions of the offloading process to find the NE solution of the users' resource purchasing strategies and server' resource pricing strategy. Then, a Runge-Kutta-based algorithm is designed to provide the numerical results for justifying the optimal solutions of the proposed pricing mechanism. The effectiveness of the proposed mechanism is also verified by
a comparison between the numerical and system-level simulation results.

\subsubsection{Matching Games}
The studies in~\cite{yu2024computation,fang2022matching,wang2022computation} employs the matching game models in which 
the players (e.g., the users and satellites) are matched to each other according to their utilities or preferences.  
For example, the authors in~\cite{yu2024computation} formulate a many-to-one matching game in which multiple ground users can be matched to one satellite. Specifically, each user aims to select a satellite for offloading its tasks with the objective of maximizing its utility function which is designed as a weighted sum of the task execution satisfaction, execution time, 
and energy consumption. Meanwhile, each satellite can accept multiple task execution requests from the users and 
use a pricing strategy to determine the corresponding prices for the users. 
The utility function of the satellite is designed to maximize its received payment while minimizing its energy usage cost. 
First, a many-to-one matching algorithm is proposed for the users to determine their offloading decision based on their preference lists. Second, a two-stage dynamic game with complete information is developed to optimize the computing resource price of the satellites and the users' offloading volume. In the first stage, the satellites set their price. Then, the users determine their offloading volume accordingly in the second stage. 
This two-stage process is repeated until the users and satellites can reach the NE. 
The simulation results are presented to verify the convergence performance of the proposed algorithm. 
However, the superior of this study is not well justified due to lack of comparison with the existing studies. 
Similar to~\cite{yu2024computation}, the authors in~\cite{wang2022computation} also formulate a matching game to match the users with each satellite. Then, a coalition game is used that allows each user to join or leave coalitions to minimize its cost. The proposed game approach helps to reduce the user cost up to $15$\%, compared with the greedy scheme where each user selects the satellite with the strongest signal.

Moreover, in~\cite{fang2022matching}, the LEO satellites can offload their computing tasks to GEO satellites. A MINP optimization problem is formulated to find the optimal offloading decisions for the LEO satellites as well as the optimal communication and computing resource allocation for the GEO satellites with the objective of 
minimizing the task execution delay and LEO satellites' energy consumption. Due its NP-hard nature, solving the formulated optimization problem is challenging, especially the system includes a large numbers of satellites and communication sub-bandwidths. Thus, the offloading problem of LEO satellites is reconstructed as a two-side many-to-one matching problem 
in which each LEO satellite can only select one GEO satellites for offloading. Then, the GEO satellite can decide to accept or reject the task offloading from the LEO satellite according its own preference which is calculated based on the task execution delay and energy cost. Simulation results show that the proposed offloading algorithm can reduce the system overhead and 
running time by about $8.8$\% and $55.53$\%, respectively, compared with the heuristic task offloading algorithm proposed in~\cite{tran2018joint}.

\subsection{Auction-Based Mechanisms}

\begin{figure}
    \centering
    \includegraphics[width=\linewidth]{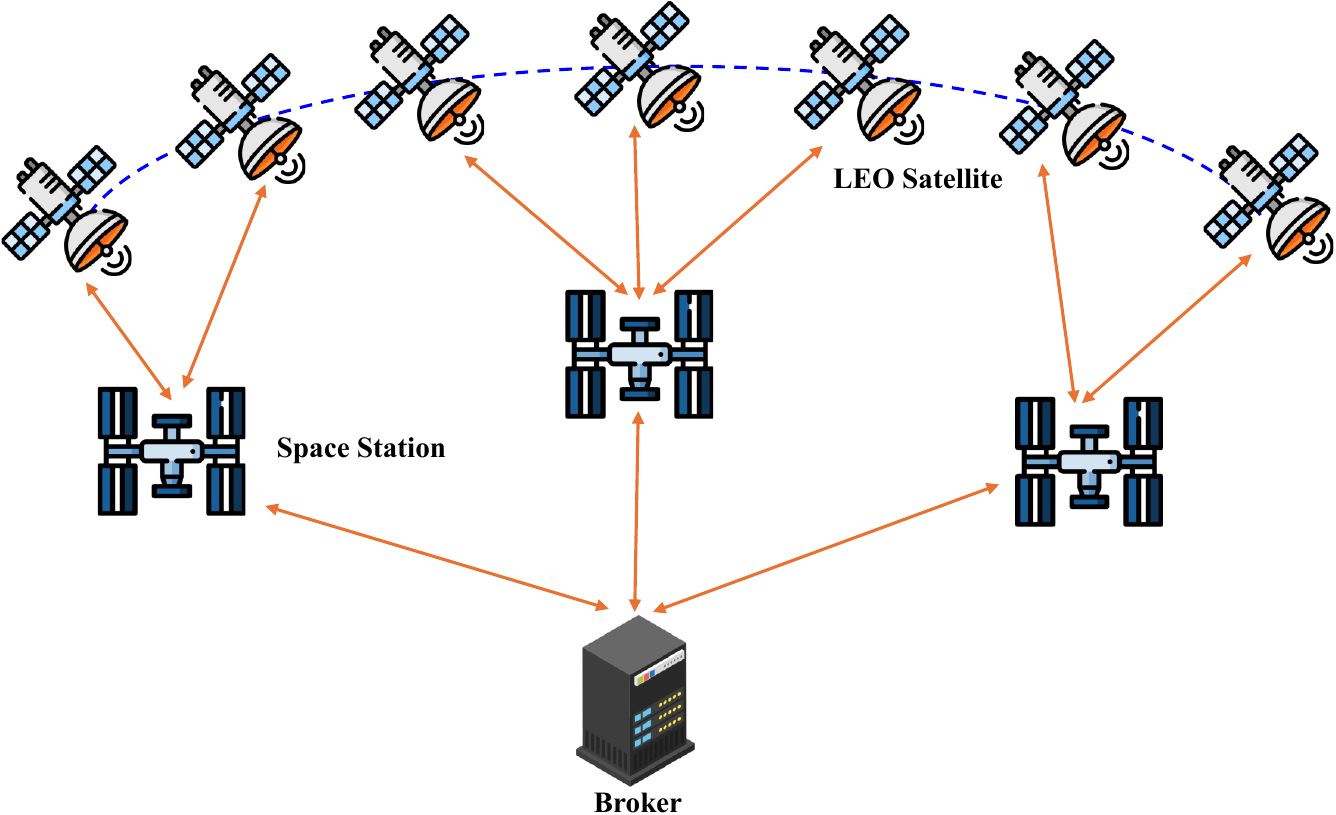}
    \caption{A space station-based LEO satellite system~\cite{jin2020double} consists of multiple LEO satellites which pay certain prices to offload their computation tasks to the powerful space stations for high-speed task processing.}
    \label{fig:space-station}
\end{figure}

The studies in~\cite{jin2020double,huang2024profit,li2021double} propose various auction-based approaches in which 
the bidding mechanisms or auction rules are employed for the participants to compete for the computing resources of the satellites and space stations. For instance, the authors in~\cite{jin2020double} consider a space station-based LEO satellite system 
as illustrated in Fig.~\ref{fig:space-station}, in which each satellite pays a certain price to offload its computation tasks 
to a powerful space station for high-speed processing. The computation resource pricing problem 
is modeled a double-auction model where the satellites and space stations play as the buyers and sellers, respectively, 
to submit their bidding and asking prices to a broker simultaneously. 
%Specifically, the satellite has a different biding price for each space station, depending on the station's quality of service to the satellite. 
Then, the broker decides the transaction price of every satellite and space station. 
%Both the satellites and space stations aim to find the optimal policies to set their bidding and asking prices, 
%respectively, . 
An experience-weighted attraction (EWA) algorithm is proposed for the market participants (i.e., satellites and space stations) the learn the optimal bidding/asking pricing policies such that their total revenue can be maximized. 
Specifically, the EWA algorithm combines the belief learning and reinforcement learning to allow the participants as the learning agents to judge their prices based on other participants' prices and their own past pricing experience, 
respectively, and then adjust their prices through multiple auction iterations. 
The simulation results demonstrate the proposed algorithm can enable a small-scale auction system 
consisting of 2 satellites and 2 space stations to reach the Nash equilibrium after 800 iterations.

In \cite{huang2024profit}, a reverse auction is formulated, in which the satellites provide the computing services to a ground BS. Specifically, as the sellers, the satellites submit their asking prices to the BS. 
Then, the BS selects the satellites for the task execution and propose the corresponding prices to maximize the total profit of the satellites. Here, the satellite profit is defined as the difference between the price and the resource cost for the task execution. Then, the satellite selection is solved by a dynamic programming algorithm, while the prices for the satellites 
are determined based on the asking prices of other satellites. The pricing is generally similar to that of the reverse Vickrey auction to guarantee the truthfulness and individual rationality of the auction. 
Simulation results show that guaranteeing the individual rationality motivates more satellites to join in the market 
that improves the task completion rate of the system up to $16$\%.

One feature of satellite networks is their wide coverage. This feature enables them to work as a central entity for resource trading between ground users and ground edge servers as presented in~\cite{li2021double}. Therein, a double auction is used that match each user (i.e., buyer) and each edge server (i.e., seller). The double auction is similar to the standard double auction. However, a critical price is set based on the sellers' asks, and edge servers with asks less than this price are selected to join in the auction. To solve the user-edge server matching, the FPTAS~\cite{ergun2002improved} as an approximate algorithm with a polynomial time is used. Meanwhile, the payments to the users are determined based on the critical price. By using the FPTAS and the critical price, the proposed double auction achieves computational
efficiency, individual rationality and incentive compatibility. 
However, by using the critical price, the number of edge servers joining in the resource trading decreases. As a result, the number of successful trades may decrease, and the users, i.e., buyers, may experience high payments.

\subsection{Other Mechanisms}

The studies in~\cite{cheng2024energy,zhang2024cost,wang2021profit} employ the optimization-based approaches 
to optimize the offloading and/or resource allocation in the satellite networks.
For example, the authors in~\cite{cheng2024energy} consider a satellite edge computing (SEC)-assisted satellite-terrestrial integrated network (STIN) in which an user device can offload its tasks to the satellites or a remote cloud center via the feeder satellite-ground links. A stochastic optimization problem is formulated to find the optimal solutions for the task offloading ratio and transmit power level of the user devices, and the computing resource allocation of the satellites and cloud 
with the objective of minimizing the long-term average of the overall task completion time, subject to the energy constraints of the LEO satellites and user devices. Solving the formulated optimization problem is challenging because 
it is NP-hard and requires the accurate future information about the channel state and stochastic packet arrivals. 
Thus, an online dynamic offloading strategy (DOS) is proposed to learn the optimal solution by 
leveraging the Lyapunov optimization theory to transform the original problem into multiple one-slot 
optimization problems. Then, the one-slot optimization problem is further decomposed into the task offloading and resource allocation subproblems which are convex and can be iteratively solved using the conventional optimization techniques. 
The simulation results show that the proposed DOS approach can reduce the average task completion time and task dropping rate 
by up to $2.1\times$ and $9.4\times$, respectively, compared with the four existing offloading strategies.

%The study in~\cite{zhang2024cost} proposes a 5G-based cost-effective hybrid computation offloading 
%(CE-HCO) framework which can enable the IoT devices in remote regions without terrestrial computing network infrastructure %offload their computation tasks the ground clouds and LEO satellites via the terrestrial-satellite terminals (TSTs) 
%equipped with multiple independent antenna apertures. 
Similarly, the offloading problem in~\cite{zhang2024cost} is formulated as a mixed-integer nonlinear programming (MINLP) problem which aims at finding the optimal offloading decisions for the IoT devices with the objective of minimizing 
the total offloading cost (i.e., resource purchasing cost) of using the satellite computing resources 
while satisfying the user's maximum task execution latency, satellite's energy and resource capability constraints. 
Then, the penalty and successive convex approximation (SCA) methods are adopted to transform the formulated NP-hard optimization problem into a tractable convex problem. An algorithm is developed to find the near-optimal solution 
by iteratively solving the reduced optimization problem until the convergence condition is met or 
the total numbers of iterations reach its pre-defined maximum threshold.    
Simulation results show that the proposed algorithm can reduce the required convergence time 
by about $2.6\times$, compared with the multi-objective particle swarm optimization (MOPSO) algorithm.
Moreover, it also increases the task success rate by about 90\% while reducing the offloading cost by about $1.83\times$, 
compared with the two offloading approaches proposed by an existing work~\cite{zhang2023energy}.

%Moreover, The study in~\cite{wang2021profit} investigates the offloading and resource allocation problem in a %satellite-terrestrial double edge computing network in which the MEC provider (i.e., owner of the ground and satellite servers) %charges the IoT users 
%for task processing with a certain fee according to the required computing resource and data size of the task. 
Moreover, in~\cite{wang2021profit}, a mixed-integer nonlinear programming problem is formulated to find the optimal offloading decision and resource allocation for the IoT user and MEC provider, respectively, with the objective of maximizing the profit of the MEC provider, while meeting the constraints on the task execution latency, communication and computing resource of the ground and satellite servers. Specifically, the profit of the MEC provider is defined as the difference between its received reward 
and satellite's energy usage cost. The formulated optimization problem is decomposed into two sub-problems for optimizing the offloading and resource allocation strategies separately. Finally, a joint Lagrange-based optimization algorithm is proposed to solve these two subproblems. The simulation results demonstrate that the proposed algorithm can increase the MEC provider's profit by up to about 80\%, compared with three offloading and resource allocation algorithms.

Unlike the optimization-based approaches in~\cite{cheng2024energy,zhang2024cost,wang2021profit}, 
the study in~\cite{yang2024joint} employs a DRL-based approach  to solve the joint satellite selection and offloading problem in a LEO ubiquitous edge computing (UEC) network. 
In the considered network, a terrestrial user (TU) can offload its delay-sensitive (DS) computation task to a computation satellite (CS) via an access satellite (AS) in each time slot.  
%An integer programming problem is formulated to find the optimal selection of the AS and CS for the TUs's task offloading
%with the main objective of maximizing the total number of successfully transmitted tasks
%while minimizing the total satellite access handover overhead for the intersatellite signalling interaction and %satellite-terrestrial connection establishment. 
%It is challenging to solve the formulated optimization problem using the conventional optimization techniques 
%because it is NP-hard and has a large search space. 
A Markov decision process (MDP) is formulated with two agents used for the TU to select the AS and CS, respectively, for its task offloading. Specifically, the AS agent considers the previous AS and AS selections, received signal strength, available bandwidth, estimated handover overhead of AS candidates to select the appropriate ASs with the objective of minimizing the total of the satellite-terrestrial and inter-satellite communication delays. Meanwhile, the CS agent aims to select the optimal CS based on 
the previous CS selection, computation capability of each CS candidate, and estimated AS-CS propagation delay such that the total of the inter-satellite and computation processing delays can be minimized. Then, a DRL-based learning approach, 
called alternating dueling double-deep network (ADDQN) is proposed for two agents to learn their optimal AS and CS selection 
policies. The simulation results indicate that the proposed approach 
can increase the number of successfully completed tasks by about $6.8$\% and reduce the satellite overhead 
by about $15\times$, compared with the four heuristic and DRL algorithms.

%\cite{li2024game}: Do not review because this paper does not propose any game theory-based offloading algorithm as stated by its title. Such a low-quality paper.  
%The authors in~\cite{li2024game} consider a cloud-edge collaboration-based maritime area detection system 
%in which the gateway ship act as an edge server to gather and process a portion of the maritime monitoring data collected 
%by the UAVs, and then forward the remanning data to the shore-based powerful cloud center 
%via the LEO satellites viewed as the data transmission relay nodes.

\subsection{Lessons Learned}
Table~\ref{tab:offloading} provides a summary of incentive mechanisms for computation offloading in satellite networks. From the above review, the existing studies widely adopt the game-theoretic 
mechanisms to optimize the offloading decisions, resource allocation and pricing in the satellite networks. 
These mechanisms provide the decentralized frameworks in which the users, edge servers, and satellites 
determine the offloading destinations, allocate the computing resources, and set the appropriate prices 
to optimize their utilities related to their task latency, energy consumption, and revenue. 
However, many of these approaches rely on the iterative algorithms which may require multiple rounds of information exchange between the players to reach the equilibrium point. 
This requirement may result in the high communication overheads, especially over the inter-satellite links.  
Moreover, in highly dynamic satellite networks, the satellite constellation topology and traffic may change rapidly. 
Thus, the equilibrium point which is computed from the outdated local information may become suboptimal/unstable. 
The software defined satellite networks offer a promising solution to address these issues~\cite{jiang2024game}. 
Specifically, a centralized software defined network (SDN) controller can adopted to aggregate the global network state 
across the satellite constellations and broadcast the local information with reduced coordination rounds and overheads. 

%=======================================================
\section{Privacy and Security}\label{Section 4}

%=======================================================

\begin{figure}[htbp]
%\vspace{-1em}
\centering
\includegraphics[width=\linewidth]{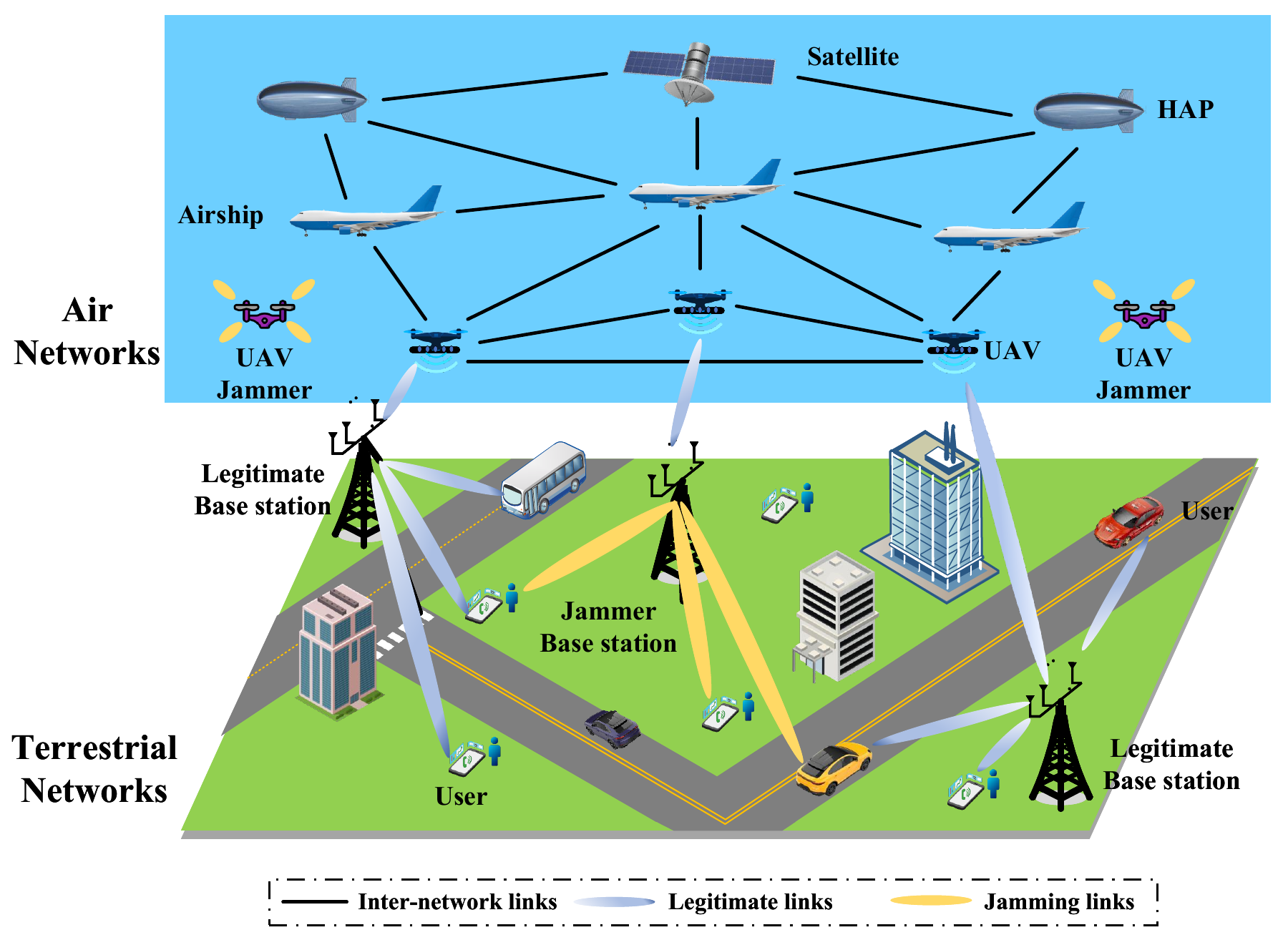}
%\vspace{-2em}
\caption{Illustration of integrated air-terrestrial networks with jammer BSs and UAVs.}
\label{Figure_Security}
%\vspace{-2em}
\end{figure}

Given the long distances between satellites and ground users, air and integrated air-terrestrial networks are increasingly exposed to security threats such as eavesdropping, jamming, spoofing, and privacy breaches, which are further exacerbated by dynamic topologies and the multi-actor nature of these systems, as illustrated in Fig. \ref{Figure_Security}. Conventional optimization schemes are insufficient to address these challenges, as they typically assume cooperative entities and full controllability. In practice, multiple stakeholders (e.g., satellites, ground stations, and IoT devices) have heterogeneous objectives and may behave selfishly. Incentive mechanisms, grounded in game theory, capture these strategic interactions by modeling both legitimate users and adversaries as rational decision-makers. Unlike static optimization, which may fail under asymmetric information or misaligned interests, incentive-based approaches yield equilibrium solutions that remain stable even when participants act in their own interest. This enables scalable and adaptive security provisioning that balances cost, quality-of-service, and protection levels in dynamic satellite–terrestrial environments. It is therefore essential to employ incentive mechanisms that efficiently allocate scarce spectrum and computational resources while aligning the objectives of diverse participants. By incorporating methods such as game theory, auctions, contracts, coalition formation, and decentralized learning, structured solutions can be devised to encourage cooperation, manage privacy–performance trade-offs, and mitigate adversarial actions across heterogeneous network layers. \textcolor{black}{While both cooperative and noncooperative games have been employed for resource allocation and security provisioning, their applicable boundaries in satellite scenarios remain distinct. Cooperative games, which rely on coalition formation and collective utility maximization, assume that participating nodes can negotiate and share information transparently. However, in heterogeneous satellite–terrestrial systems, entities such as LEO constellations, ground gateways, and third-party service providers often have conflicting commercial or operational objectives. The lack of a trusted coordinator and the presence of asymmetric information make full cooperation impractical. In such cases, noncooperative game models, including Stackelberg, auction, and repeated games are more appropriate, as they capture competition, self-interest, and strategic interaction under partial observability. Conversely, cooperative games are better suited for intra-operator coordination or trusted coalition environments, where the interests of participating nodes are aligned. Clarifying this distinction is crucial for selecting appropriate incentive mechanisms under different network ownership and security configurations.} Accordingly, this section reviews recent studies that leverage incentive-driven resource allocation strategies to achieve security objectives under diverse operational and architectural constraints.

%%%%%%%%%%%%%%%%%%%%%%%%%%%% Figure_Security %%%%%%%%%%%%%%%%%%%%%%%%%%%%%%%
%%%%%%%%%%%%%%%%%%%%%%%%%%%%%%%%%%%%%%%%%%%%%%%%%%%%%%%%%%%%%%%%%%%%%

\subsection{Stackelberg Game-based Resource Allocation}
Several works employed Stackelberg game frameworks to allocate resources in satellite and satellite-terrestrial systems. The authors in \cite{cai2022security} investigate a satellite-terrestrial IoT network in which remote users purchase encryption services from a provider to protect their data privacy. The Stackelberg Game is known to be able to characterize the resource exchange process between the satellite and terrestrial networks. Therefore, the interaction is modeled as a Stackelberg game, where the provider (leader) sets prices and users (followers) choose their security configurations. To handle the sequential and asymmetric nature of the system, the authors leverage the MDP and conceive a multi-agent reinforcement learning algorithm (WoLF-PHC) to find near-Nash equilibrium strategies. It can be observed that the proposed WoLF-PHC achieves better convergence and is more effective in meeting user privacy requirements compared to Q-learning. Specifically, the price achieved by the proposed WoLF-PHC method is twice that of the Q-learning method, with a higher security configuration level. The converging point of the WoLF-PHC is nearly identical to the Nash equilibrium point. Consequently, \cite{li2022secure} and \cite{han2020spatial} balance secure spectrum and energy efficiency using Stackelberg hierarchies between primary and secondary users or eavesdroppers. A cognitive satellite-terrestrial network employing two-way relaying and friendly jamming to enhance security against eavesdropping is proposed in  \cite{li2022secure}. It is illustrated that the tradeoff between secure spectrum efficiency and secure energy efficiency is formulated as a Stackelberg game, with the satellite network as the leader and the terrestrial network as the follower. To better capture realistic scenarios, the authors model varying activity states of primary users and eavesdroppers upon exploiting adaptable frame structures. An iterative joint spectrum and energy allocation optimization algorithm is proposed, yielding improved utility by 28.5\% compared to the fixed allocation scheme and faster convergence with 25 times fewer iterations compared to conventional fixed allocation schemes.

In \cite{han2020spatial}, a hierarchical Stackelberg game, along with a deep reinforcement learning scheme is employed to develop an anti-jamming strategy tailored for highly dynamic satellite networks, where the jammer is the leader and the user is the follower. Specifically, the Stackelberg anti-jamming routing game is utilized to model the communication countermeasures employed by jammers and satellite users. The proposed method leverages deep learning to reduce the decision space and reinforcement learning to achieve fast adaptive routing under smart jamming, yielding an iterative solution that converges to a stable equilibrium. Compared to conventional schemes, the proposed approach is capable of striking lower routing costs by 57\% and improved robustness under targeted jamming. Similarly, in \cite{liao2022secure}, a two-layer Stackelberg game model is proposed to secure backhaul links in satellite-UAV integrated networks against full-duplex attackers capable of simultaneous jamming and eavesdropping. In the first layer, the attacker and the source UAV are the leader and the follower, respectively, while the cooperative UAV and the source UAV are utilized as the leader and the follower, respectively. It captures the distinctive exposure of satellite-UAV channels and the limited cooperative behavior of UAV nodes, which differs from conventional terrestrial models. The first layer models the confrontation relationship between the UAV network and the attacker, where the transmission power strategy of UAVs can be adjusted based on the attacker's jamming strategy. Moreover, the second layer modeled a cooperative UAV setting, where unit prices for jamming services are set, incentivizing participation despite the inherent selfishness. Then, the three-stage optimal response iterative algorithm is proposed to solve the game equilibrium iteratively, showing improved secure rates by 50\% compared to he non-cooperative jamming and higher cooperative UAV utility by 7\% and 27\% given the self-interference factor of $10^{-3}$ compared with existing pricing and bargaining methods, respectively. 

The concept of \cite{liao2022secure} is later extended to covert communication in large-scale multi-tier networks \cite{feng2024covert}. To consider the wide-area coverage and dense multi-tier structure unique to satellite backhauls, the authors in \cite{feng2024covert} formulate a two-stage Stackelberg game to secure large-scale multi-tier LEO satellite networks against detection by vigilant adversarial terrestrial BSs, directly addressing. Specifically, terrestrial BSs are modeled as noncooperative followers to minimize detection errors via energy detection. At the same time, the LEO satellite network acts as the leader, performing power control to maximize utility under covert constraints. The proposed design leverages stochastic geometry with binomial point process models and mean-field methods to handle the spatial distribution of satellites, UAVs, and BSs, which distinguishes it from conventional small-scale terrestrial covert schemes. A bi-level algorithm combining SCA and golden-section search has been developed to achieve the Stackelberg equilibrium, as validated by numerical results that demonstrate how interference and satellite activation probabilities jointly impact covert performance. In practical secure transmission in CSTNs, the ownership of satellite and terrestrial networks is separate, and these two networks cannot cooperate to their mutual benefit. Therefore, a Stackelberg game-based secure transmission strategy for CSTNs is proposed in \cite{wen2022stackelberg}. Explicitly, the satellite network is modeled as the leader that recruited a terrestrial node as the follower, where the follower is employed to assist anti-jamming. With the help of the follower that can be compensated through allocated spectrum access time, the secrecy rate of the lead satellite network is maximized at the cost of overall transmit energy. The approach incorporates the impact of imperfect CSI on the eavesdropper link, reflecting the uncertain and broad coverage characteristics of satellite systems. Moreover, the authors also derive the Stackelberg equilibrium via backward induction, yielding optimal time and power allocation that achieved a tradeoff between secrecy rate and terrestrial transmission utility. 

The study in \cite{zhang2024dynamic} combines Stackelberg and matching games to allocate anti-jamming strategies dynamically under evolving SIoT interference. Specifically, based on the dynamic relationship between jammer satellites and terrestrial cells, the study in \cite{zhang2024dynamic} develops a hierarchical anti-jamming Stackelberg game (HASG) to address jamming from LEO satellites by utilizing many-to-one and one-to-one matching to construct stable anti-jamming strategies, where the jamming satellite is the leader and the ground cell is the follower. The proposed HASG reduces 90\% computational complexity of the exhaustive search when using eight satellites while attaining nearly identical transmission rates. The proposed HASG is designed for dynamic adversarial interactions between jamming satellites and terrestrial users, supported by convergence and equilibrium proofs. In \cite{liao2023irs}, considering an intelligent reflecting surface (IRS)-assisted integrated satellite-UAV-terrestrial (SUT) IoT network, a Stackelberg game framework is proposed to address the joint optimization of hybrid beamforming design of the satellite, UAV, and IRS against a smart jammer under imperfect CSI. Specifically, the satellite and UAV, as well as the IRS and the jammer, are modeled as the leader and the follower, respectively. The method exploits an AoA-based discretization and use-and-then-forget strategy to handle unknown jamming channels and derives closed-form suboptimal jamming powers using the Cauchy–Schwarz inequality. An alternating optimization with penalty functions is proposed to solve the coupled leader subproblem, achieving convergence with reduced transmit power by 19.25\% compared to the conventional MRT beamforming scheme. By taking into account the unique features of UAV networks, including open wireless exposure and decentralized UAV-to-UAV topology, a stochastic Stackelberg game is proposed in \cite{yin2024uav} to counter intelligent jamming. The proposed paradigm models the terrestrial jammer as the leader to optimize the jamming channel and power. At the same time, UAV pairs act as followers, collaboratively selecting transmission channels and power levels to maximize throughput while considering power costs. The proposed mechanism employs federated deep Q-learning across UAV pairs. It leverages a satellite server to aggregate local models without centralizing raw data, thereby preserving privacy while enhancing coordination in the face of shared jamming threats. The proposed approach demonstrates that federated learning can nearly achieve the centralized performance and promote anti-jamming efficiency by 23.3\% over decentralized learning.

In the area of computation offloading and privacy-preserving allocation, \cite{gong2024orbit} and \cite{gong2024computation} employ Lyapunov optimization coupled with federated learning to manage stochastic satellite-terrestrial task placement while ensuring privacy guarantees. In \cite{gong2024orbit}, a two-layer Stackelberg game, combined with a deep federated meta-reinforcement learning scheme, has been proposed to orchestrate cycle frequency, channel assignment, and block size in satellite-terrestrial twin networks. The satellite cloud server
and terrestrial users are denoted as the leader and the followers, respectively. This approach incorporates Lyapunov stability theory to decouple long-term queues into short-term resource decisions. The proposed strategy can achieve an adaptive solution under dynamic tasks and varying satellite positions. Compared with benchmarks, the proposed method demonstrates higher throughput by 10\% at most and reduced authentication overhead by 78\%. To carry out computation offloading, mitigate channel interference, and promote the privacy of satellite-terrestrial digital twin networks, a blockchain-aided two-stage Stackelberg game has been proposed in \cite{gong2024computation}, where cloud providers and multiple digital twins are modeled as leaders and followers, respectively. In contrast to the conventional terrestrial systems, the proposed model is capable of handling stochastic task arrivals, time-varying LEO satellite locations, and cross-layer interference. Considering the unique constraints of computational and connectivity in satellite-terrestrial integration networks, the authors design a Lyapunov-based transformation to decouple long-term queues. They apply a multi-agent deep federated reinforcement learning framework to optimize CPU frequency, channel allocation, offloading, block size, and cloud pricing jointly. The method incorporates federated aggregation and issuing to preserve privacy across untrusted digital twins under satellite connectivity while ensuring transaction verification via blockchain.

\subsection{Contract-based Incentive Mechanisms}
Some studies instead adopted contract-based incentive mechanisms. In cooperative localization systems, the privacy cost occurs when the nodes transmit their location information. Consequently, the work in \cite{shi2021clap} proposed an incentive mechanism to evaluate the tradeoff between the cooperative node privacy and the target localization accuracy. Explicitly, the proposed scheme is based on contract theory to ensure the localization performance of cooperative nodes while minimizing the target payment. It transformed the original nonconvex optimization into a convex problem, providing a closed-form solution that minimized payments under localization accuracy constraints.

\subsection{Deep Learning-based Incentive Mechanisms}
\begin{figure}[htbp]
%\vspace{-1em}
\centering
\includegraphics[width=\linewidth]{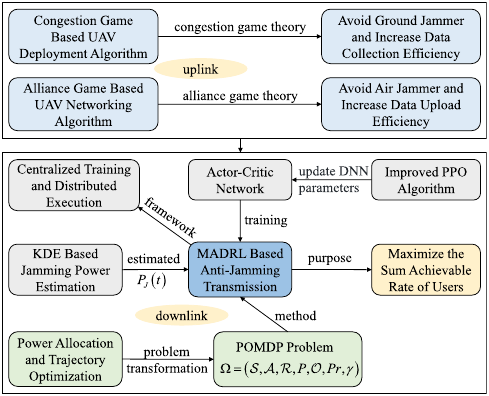}
%\vspace{-2em}
\caption{The game theory with MADRL-aided anti-jamming transmission scheme proposed in \cite{liao2024game}.}
\label{Figure_Security_Game_DRL}
%\vspace{-2em}
\end{figure}

A different approach to resource coordination appeared in coalition formation and reinforcement learning frameworks. Upon considering the wide coverage and sparse infrastructure that distinguish air-terrestrial systems from purely terrestrial networks, the work in \cite{liao2024game} develops an anti-jamming transmission strategy for integrated air-terrestrial networks by combining game theory with MADRL, which is shown in Fig. \ref{Figure_Security_Game_DRL}. By leveraging these models to counteract both terrestrial and aerial jammers while maximizing uplink data collection and reliability, the proposed method formulates UAV swarm deployment as a congestion game and dynamic subnet formation as a coalition game. In downlink scenarios, the joint trajectory and power optimization problem is modeled as a partially observable Markov decision process and is solved by a MADRL framework under centralized training at high-altitude platforms. The MADRL framework alleviates computational burdens on UAVs and ensures distributed execution. Moreover, an improved proximal policy optimization algorithm with adaptive clipping is proposed to accelerate convergence under large state spaces. Simulation results show that the proposed scheme can achieve 16.6\% higher total collected data compared to the Pareto scheme in terms of the UAV-HAP links.
%%%%%%%%%%%%%%%%%%%%%%%%%%%% Figure_Security %%%%%%%%%%%%%%%%%%%%%%%%%%%%%%%
%%%%%%%%%%%%%%%%%%%%%%%%%%%%%%%%%%%%%%%%%%%%%%%%%%%%%%%%%%%%%%%%%%%%%

\subsection{Repeated Games and Groves Mechanisms}
Beyond these learning-driven methods, specific works use repeated games or Groves mechanisms to enforce fairness and efficiency in resource scheduling. To coordinate user behavior and mission scheduling, a repeated game-based resource allocation approach for data relay satellite systems is proposed in \cite{wang2018high}, addressing the limited inter-satellite beam resources and visibility window constraints unique to geosynchronous relay systems. Specifically, a payoff structure incorporating resource satisfaction and scheduling success rates is proposed. Then, the Nash equilibrium under one-shot competition is derived, showing that users would selfishly request maximum resources. A repeated game with punishment and forgiveness is conceived to incentivize long-term cooperation, supported by analytical conditions on discount factors and punishment duration. A cheat-proof mechanism, based on Groves mechanisms, further encourages users to submit their actual resource requirements, leveraging the transparent operational agreements and security protocols characteristic of relay satellite systems.

\subsection{Direct Pricing Strategies}
Some studies adopted direct pricing without hierarchical Stackelberg games. In  \cite{zhang2023value}, a distributed noncooperative game mechanism is proposed to minimize the cost of information value (CoIV) in priority-aware irregular repetition slotted ALOHA (IRSA), namely value-optimal enhanced IRSA (V-IRSA), for satellite-integrated IoT (S-IoT). In contrast to conventional AoI metrics, the proposed scheme introduces a CoIV-based utility that evaluates the importance of packets from the user's perspective and derives closed-form expressions of CoIV functions under both linear and exponential decay scenarios. A differential evolution algorithm has been employed to optimize the degree distributions, with convergence to a Nash equilibrium demonstrated under quasi-concavity. The scheme prioritizes high-importance nodes effectively, achieving improved throughput and reduced CoIV over standard IRSA. To address the sparse terrestrial infrastructure and high mobility unique to satellite scenarios, a privacy-aware spectrum pricing and power control framework for LEO satellite IoT is proposed in \cite{shen2024privacy}. The proposed framework formulates pricing as an MDP solved by local DDQN agents at satellites, with computation offloaded to terrestrial servers to overcome the limited onboard processing capabilities of satellites. Moreover, the proposed approach integrates a federated learning scheme coordinated via a reputation-based blockchain, while leveraging the decentralized satellite network to avoid the single-point failures common in terrestrial federated learning setups. This design incentivizes honest participation by linking spectrum leasing and power compliance to reputation, which governs the aggregation and distribution of rewards.

\begin{table*}[ht]
\caption{Summary of Incentive Mechanisms For Privacy and Security.}
\renewcommand{\arraystretch}{1.2}
\centering
\begin{tabular}{|c| p{2.7cm}| p{4.0cm}| p{4.2cm}| p{4.2cm}|}
\hline
\textbf{Ref.} & \textbf{Scenario} & \textbf{Problem focus} & \textbf{Methodology} & \textbf{Players} \\
\hline
\cite{han2023anti} & Satellite IoT & Anti-jamming & Game theory & Satellites, BSs, users and jammers \\
\hline
\cite{han2020spatial} & Internet of Satellites & Routing optimization and anti-jamming & Stackelberg game + learning & Satellites, RISs and users \\
\hline
\cite{feng2024covert} & LEO satellite networks & Covert communication & Stochastic geometry + game theory & Satellites, BSs, users and UAVs \\
\hline
\cite{abdrabou2024game} & LEO satellite networks & Physical layer authentication & Hypothesis testing + game theory & BSs, spoofers and satellites \\
\hline
\cite{huang2025consolidated} & Satellite-enabled IoT & Cooperative
defense & Flow pricing + learning & Satellites, UAVs and users \\
\hline
\cite{shi2021clap} & Cooperative localization & Privacy vs. accuracy tradeoff & Incentive mechanism + optimization & Nodes and targets \\
\hline
\cite{liao2023irs} & RIS-assisted SAGIN & Anti-jamming & Hybrid beamforming + game theory & Satellites, UAVs, RISs and jammers \\
\hline
\cite{yin2024uav} & UAV networks & Anti-jamming & Game theory + learning & UAVs and jammers \\
\hline
\cite{liao2024game} & SAGIN & Anti-jamming + trajectory optimization & Game theory + Markov decision process + learning & UAVs and HAPs \\
\hline
\cite{wang2018high} & Data relay satellite systems & Mission scheduling + user behavior coordination & Game theory + cheatproof mechanism & Satellites, BS and users \\
\hline
\cite{gong2024orbit} & SAGIN digital twin networks & Security authentication & Stackelberg game + learning & LEO satellites, BSs and users \\
\hline
\cite{gong2024computation} & SAGIN digital twin networks & Computation + privacy protection & Game theory + blockchain + Lyapunov stability theory + learning & Satellites, BSs and users \\
\hline
\cite{zhang2023value} & Satellite-Integrated IoT & Diversified access & Game theory + learning & BSs and ground nodes \\
\hline
\cite{shen2024privacy} & LEO satellite IoT & Spectrum resource management & Blockchain + game theory + learning & Satellites, BSs and users \\
\hline
\cite{liu2021decentralized} & Space-ground integrated network & Anonymous authentication & Blockchain + billing & Satellites, ground stations and users \\
\hline
\end{tabular}
\label{tab:privacy}
\end{table*}

\subsection{Other Game Approaches}
A few studies, instead, focus on anti-jamming, detection, or authentication problems. Unlike dense terrestrial infrastructures, large-scale LEO satellite deployments are vulnerable to spoofing attacks. In \cite{gong2020pursuit}, a pursuit-evasion differential game (PE Game) for continuous-thrust satellites has been investigated from the reachable domain (RD) perspective. Consequently, a concept of hyper RD (HRD) is introduced to analyze the PE Game problem, and a sufficient condition of PE Game capture is conceived based on the HRD of players. Moreover, based on the Clohessy-Wiltshire model, an analytical expression for the reachable domain is derived from leveraging the RD tool. The simulation results demonstrate that the proposed Clohessy-Wiltshire model can efficiently evaluate capture scenarios without resorting to traditional shooting methods. Similar double auction designs are presented in \cite{xia2023incentive}, which focused on delay-aware satellite edge learning. A game-theoretic spoofing detection scheme for space information networks is proposed in \cite{abdrabou2024game}. The method models physical layer authentication as a zero-sum game where the terrestrial station selects the detection threshold while the spoofing attack probability is optimized. Moreover, Doppler frequency spread and received power are exploited to distinguish satellite trajectories. Consequently, it can switch between Doppler and power attributes across different elevation angles, thereby improving authentication rates. Compared to the single-layer defense in satellite-enabled IoT, cooperative defense involves IoT devices, access networks, and satellite transmission networks that achieve better fine-grained defense. Specifically, given the testing threshold $\tau=0.5$, the scheme improves the average authentication rate by 14.5\% compared to the combined authentication schemes. A tripartite security game framework for cooperative defense against cross-domain attacks in satellite-enabled IoT scenarios is proposed in \cite{huang2025consolidated}, explicitly addressing the open problem of quantifying cross-domain impacts and incentive misalignment among different components. The framework evaluates the impacts of different attack and defense schemes in satellite and terrestrial domains. Specifically, in the satellite domain, a flow pricing method is developed to incentivize IoT network operators to prevent the infiltration of malicious information. Efficient learning algorithms are developed to enable IoT network operators to identify optimal flow sampling strategies and satellite service providers to effectively set flow prices. Compared to deviated strategies-based schemes, the maximum utility increases by 11.9\% when leveraging the Stackelberg–Nash equilibrium. The framework conceives existence conditions for a consolidated Stackelberg–Nash equilibrium and develops finite-step learning algorithms to reach stable strategies under partial information. An anti-jamming strategy for satellite IoT systems is proposed in \cite{han2023anti}, which is based on the derived transmission rate of SIoT systems and exploits beam switching and beam angle optimization. The scheme utilizes a suitable satellite to provide sufficient spatial diversity, ensuring coverage. Moreover, game theory is employed to establish a stable matching between satellites and terrestrial cells, taking into account externalities, thereby reducing complexity compared to an exhaustive search while achieving nearly equivalent transmission rates. A variable metric method then refines the satellite beam angles to further maximize network throughput. The method leverages the spatial diversity of satellite clusters to dynamically reassign coverage, which mitigates jamming attacks without relying on conventional schemes that increase satellite energy costs. The authors in \cite{shen2020enhanced} conceive a modified conditional generative adversarial network (GAN) framework that incorporates general-sum game structures and Fictitious Play to stabilize training for satellite behavior discovery. Specifically, the noise-injected loss function is employed to achieve the persistence of excitation, leading to improved convergence compared to conventional GAN training. The paradigm combines a multilayer conditional generator and discriminator with dedicated embedding and convolutional blocks, utilizing adaptive gradient updates for iterative optimization. It is demonstrated that a 93\% classification consistency is achieved with a $0.002$ Frechet inception distance, indicating the effectiveness of the GAM training scheme with a noise-injected cost function. In \cite{wu2023hybrid}, the interception problem is formulated as a zero-sum differential game between two players: the pursuer (interceptor) and the evader (target satellite). Each player controls the orientation of its continuous low-thrust propulsion via two angular parameters, namely the azimuth and elevation of the thrust vector, which together define the control strategies. The pursuer seeks to minimize interception time and fuel consumption, while the evader aims to maximize survival time or exhaust the pursuer’s resources. By applying first- and second-order optimality conditions, a finite set of feasible control modes is derived for each side, and the saddle-point solution yields the bilateral optimal strategy. This equilibrium ensures that neither player can improve its outcome by unilateral deviation, thereby characterizing the optimal pursuit–evasion trajectories under given boundary and dynamic constraints. The scheme maintains orbital safety constraints while achieving better convergence than unilateral control benchmarks. The authors in \cite{liu2021decentralized} propose a decentralized anonymous authentication for space-terrestrial integrated networks (SGINs), yielding fast handover authentication and fair billing procedures. Specifically, the scheme delegates terrestrial authentication tasks to satellites to reduce authentication delays caused by space-terrestrial propagation, which differs from conventional terrestrial networks. It incorporates threshold-based credential aggregation to enable seamless roaming across multiple satellite operators without the need for separate accounts. More recently, in \cite{barkatsa2025coordinated}, by integrating a contribution-based aggregation algorithm with a Bayesian power control game, a unified framework for defending wireless federated learning against coordinated jamming and poisoning is proposed. Explicitly, a Shapley-inspired index is designed to evaluate node contributions, yielding linear complexity and preserving robustness to aggregation despite poisoned updates, which is a crucial enhancement for resource-constrained networks with variable participation, such as satellite-based IoT networks. The formulated Bayesian game addresses uncertainty in node behaviors without requiring full type disclosure while leveraging incomplete information to adapt power control under probabilistic beliefs. The scheme demonstrates resilience under diverse attacks and maintains global model accuracy with lower power and time overhead compared to the benchmark. However, this paper does not consider multi-cluster coordination and ignores cross-domain attacks beyond jamming and poisoning. 

\subsection{Lessons Learned}
Table~\ref{tab:privacy} summarizes recent incentive mechanisms addressing privacy, secrecy, and anti-jamming strategies among heterogeneous participants. Stackelberg game frameworks have been widely adopted in satellite and satellite--terrestrial networks for spectrum sharing, energy efficiency, anti-jamming, covert communication, and computation offloading~\cite{cai2022security,li2022secure,han2020spatial,liao2022secure,feng2024covert,wen2022stackelberg,zhang2024dynamic,abdrabou2024game,liao2023irs,yin2024uav,gong2024orbit,gong2024computation}. These studies leverage hierarchical leader--follower interactions, often integrated with advanced learning techniques~\cite{liao2024game}, and consistently demonstrate superior convergence, robustness, and utility compared to conventional optimization schemes~\cite{wang2018high,zhang2023value,shen2024privacy,gong2020pursuit,xia2023incentive}. \textcolor{black}{By specifying the incentive schemes and corresponding secure resource management problems in satellite networks, several key observations can be drawn as follows.
\\
\textbf{1) Stackelberg game--based mechanisms:} These approaches are best suited for \textit{hierarchical and asymmetric resource management} problems, where clear leader--follower relationships exist (e.g., satellite vs. terrestrial users, or jammer vs. defender). They effectively model sequential decision processes, enabling adaptive responses to dynamic threats such as jamming and eavesdropping. When combined with deep or federated reinforcement learning, Stackelberg formulations can further handle stochastic satellite mobility and uncertain channel states. Nevertheless, their effectiveness depends on reliable feedback and may degrade due to slow convergence in large-scale multi-agent environments.
\\
\textbf{2) Contract-based mechanisms:} These schemes are particularly advantageous for \textit{privacy-preserving cooperation} and \textit{localization-related resource sharing}, where agents hold private information. By designing incentive-compatible contracts, such mechanisms mitigate information asymmetry between coordinators and participants, ensuring truthful engagement while minimizing operational costs. However, their performance is sensitive to inaccurate system modeling and may deteriorate under highly dynamic network conditions.
\\
\textbf{3) Deep learning and coalition-based incentive schemes:} These methods are more appropriate for \textit{decentralized and large-scale coordination} tasks, such as distributed anti-jamming and computation offloading among UAV or LEO swarms. The integration of MADRL enables distributed decision-making with limited information exchange, while coalition or congestion games capture evolving group behaviors. Although these schemes offer scalability and resilience, they typically incur high computational complexity and substantial training data requirements.
\\
\textbf{4) Repeated games and Groves mechanisms:} These approaches are well suited for \textit{long-term cooperation and fair scheduling} in systems with recurring interactions, such as inter-satellite relay networks. They promote sustainability by penalizing selfish actions and rewarding cooperative behavior over time. Their analytical tractability allows formal equilibrium and fairness analyses, though they exhibit limited adaptability in highly dynamic or one-shot scenarios.
\\
Despite these advances, most existing approaches rely on idealized assumptions such as rational adversaries, accurate channel state information, and synchronized interactions, which seldom hold in practice. Current models are typically confined to simplified two-layer structures, limiting scalability to multi-satellite, multi-UAV, and large-user scenarios. Moreover, coupling game theory with machine learning introduces substantial training overhead and instability in non-stationary environments, reducing deployment feasibility. Finally, many studies address a single optimization objective without jointly considering cross-layer security, privacy, and heterogeneous threats. Therefore, achieving scalable, resilient, and privacy-aware resource allocation in dynamic satellite networks remains an open research challenge.}

%==========================================================================
\section{Other Emerging Issues in Satellite Networks}
\label{section:mixed}
%==========================================================================
Apart from the issues discussed above, the growing complexity and heterogeneity of satellite-terrestrial networks imposes other emerging resource management issues. \textcolor{black}{This section is dedicated to promising yet relatively nascent areas—such as routing, caching, and data offloading. As the literature in these emerging areas remains limited, yet their significance cannot be overlooked, we have consolidated their discussion to provide a pivotal and forward-looking perspective. 
% We explore how incentive mechanisms have emerged as a powerful, unifying tool to solve these diverse coordination problems under selfish or decentralized settings.
The section is structured into focused subsections: Section \ref{section5-a} on Intelligent Routing and Traffic Reallocation, Section \ref{section5-b} on Caching and Access Control, Section \ref{section5-c} on Topology-Aware Resource and Position Optimization, Section \ref{section5-d} on Data Offloading and Section \ref{section5-e} on Relay Selection.} To meet the performance, scalability, and responsiveness requirements of 6G SAGIN \cite{10816375}, incentive mechanisms have emerged as a powerful tool for enabling cooperative behavior under selfish or decentralized settings. {\color{black}This section explores recent advances where economic or game-theoretic incentives are used to solve diverse coordination problems, ranging from routing and traffic reallocation, learning-based caching and access control, topology-aware resource optimization, to data offloading and relay selection. Since this section consists of miscellaneous issues, we first organize by issues and then discuss how different incentive mechanisms are used to solve the issues.}

%A summary of incentive mechanisms for the emerging issues in satellite networks is shown in Table~\ref{tab:emerging}.

%===================
\subsection{Intelligent Routing and Traffic Reallocation}\label{section5-a}
\begin{figure}[t]
%\vspace{-1em}
\centering
\includegraphics[width=\linewidth]{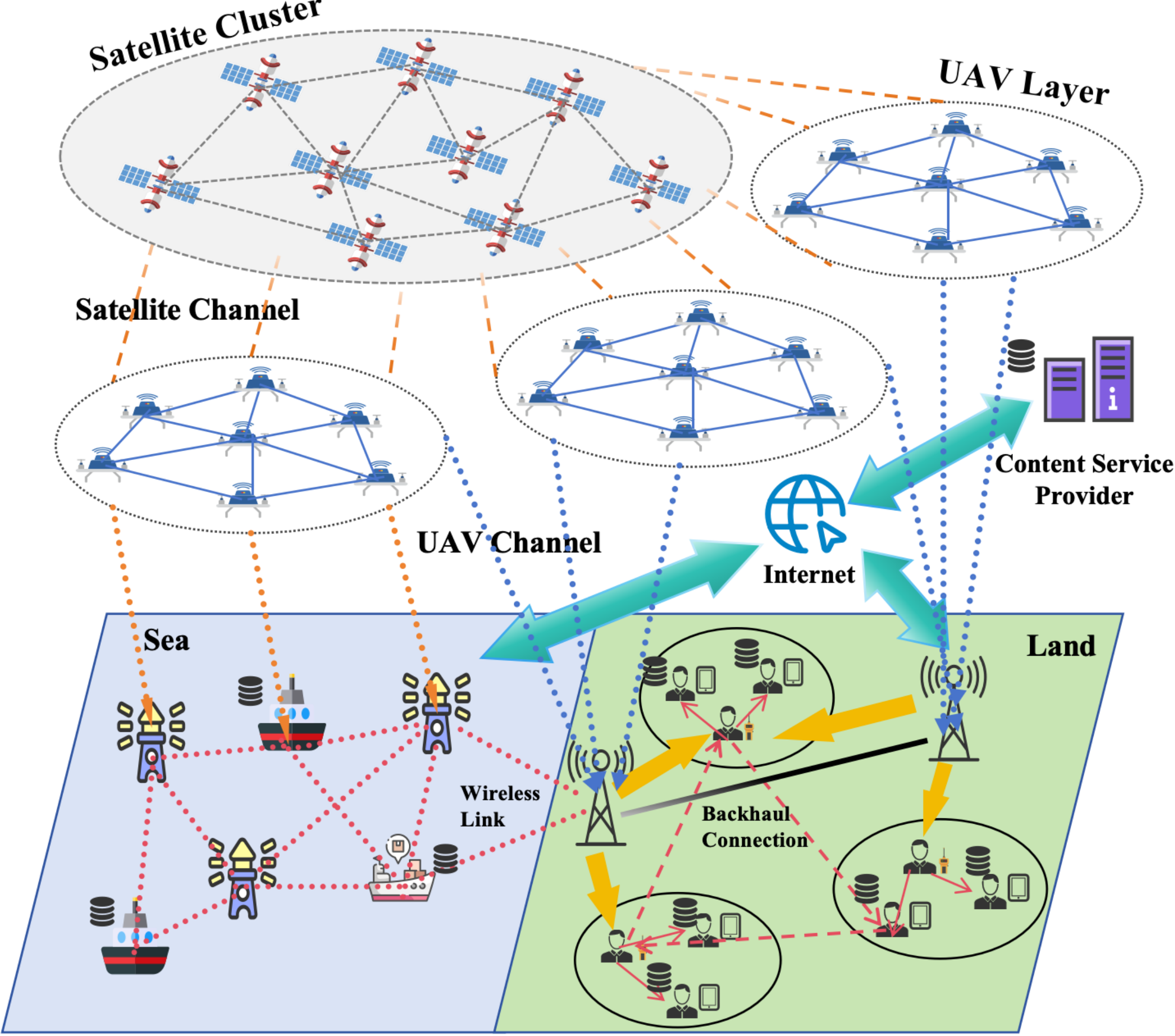}
\vspace{-2em}
\caption{Illustration of integrated air-terrestrial-maritime networks with offloading and coordination.}
\label{air-terrestrial-maritime}
%\vspace{-2em}
\end{figure}

%===================
Satellite networks contain hybrid wireless connected devices whose packets require intelligent routing to the receivers (as shown in Fig.\ref{air-terrestrial-maritime}). Routing and traffic scheduling in these networks necessitate intelligent decision-making due to dynamic topologies, heterogeneous node capabilities, and stringent QoS demands. The introduction of incentive mechanisms, especially game-theoretic models, provides a distributed and adaptive framework to guide nodes’ behaviors toward globally efficient routing strategies. These mechanisms encourage cooperation among satellites or between satellites and terrestrial/maritime agents, enabling effective traffic balancing, congestion avoidance, and QoS-aware service delivery.

\subsubsection{Non-Cooperative Games}
In \cite{du2022game}, the authors investigate QoS-driven traffic scheduling in a multilayer software-defined satellite network. They categorize traffic flows into three classes: delay-sensitive (Class A), bandwidth-sensitive (Class B), and best-effort (Class C). To balance throughput and fairness, a hybrid scheduling framework is proposed, combining a supermodular game with priority queuing, where Class A traffic is given absolute precedence. In this non-cooperative game, the players are the Class B and Class C traffic aggregates. Each player aims to maximize its own utility function by competitively adjusting its scheduling strategy, which relates to its sending rate or priority level. The utility functions are defined as a weighted trade-off between priority and arrival rate. The equilibrium can be reached through iterative best-response algorithms, where no player can unilaterally improve its utility. Additionally, their in-orbit-first routing algorithm minimizes inter-orbit hops for high-priority traffic to reduce delay.

\subsubsection{Coalitional Games}
Unlike the intra-satellite focused work in \cite{du2022game}, the authors in \cite{kim2022cooperative} consider a broader integrated satellite–maritime network (ISMN) scenario, where routing decisions involve multiple types of network entities, including marine vessels, floating buoys, low Earth orbit (LEO) satellites, and terrestrial base stations (TBSs). They introduce a three-phase coalitional game–based multi-path routing strategy in which these entities act as players forming routing coalitions. The strategy dynamically groups agents into routing coalitions based on traffic types (real-time vs. non-real-time) and network state. Simulation results demonstrate that the proposed coalitional game approach significantly enhances system performance: it improves overall throughput by 15–20\% compared to baseline schemes such as INPM, SMUC, and MRCA, particularly under medium to high traffic loads.
Both studies \cite{du2022game} and \cite{kim2022cooperative} demonstrated how game-theoretic incentives can effectively guide distributed routing behaviors under complex satellite network environments. While \cite{du2022game} emphasizes QoS-driven prioritization within a hierarchical satellite system using a non-cooperative supermodular game, \cite{kim2022cooperative} focuses on collaboration across hybrid satellite–maritime infrastructures using a coalitional game. However, a key challenge remains in harmonizing multi-domain routing incentives across satellite-terrestrial-maritime boundaries under real-time constraints. Future work could explore hierarchical or hybrid incentive frameworks that jointly consider routing efficiency, fairness, and cross-layer adaptability.

\subsection{Caching and Access Control}\label{section5-b}
Incentive mechanisms are increasingly vital in satellite-integrated networks for intelligent caching and access control. As content popularity becomes more dynamic and user behavior more unpredictable, satellite-ground systems must coordinate cache placement, user clustering, and resource allocation in real-time. Incentives such as payment schemes or cooperative gains encourage users and intermediate nodes (e.g., UAVs or D2D devices) to contribute resources like cache space and communication bandwidth. Learning-based game approaches provide scalable ways to guide distributed decisions and reduce latency and operational cost.

\subsubsection{Auction Mechanisms}
The authors in \cite{zhou2021incentive} tackle the dual challenge of limited caching capacity and lack of user motivation in cellular D2D networks by proposing an incentive-driven deep reinforcement learning framework (IDQNM). The primary objective of their model is to maximize the saving cost of the Content Service Provider (CSP), which reflects the net reduction in cellular transmission costs achieved through D2D offloading, after accounting for payments to participating users. In this framework, the CSP acts as the central controller and auctioneer, while mobile users serve as bidders offering caching and data forwarding services in a reverse auction. The approach integrates a deep Q-network (DQN) for adaptive content caching with a VCG payment rule to ensure truthful bidding and individual rationality. Simulation results demonstrate that the proposed method significantly outperforms baseline strategies: under real mobility traces, IDQNM improves the CSP’s saving cost by approximately 20–30\% compared to decay-based, greedy, and random caching methods, particularly as the number of nodes and content tolerance delays increase. However, the efficiency of this approach is sensitive to user behavior predictability—if preference prediction is uncertain, the incurred caching overhead may offset the benefits.

\subsubsection{Non-Cooperative Games}
Unlike the single-layer D2D focus in \cite{zhou2021incentive}, the authors in [129] consider a broader 6G satellite–UAV–terrestrial network. They propose a real-time distributed optimization framework in which the overall latency minimization problem is decomposed into three subproblems: user clustering, cache placement, and power allocation. These are addressed respectively through a non-cooperative game among ground users for clustering, a genetic algorithm executed by the UAVs for cache decision-making, and a low-complexity estimation technique for power allocation. The Nash equilibrium for the clustering game is reached through an iterative best-response algorithm, where each user selfishly selects the UAV that minimizes its own latency until no user can improve its utility unilaterally. While effective in latency reduction, their iterative decomposition raises concerns regarding the proper decoupling of variables in dynamic environments. The authors in \cite{tang2024cooperative} advance spatial caching by proposing a cooperative caching scheme in satellite-terrestrial integrated networks (STINs). The LEO satellites serve as the primary actors executing distributed caching decisions. They incorporate a ridge regression-based prediction model to support distributed LEO caching, using a non-cooperative game to guide cache utility optimization within clustered cooperative areas. Each satellite, based on others' current strategies, updates its own caching decisions to maximize the combined utility of itself and its cooperative neighbors, with convergence guaranteed by a potential function. A noted limitation, compared to the approach in \cite{nguyen2023real}, is a lack of adaptability to fast-changing user mobility, leaving a gap in dynamic cluster formation.

\subsubsection{Cost-Minimization Framework}
In a similar multi-tier scenario, the authors in \cite{mishra2024minimizing} develop a cost-minimization framework for CSPs using UAV-based caching, with the goal of reducing service latency and operational expenses. They adopt user clustering and cache placement strategies similar to \cite{nguyen2023real}, using a non-cooperative game approach for user clustering and a genetic algorithm for cache placement, but notably simplify the optimization by assuming fixed transmission power. Experimental results demonstrate that their approach significantly reduces CSP costs by up to 28\% compared to baseline strategies without collaborative caching, while also decreasing average service latency by approximately 22\%. This assumption, while practical, overlooks the impact of varying energy consumption and link quality in long-range satellite–UAV–ground links, pointing to an opportunity for more adaptive resource coupling.

These studies in \cite{zhou2021incentive, nguyen2023real, tang2024cooperative, mishra2024minimizing} demonstrate the effectiveness of combining Vickrey–Clarke–Groves rule, the auction and the non-cooperative games with learning-based strategies for caching in space–air–ground networks. From D2D auctions to multi-layer cooperative caching, incentives align distributed decisions with global efficiency goals. However, open challenges persist in modeling the interdependencies among caching, power allocation, and user mobility. Future research could explore adaptive, mobility-aware incentive frameworks that jointly consider content popularity prediction, energy budgets, and cross-layer performance constraints.

%===================
\subsection{Topology-Aware Resource and Position Optimization}\label{section5-c}
%===================
Incentive mechanisms in satellite networks are not only useful for task execution or resource exchange, but also for encouraging cooperation in dynamic topology optimization, especially for position adjustment and resource coordination in multi-satellite constellations. As the number of satellites increases and topology becomes more dynamic, intelligent coordination schemes are necessary to reduce control overhead, maintain coverage robustness, and enable fuel-efficient operations. Game-theoretic models and learning-based optimization methods are frequently used to ensure fair collaboration among satellites while balancing constraints like energy, latency, and inter-satellite interference.

\subsubsection{Coalitional Games}
The authors in \cite{liu2021reliable} address clustering in large-scale LEO satellite networks, where frequent topology changes make stable communication and coordination difficult. They formulate a coalition formation game in which the satellites themselves act as autonomous players, making distributed decisions to form clusters and select cluster heads (CHs). The primary objective is to minimize protocol signaling overhead by optimizing the network structure through cluster organization. Unlike resource-sharing mechanisms, their focus is on organizing the network structure itself. Each satellite operates as an independent agent, continuously assessing whether transitioning to another cluster improves the utility function that jointly considers cluster reliability and management overhead. This process continues until no satellite can unilaterally improve its utility by changing clusters, thereby achieving the Nash equilibrium. While the scheme effectively reduces the control message burden, it omits other sources of overhead such as energy consumption for constant state monitoring and computational costs of decision-making.

\subsubsection{Differential Games}
Unlike the clustering-centric approach in \cite{liu2021reliable}, the authors in \cite{ni2023fault} focus on fault-tolerant regional coverage in multi-satellite systems by leveraging differential graphical game theory. In their approach, each satellite acts as an autonomous player that makes distributed decisions to achieve cooperative coverage control. They model a coupled position–attitude control mechanism to enable graceful degradation of coverage in case of satellite failures. A key contribution is their single-network approximate dynamic programming controller, which ensures fast adaptation with low computation complexity. Similar to \cite{liu2021reliable}, this work adopts a distributed strategy, but places greater emphasis on maintaining service quality under failure conditions rather than minimizing control overhead. Similar to \cite{ni2023fault}, the authors in \cite{amozegari2024co} also employ differential game theory, but extend its application to the co-location control of GEO satellites. In their approach, each satellite acts as an independent decision-maker that optimizes its own station-keeping maneuvers under both cooperative and non-cooperative game modes. They jointly optimize fuel consumption, collision avoidance, and signal interference in a highly constrained orbital environment. While their approach is broader in scope—considering both physical and communication-layer constraints—it assumes precomputed maneuvers and does not incorporate real-time feedback control. This limits the system’s ability to handle dynamic uncertainties like orbital drift or communication delays.

These works \cite{liu2021reliable, ni2023fault, amozegari2024co} collectively reveal the importance of intelligent, incentive-compatible mechanisms for topology and position control in dynamic satellite constellations. Whether clustering, fault tolerance, or co-location are concerned, game-theoretic frameworks provide structured negotiation and coordination among autonomous agents. However, current studies tend to isolate aspects such as energy cost, delay tolerance, or system observability. Future research should consider joint modeling of control signaling, computation energy, and delay-aware learning to achieve more robust, real-time closed-loop optimization in complex space systems.

\begin{figure}[t]
\vspace{-1em}
\centering
\includegraphics[width=\linewidth]{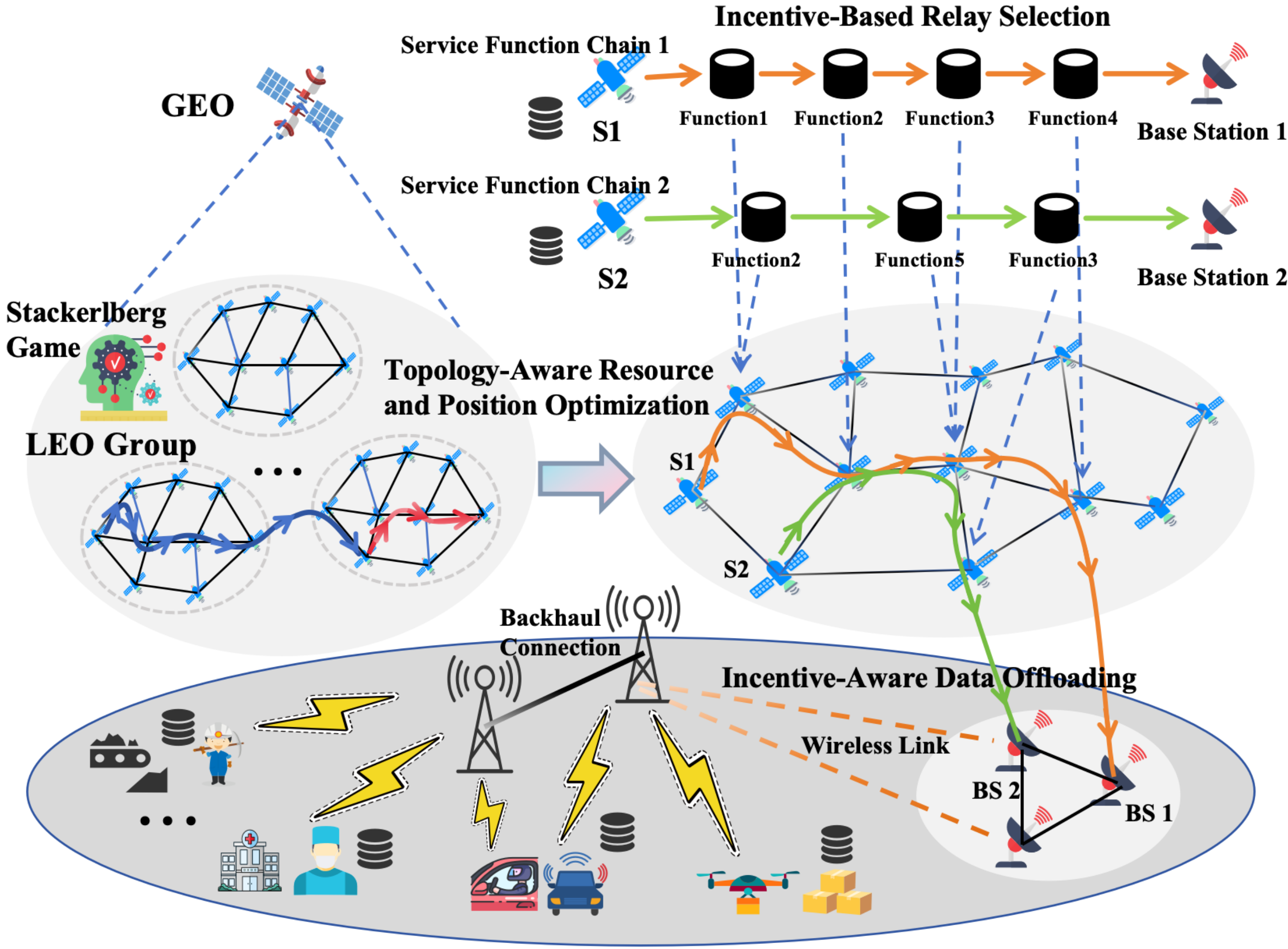}
\vspace{-2em}
\caption{Illustration of integrated air-terrestrial networks with topology and relay optimization.}
\label{air-terrestrial topology}
%\vspace{-2em}
\end{figure}

%=================================
\subsection{Data Offloading}\label{section5-d}
%=================================
Data offloading in satellite-terrestrial networks has emerged as a crucial solution for mitigating the limitations of terrestrial backhaul and enhancing service coverage \cite{11023214}, particularly in ultra-dense or geographically remote scenarios (as shown in Fig.\ref{air-terrestrial topology}). However, due to heterogeneous ownership, cost asymmetries, and diverse service objectives among entities (e.g., satellite operators, ground service providers, and users), incentive-compatible mechanisms become essential to ensure effective offloading. Game-theoretic models, especially Stackelberg games, Bayesian games, bargaining games, and coalition games, have become standard tools for aligning offloading strategies with economic rationality, topology dynamics, and QoS constraints.

\subsubsection{Bayesian Games}
The authors in \cite{anastasopoulos2012feedback} pioneer the idea of incentive-aware feedback suppression in satellite multicast networks by modeling the decision-making process of users as a Bayesian game. In this framework, each user acts as an independent player who strategically decides whether to send a feedback message based on private information such as residual battery power and round-trip time. The Nash equilibrium is reached by deriving a threshold strategy: each user sends a feedback message only if their normalized cost falls below a certain equilibrium cutoff value, which is uniquely determined by solving a fixed-point equation that incorporates the distributions of energy and delay costs across all users. While this work does not directly offload data, it minimizes unnecessary return traffic, effectively reducing load and preserving bandwidth—a form of implicit offloading. Simulation results demonstrate that the proposed scheme significantly suppresses feedback messages, limiting them to fewer than 30 even with \(10^6\) users under high battery conditions, while maintaining near-zero feedback latency, outperforming traditional timer-based approaches. However, unlike later works, it does not consider dynamic traffic patterns or mobility, which are central in modern offloading scenarios.

\subsubsection{Stackelberg Games}
Expanding into bidirectional operator interactions, the authors in \cite{deng2020ultra} develop a Stackelberg game framework for inter-operator offloading between terrestrial and satellite systems. Similar to \cite{anastasopoulos2012feedback}, this work models asymmetric information and incentive-driven behavior, but focuses on spectrum leasing and pricing. In this hierarchical model, the satellite operator acts as the leader (first-mover) to set prices or leasing terms, while the terrestrial operator acts as the follower, adjusting user associations accordingly. The Stackelberg equilibrium is achieved through backward induction: the follower’s best response is first derived under fixed leader strategies, and then the leader optimizes its strategy anticipating the follower’s reaction. This iterative process continues until neither party has an incentive to deviate, ensuring convergence to a stable equilibrium. However, unlike \cite{wang2022dynamic}, this model lacks real-time adaptability to traffic burstiness and mobility, which limits its scalability. The authors in \cite{wang2022dynamic} advance the modeling paradigm by introducing a Stackelberg Mean Field Game (SMFG) to manage massive user offloading in ultra-dense LEO networks. This framework cleverly combines the hierarchical structure of a Stackelberg game (satellites vs. a large population of users) with the scalability of a Mean Field Game to handle the interactions within the massive user population. Unlike \cite{deng2020ultra}, this framework captures both inter-agent influence and dynamic pricing, allowing satellites and users to make decentralized yet coordinated decisions. However, the current model assumes homogeneous service demands, whereas real-world systems often involve delay-sensitive, privacy-critical, or computationally heavy offloading tasks. This assumption constrains the model’s applicability to multi-service 6G contexts. Different from terrestrial–satellite coordination, the authors in \cite{chen2023remote} focus on intra-satellite offloading from LEO to GEO nodes for remote sensing data. They formulate a two-stage hybrid strategy combining data partitioning and a bargaining game-based GEO resource allocation mechanism, specifically a two-way bargaining game under dynamic topology (TWBGS-DT). In this scheme, LEO satellites act as buyers seeking cache space, GEO satellites serve as sellers offering storage resources, and a centralized SDN controller acts as the market administrator facilitating negotiations over pricing and cache allocation. The authors also introduce a non-uniform time slot scheduling mechanism (NUTSDM) to handle link dynamics. In contrast to the static assumptions in \cite{deng2020ultra}, this work emphasizes topological shifts and path diversity, though it does not explicitly integrate differentiated service guarantees or time-critical priorities.

\subsubsection{Distributed decision-making}
Lastly, the authors in \cite{qin2023service} address data delivery and resource orchestration via service function chaining (SFC). In their model, each SFC acts as an independent player, selecting a multi-hop routing and NF embedding strategy to minimize its own service completion latency. While this approach provides flexibility and scalability, unlike \cite{wang2022dynamic} and \cite{chen2023remote}, it lacks an explicit game-theoretic incentive mechanism to encourage data offloading cooperation. Economic considerations for satellite nodes or intermediate service agents are implicitly assumed rather than optimized through a formal game model, which could limit deployment in competitive multi-operator environments.

In summary, recent advances in data offloading mechanisms for satellite-terrestrial networks demonstrate the effectiveness of incentive-driven strategies in aligning resource utilization, service quality, and economic interests. Models ranging from Bayesian games and Stackelberg games to bargaining games and coalition-based SFC address various stakeholder interactions, topology changes, and scale challenges. However, future research should focus on heterogeneous and time-sensitive offloading, where diverse service classes require differentiated pricing, latency-aware bargaining, and reputation-aware cooperation. Additionally, multi-agent incentives under partial observability and long-term fairness remain underexplored but are critical for sustainable offloading in dense and dynamic 6G satellite infrastructures.

\subsection{Relay Selection}\label{section5-e}
In hybrid satellite-terrestrial networks (HSTNs), relay selection plays a crucial role in maintaining reliable transmission, especially when direct satellite links are degraded or unavailable \cite{10938344}. The involvement of terrestrial relays introduces the possibility of data offloading from satellites to the ground infrastructure, which can significantly improve spectrum efficiency and service coverage. However, effective relay participation relies on incentive-compatible mechanisms that ensure secondary users, acting as relays, receive fair compensation or utility gains in return for their resources and cooperation. Auction-based models, especially truthful mechanisms like VCG auction, are an effective tool to coordinate such relay assignments in a distributed yet efficient manner.

\subsubsection{Auction Mechanisms}
The authors in \cite{zhang2020vcg} propose a VCG auction–based relay selection framework for hybrid satellite-terrestrial overlay networks, in which the primary satellite network acts as the auctioneer and the terrestrial secondary relays serve as bidders. The terrestrial relays are incentivized to assist satellite users through cooperative spectrum sharing in exchange for spectrum access. Unlike traditional matching-based relay selection schemes that rely on fixed link metrics, this work incorporates economic valuation and bid-driven cooperation, ensuring truthful participation while achieving high spectrum utilization. The winner selection is formulated as an assignment problem and efficiently solved by the Hungarian algorithm. In the VCG mechanism, truth-telling is a dominant strategy for each bidder, leading to a Nash equilibrium where no participant can gain by unilaterally deviating from their true valuation. Simulation results demonstrate that the proposed auction mechanism significantly outperforms baseline methods such as location-based and random relay selection while maintaining fairness among secondary users through efficient time-slot redistribution. However, compared to dynamic offloading models such as \cite{wang2022dynamic} in the previous subsection, this relay selection mechanism operates in static, block-based auctions, lacking the ability to adapt to real-time channel variations, user mobility, or service heterogeneity. Moreover, the current model assumes each relay assists only one primary receiver per auction round, overlooking multi-relay chaining or load balancing under high-demand scenarios. In summary, incentive-compatible relay selection mechanisms such as the VCG auction offer strong potential for enabling trustworthy and efficient relay cooperation in satellite-terrestrial networks. They strike a balance between offloading transmission duties and rewarding relays for their contributions. 

Across the five categories reviewed, a common thread emerges: the integration of incentive mechanisms is not merely a theoretical construct, but a practical necessity for the efficient operation of large-scale, heterogeneous 6G SAGIN. These mechanisms enable decentralized coordination, fair resource allocation, and adaptive cooperation among rational agents, be it satellites, terrestrial operators, or end users. While substantial progress has been made, several cross-cutting challenges remain. First, many models adopt static or snapshot-based optimization, which fails to capture the temporal dynamics inherent in satellite mobility, user behavior, and traffic bursts. Second, existing works often assume homogeneous services and user types, overlooking the increasing service diversity and QoS heterogeneity in future networks. Third, the majority of auction or game-theoretic frameworks emphasize truthfulness and equilibrium, but lack integration with learning-based adaptation to handle incomplete information or evolving environments.

\section{Conclusions and Future Research Directions}\label{section:concl_future}

\textcolor{black}{This paper has presented a comprehensive review of applications of incentive mechanisms for resource management in satellite networks. These mechanisms are categorized by methodological paradigms such as game theory, auction, matching, and learning. To establish a comprehensive and coherent understanding across these paradigms, we introduce a unified comparison framework that benchmarks all reviewed mechanisms under consistent evaluation dimensions. The selected dimensions, involving communication overhead, convergence speed, fairness, and scalability are fundamental to the efficiency and practicality of incentive-driven satellite resource management. Table~\ref{tab:comparison} summarizes the key characteristics and trade-offs among these mechanisms. It can be observed that each incentive mechanism exhibits distinctive trade-offs. Stackelberg and cooperative games are well-suited to hierarchical or collaborative optimization but require extensive information exchange to achieve equilibrium. Auction and matching mechanisms provide faster convergence and high fairness under truthfulness or stability guarantees while maintaining relatively low signaling costs. Learning-based frameworks offer strong adaptability and scalability for highly dynamic satellite environments but typically converge more slowly due to the exploration–exploitation process. This unified framework bridges the methodological boundaries across the above Sections, forming a coherent foundation for comparative evaluation and future hybrid incentive designs in large-scale satellite networks. Furthermore, based on our findings, we outline promising directions for future research as follows.}

\begin{table*}[htbp]
\centering
\footnotesize
\renewcommand{\arraystretch}{1.4}
\caption{\textcolor{black}{Unified Comparison of Incentive Mechanisms for Satellite Networks}}
\label{tab:comparison}
\begin{tabular}{!{\vrule width0.6pt}c|c|c|c|c|c|c!{\vrule width0.6pt}}
\hline
\textbf{Mechanisms} & \textbf{Typical Framework} & \textbf{Overhead} & \textbf{Convergence} & \textbf{Fairness} & \textbf{Scalability} & \textbf{Refs.} \\
\hline
Stackelberg game & \makecell{Hierarchical leader-\\follower optimization} & \makecell{Medium-high\\(iterative exchange)} & \makecell{Fast} & \makecell{Moderate\\(leader-oriented)} & High & \cite{zhang2022joint,zhu2020two} \\
\hline
Cooperative game & \makecell{Coalition/potential game\\for joint optimization} & \makecell{High\\(players cooperation)} & Moderate & \makecell{High\\(collective utility)} & Medium & \cite{wang2022qos, du2019coalitional} \\
\hline
Non-Cooperative game & \makecell{Independent strategy learning\\ toward Nash equilibrium} & Low & Moderate & Low-moderate & High & \cite{wang2019game,chen2020correlated} \\
\hline
Auction-based scheme & \makecell{Vickrey/double\\auction mechanism} & Low-medium & Fast & \makecell{High\\(truthful bidding)} & High & \cite{jin2020double, huang2024profit} \\
\hline
Matching game & \makecell{Stable pairing via\\many-to-one mapping} & Medium & Fast & \makecell{High\\(stability criterion)} & Medium & \cite{yu2024computation,fang2022matching} \\
\hline
Learning-based scheme & \makecell{Reinforcement/DRL-\\enhanced incentive design} & \makecell{Low\\(local updates)} & Adaptive & \makecell{Moderate\\(reward-dependent)} & High & \cite{yuan2023game,yang2024joint} \\
\hline
\end{tabular}
\end{table*}

\subsection{Lightweight Constraint-aware Inference for Satellite Networks}

A promising research direction is to focus on the co-design of lightweight, resource-aware agents capable of performing split inference between satellite and ground infrastructure. These agents must operate under strict constraints on computation, energy, memory, and spectrum, with each resource treated as a first-class, tradable element within the incentive layer \cite{zhang2025toward111}. Mechanisms such as inference-time auctions, energy-budget contracts, and token-based prioritization can be employed to allocate these limited resources across competing tasks. This approach aligns the operational cost of AI with overall network objectives, supporting realtime responsiveness, graceful degradation during overload, and autonomous functionality when backhaul connectivity is constrained. Performance indicators may include deadline satisfaction rates, task success normalized by energy or spectrum usage, adaptability to varying inference splits, and the stability of end-to-end resource allocation under preemptive or interruptible execution.

\subsection{Learning-Based Mechanism Design for Dynamic SAGIN}
Another promising direction lies in the automated design of incentive mechanisms for nonstationary and partially observable SAGIN. This involves learning mechanism-generation policies via self-play and model-based reinforcement learning in digital twins that simulate orbital dynamics, link intermittency, and strategic user behavior \cite{11049053}. Unlike static, hand-crafted schemes, such methods enable dynamic auctions, adaptive contracts, and congestion-aware pricing that can evolve with changing environments. Robustness and fairness can be embedded directly into the training loop through population-based methods and adversarial testing. This direction enables rapid adaptation without manual redesign. Relevant evaluation criteria include social welfare, energy and cost efficiency, re-optimization latency, robustness to incomplete information or adversarial behavior, and service fairness across network entities.

\subsection{Security-Aware Incentive Design for Satellite Networks}
A third research direction focuses on developing incentive mechanisms that are resilient to adversarial behaviors and untrustworthy participants in satellite networks. As threats such as jamming, spoofing, misreporting, and collusion become more prevalent in space–air–ground systems, incentive design must integrate security primitives including cryptographic verification, trust and reputation systems, and anomaly or misbehavior detection \cite{tang2022blockchain}. Learning-based approaches can dynamically adapt incentives in response to evolving threat patterns or detected anomalies. Furthermore, mechanisms must balance the trade-off between system security and efficiency, ensuring that protection does not degrade performance under normal conditions. Evaluation metrics may include robustness under targeted attacks, accuracy and latency of threat detection, performance degradation under adversarial scenarios, and resource overhead introduced by security enforcement.

\subsection{Secure Multi-domain Resource Exchanges}
A final research direction concerns the design of incentive mechanisms that enable secure and efficient resource exchanges across heterogeneous satellite–terrestrial ecosystems, including LEO constellations, terrestrial infrastructures, and HAPS platforms. These mechanisms must handle asymmetric information, intermittent connectivity, and diverse quality-of-service requirements, while ensuring fairness, verifiability, and robustness against adversarial behaviors. Promising approaches include the use of lightweight cryptographic primitives, privacy-preserving bidding protocols, and collusion-resistant auction designs to safeguard exchanges without introducing excessive overhead. Future work should also consider the integration of reputation and trust systems with cross-layer defenses, allowing resource markets to remain resilient in highly dynamic and adversarial environments. Evaluation metrics may include fairness across domains, resilience to manipulation, efficiency of exchange under connectivity disruptions, and security overhead relative to achieved performance.

%==========================================================
%\cite{jiang2023reinforcement} ,\cite{liu2024high}, \cite{al2023artificial} \\

%==================

%\begin{thebibliography}{100}
\bibliographystyle{IEEEtran}
\bibliography{REF}

\begin{thebibliography}{100}
\providecommand{\url}[1]{#1}
\csname url@samestyle\endcsname
\providecommand{\newblock}{\relax}
\providecommand{\bibinfo}[2]{#2}
\providecommand{\BIBentrySTDinterwordspacing}{\spaceskip=0pt\relax}
\providecommand{\BIBentryALTinterwordstretchfactor}{4}
\providecommand{\BIBentryALTinterwordspacing}{\spaceskip=\fontdimen2\font plus
\BIBentryALTinterwordstretchfactor\fontdimen3\font minus \fontdimen4\font\relax}
\providecommand{\BIBforeignlanguage}[2]{{%
\expandafter\ifx\csname l@#1\endcsname\relax
\typeout{** WARNING: IEEEtran.bst: No hyphenation pattern has been}%
\typeout{** loaded for the language `#1'. Using the pattern for}%
\typeout{** the default language instead.}%
\else
\language=\csname l@#1\endcsname
\fi
#2}}
\providecommand{\BIBdecl}{\relax}
\BIBdecl

\bibitem{10679152}
R.~Zhang, H.~Du, Y.~Liu, D.~Niyato, J.~Kang, Z.~Xiong, A.~Jamalipour, and D.~In~Kim, ``Generative ai agents with large language model for satellite networks via a mixture of experts transmission,'' \emph{IEEE Journal on Selected Areas in Communications}, vol.~42, no.~12, pp. 3581--3596, 2024.

\bibitem{kodheli2020satellite}
O.~Kodheli, E.~Lagunas, N.~Maturo, S.~K. Sharma, B.~Shankar, J.~F.~M. Montoya, J.~C.~M. Duncan, D.~Spano, S.~Chatzinotas, S.~Kisseleff \emph{et~al.}, ``Satellite communications in the new space era: A survey and future challenges,'' \emph{IEEE Communications Surveys \& Tutorials}, vol.~23, no.~1, pp. 70--109, 2020.

\bibitem{giordani2020non}
M.~Giordani, M.~Polese \emph{et~al.}, ``Non-terrestrial networks in the 6g era: Challenges, opportunities, and open issues,'' \emph{IEEE Communications Magazine}, vol.~58, no.~9, pp. 62--68, 2020.

\bibitem{sui2025multi}
Z.~Sui, Q.~Luo, Z.~Liu, M.~Temiz, L.~Musavian, C.~Masouros, Y.~L. Guan, P.~Xiao, and L.~Hanzo, ``Multi-functional chirp signalling for next-generation multi-carrier wireless networks: Communications, sensing and {ISAC} perspectives,'' \emph{arXiv preprint arXiv:2508.06022}, 2025.

\bibitem{routray2019satellite}
S.~K. Routray and H.~M. Hussein, ``Satellite based iot networks for emerging applications,'' \emph{arXiv preprint arXiv:1904.00520}, 2019.

\bibitem{peng2018review}
Y.~Peng, T.~Dong, R.~Gu, Q.~Guo, J.~Yin, Z.~Liu, T.~Zhang, and Y.~Ji, ``A review of dynamic resource allocation in integrated satellite and terrestrial networks,'' in \emph{International Conference on Networking and Network Applications (NaNA)}, 2018, pp. 127--132.

\bibitem{zhou2021machine}
D.~Zhou, M.~Sheng, Y.~Wang, J.~Li, and Z.~Han, ``Machine learning-based resource allocation in satellite networks supporting internet of remote things,'' \emph{IEEE Transactions on Wireless Communications}, vol.~20, no.~10, pp. 6606--6621, 2021.

\bibitem{10670196}
R.~Zhang, H.~Du, D.~Niyato, J.~Kang, Z.~Xiong, A.~Jamalipour, P.~Zhang, and D.~I. Kim, ``Generative ai for space-air-ground integrated networks,'' \emph{IEEE Wireless Communications}, vol.~31, no.~6, pp. 10--20, 2024.

\bibitem{11185315}
Z.~Sui, Z.~Liu, L.~Musavian, L.-L. Yang, and L.~Hanzo, ``Generalized spatial modulation aided affine frequency division multiplexing,'' \emph{IEEE Transactions on Wireless Communications}, pp. 1--1, 2025.

\bibitem{10183832}
Z.~Sui, S.~Yan, H.~Zhang, S.~Sun, Y.~Zeng, L.-L. Yang, and L.~Hanzo, ``Performance analysis and approximate message passing detection of orthogonal time sequency multiplexing modulation,'' \emph{IEEE Transactions on Wireless Communications}, vol.~23, no.~3, pp. 1913--1928, 2024.

\bibitem{10250854}
Z.~Sui, H.~Zhang, S.~Sun, L.-L. Yang, and L.~Hanzo, ``Space-time shift keying aided {OTFS} modulation for orthogonal multiple access,'' \emph{IEEE Transactions on Communications}, vol.~71, no.~12, pp. 7393--7408, 2023.

\bibitem{10217007}
Z.~Ye, S.~Yan, Z.~Sui, H.~Zhang, and S.~Sun, ``Successive interference cancellation aided bidirectional soft decision feedback equalization for {OTFS} systems,'' \emph{IEEE Wireless Communications Letters}, vol.~12, no.~12, pp. 2028--2032, 2023.

\bibitem{10129061}
Z.~Sui, H.~Zhang, Y.~Xin, T.~Bao, L.-L. Yang, and L.~Hanzo, ``Low complexity detection of spatial modulation aided {OTFS} in doubly-selective channels,'' \emph{IEEE Transactions on Vehicular Technology}, vol.~72, no.~10, pp. 13\,746--13\,751, 2023.

\bibitem{zhang2022joint}
X.~Zhang, X.~Qin, B.~Qian, T.~Ma, and H.~Zhou, ``Joint mode selection and dynamic pricing in ultra dense leo integrated satellite-terrestrial networks,'' in \emph{2022 IEEE/CIC International Conference on Communications in China (ICCC)}.\hskip 1em plus 0.5em minus 0.4em\relax IEEE, 2022, pp. 1090--1094.

\bibitem{lin2023leo}
X.~Lin, A.~Liu, C.~Han, X.~Liang, K.~Pan, and Z.~Gao, ``Leo satellite and uavs assisted mobile edge computing for tactical ad-hoc network: A game theory approach,'' \emph{IEEE Internet of Things Journal}, vol.~10, no.~23, pp. 20\,560--20\,573, 2023.

\bibitem{peng2024cloud}
Y.~Peng, X.~Guang, X.~Zhang, L.~Liu, C.~Wu, and L.~Huang, ``A cloud-edge collaborative computing framework using potential games for space-air-ground integrated {IoT},'' \emph{EURASIP Journal on Advances in Signal Processing}, vol. 2024, no.~1, p.~54, 2024.

\bibitem{wang2019game}
Y.~Wang, J.~Yang, X.~Guo, and Z.~Qu, ``A game-theoretic approach to computation offloading in satellite edge computing,'' \emph{IEEE Access}, vol.~8, pp. 12\,510--12\,520, 2019.

\bibitem{tang2024digital}
X.~Tang, Q.~Chen, R.~Yu, and X.~Li, ``Digital twin-empowered task assignment in aerial mec network: A resource coalition cooperation approach with generative model,'' \emph{IEEE Transactions on Network Science and Engineering}, 2024.

\bibitem{luong2016data}
N.~C. Luong, D.~T. Hoang, P.~Wang, D.~Niyato, D.~I. Kim, and Z.~Han, ``Data collection and wireless communication in internet of things ({IoT}) using economic analysis and pricing models: A survey,'' \emph{IEEE Communications Surveys \& Tutorials}, vol.~18, no.~4, pp. 2546--2590, 2016.

\bibitem{luong2017resource}
N.~C. Luong, P.~Wang, D.~Niyato, Y.~Wen, and Z.~Han, ``Resource management in cloud networking using economic analysis and pricing models: A survey,'' \emph{IEEE Communications Surveys \& Tutorials}, vol.~19, no.~2, pp. 954--1001, 2017.

\bibitem{zhu2020two}
X.~Zhu, C.~Jiang, L.~Kuang, Z.~Zhao, and S.~Guo, ``Two-layer game based resource allocation in cloud based integrated terrestrial-satellite networks,'' \emph{IEEE Transactions on Cognitive Communications and Networking}, vol.~6, no.~2, pp. 509--522, 2020.

\bibitem{wang2022qos}
Y.~Wang, X.~Qin, Z.~Tang, T.~Ma, X.~Zhang, and H.~Zhou, ``{QoS}-centric handover for civil aviation aircraft access in ultra-dense leo satellite networks,'' in \emph{2022 IEEE/CIC International Conference on Communications in China (ICCC)}.\hskip 1em plus 0.5em minus 0.4em\relax IEEE, 2022, pp. 1085--1089.

\bibitem{du2019coalitional}
B.~Du, R.~Xue, L.~Zhao, and V.~C. Leung, ``Coalitional graph game for air-to-air and air-to-ground cognitive spectrum sharing,'' \emph{IEEE Transactions on Aerospace and Electronic Systems}, vol.~56, no.~4, pp. 2959--2977, 2019.

\bibitem{xu2023adaptive}
X.~Xu, Q.~Wang, S.~Li, H.~Xu, H.~Zhao, and Z.~Han, ``An adaptive dual-mode task-oriented resource management strategy for geo relay systems,'' \emph{IEEE Transactions on Mobile Computing}, vol.~23, no.~5, pp. 4303--4317, 2023.

\bibitem{jin2020double}
Y.~Jin, H.~Yao, and T.~Mai, ``Double auction game-based computing resource allocation in leo satellite system,'' in \emph{International Wireless Communications and Mobile Computing}, 2020, pp. 274--279.

\bibitem{huang2024profit}
J.~Huang, R.~Xing, X.~Ma, A.~Zhou, and S.~Wang, ``Profit-aware task allocation in satellite computing,'' in \emph{2024 IEEE International Conference on Web Services (ICWS)}.\hskip 1em plus 0.5em minus 0.4em\relax IEEE, 2024, pp. 696--706.

\bibitem{li2022secure}
X.~Li, R.~Yao, Y.~Fan, P.~Wang, N.~Qi, N.~I. Miridakis, and T.~A. Tsiftsis, ``Secure spectrum-energy efficiency tradeoff based on stackelberg game in a two-way relay cognitive satellite terrestrial network,'' \emph{IEEE Wireless Communications Letters}, vol.~11, no.~8, pp. 1679--1683, 2022.

\bibitem{han2020spatial}
C.~Han, L.~Huo, X.~Tong, H.~Wang, and X.~Liu, ``Spatial anti-jamming scheme for internet of satellites based on the deep reinforcement learning and stackelberg game,'' \emph{IEEE Transactions on Vehicular Technology}, vol.~69, no.~5, pp. 5331--5342, 2020.

\bibitem{zhao2022tensor}
R.~Zhao, L.~T. Yang, D.~Liu, X.~Deng, and Y.~Mo, ``A tensor-based truthful incentive mechanism for blockchain-enabled space-air-ground integrated vehicular crowdsensing,'' \emph{IEEE Transactions on Intelligent Transportation Systems}, vol.~23, no.~3, pp. 2853--2862, 2022.

\bibitem{ma2022blockchain}
Z.~Ma, Y.~Wang, J.~Li, and Y.~Liu, ``A blockchain based privacy-preserving incentive mechanism for internet of vehicles in satellite-terrestrial crowdsensing,'' \emph{Wireless Communications and Mobile Computing}, vol. 2022, no.~1, p. 4036491, 2022.

\bibitem{shi2021clap}
X.~Shi, D.~Yu, M.~Fu, and W.-A. Zhang, ``{CLAP}: A contract-based incentive mechanism for cooperative localization balancing localization accuracy and location privacy,'' \emph{IEEE Internet of Things Journal}, vol.~9, no.~9, pp. 6678--6687, 2021.

\bibitem{huang2025consolidated}
L.~Huang, P.~Liu, X.~Chen, C.~Jiang, L.~Kuang, and J.~Lu, ``A consolidated game framework for cooperative defense against cross-domain cyber attacks in satellite-enabled internet of things,'' \emph{IEEE Internet of Things Journal}, vol.~12, no.~9, pp. 12\,853--12\,868, 2025.

\bibitem{du2022game}
J.~Du, C.~Wang, C.~Wang, F.~Jiang, and W.~Wang, ``Game theoretic traffic scheduling and hybrid qos-aware routing in software defined satellite networks,'' in \emph{2022 14th International Conference on Wireless Communications and Signal Processing (WCSP)}.\hskip 1em plus 0.5em minus 0.4em\relax IEEE, 2022, pp. 859--864.

\bibitem{deng2020ultra}
R.~Deng, B.~Di, S.~Chen, S.~Sun, and L.~Song, ``Ultra-dense leo satellite offloading for terrestrial networks: How much to pay the satellite operator?'' \emph{IEEE Transactions on Wireless Communications}, vol.~19, no.~10, pp. 6240--6254, 2020.

\bibitem{luong2018applications}
N.~C. Luong, P.~Wang, D.~Niyato, Y.-C. Liang, Z.~Han, and F.~Hou, ``Applications of economic and pricing models for resource management in 5g wireless networks: A survey,'' \emph{IEEE Communications Surveys \& Tutorials}, vol.~21, no.~4, pp. 3298--3339, 2018.

\bibitem{qiu2022applications}
H.~Qiu, K.~Zhu, N.~C. Luong, C.~Yi, D.~Niyato, and D.~I. Kim, ``Applications of auction and mechanism design in edge computing: A survey,'' \emph{IEEE Transactions on Cognitive Communications and Networking}, vol.~8, no.~2, pp. 1034--1058, 2022.

\bibitem{jiang2024game}
W.~Jiang, H.~Han, M.~He, and W.~Gu, ``When game theory meets satellite communication networks: A survey,'' \emph{Computer Communications}, vol. 217, pp. 208--229, 2024.

\bibitem{wang2018high}
L.~Wang, C.~Jiang, L.~Kuang, S.~Wu, H.~Huang, and Y.~Qian, ``High-efficient resource allocation in data relay satellite systems with users behavior coordination,'' \emph{IEEE Transactions on Vehicular Technology}, vol.~67, no.~12, pp. 12\,072--12\,085, 2018.

\bibitem{xiaogang2016survey}
Q.~Xiaogang, M.~Jiulong, W.~Dan, L.~Lifang, and H.~Shaolin, ``A survey of routing techniques for satellite networks,'' \emph{Journal of Communications and Information Networks}, vol.~1, no.~4, pp. 66--85, 2016.

\bibitem{fontanesi2025artificial}
G.~Fontanesi, F.~Ort{\'\i}z, E.~Lagunas, L.~M. Garc{\'e}s-Socarr{\'a}s, V.~M. Baeza, M.~{\'A}. V{\'a}zquez, J.~A. V{\'a}squez-Peralvo, M.~Minardi, H.~N. Vu, P.~J. Honnaiah \emph{et~al.}, ``Artificial intelligence for satellite communication: A survey,'' \emph{IEEE Communications Surveys \& Tutorials}, to appear.

\bibitem{chen2024survey}
Q.~Chen, Z.~Guo, W.~Meng, S.~Han, C.~Li, and T.~Q. Quek, ``A survey on resource management in joint communication and computing-embedded sagin,'' \emph{IEEE Communications Surveys \& Tutorials}, vol.~27, no.~3, pp. 1911--1954, 2024.

\bibitem{smith2021ran}
R.~Smith, C.~Freeberg, T.~Machacek, and V.~Ramaswamy, ``An o-ran approach to spectrum sharing between commercial 5g and government satellite systems,'' in \emph{IEEE MILCOM}, 2021, pp. 739--744.

\bibitem{yan2022electromagnetic}
P.~Yan, F.~Chu, L.~Jia, and N.~Qi, ``Electromagnetic barrier assisted dynamic spectrum access in satellite internet communication confrontation,'' \emph{IET Communications}, vol.~16, no.~18, pp. 2145--2157, 2022.

\bibitem{wen2024hierarchical}
X.~Wen, Y.~Ruan, Y.~Li, C.~Pan, M.~Elkashlan, R.~Zhang, and T.~Li, ``A hierarchical game framework for win-win resource trading in cognitive satellite terrestrial networks,'' \emph{IEEE Transactions on Wireless Communications}, vol.~23, no.~10, pp. 13\,530--13\,544, 2024.

\bibitem{xu2022hierarchical}
Q.~Xu, Z.~Su, D.~Fang, and Y.~Wu, ``Hierarchical bandwidth allocation for social community-oriented multicast in space-air-ground integrated networks,'' \emph{IEEE Transactions on Wireless Communications}, vol.~22, no.~3, pp. 1915--1930, 2022.

\bibitem{xu2022ubiquitous}
Q.~Xu, Z.~Su, R.~Lu, and S.~Yu, ``Ubiquitous transmission service: Hierarchical wireless data rate provisioning in space-air-ocean integrated networks,'' \emph{IEEE Transactions on Wireless Communications}, vol.~21, no.~9, pp. 7821--7836, 2022.

\bibitem{liu2024demand}
Y.~Liu, Y.~Wang, and Y.~Shen, ``Demand-aware distributed link allocation in a multilayer heterogeneous satellite network: A game theory approach,'' \emph{IEEE Internet of Things Journal}, vol.~11, no.~10, pp. 17\,629--17\,641, 2024.

\bibitem{liu2021research}
Z.~Liu, X.~Zha, X.~Ren, and Q.~Yao, ``Research on handover strategy of {LEO} satellite network,'' in \emph{International Conference on Big Data and Informatization Education}, 2021, pp. 188--194.

\bibitem{zhang2021potential}
X.~Zhang, B.~Zhang, D.~Guo, K.~An, S.~Qi, and G.~Wu, ``Potential game-based radio resource allocation in uplink multibeam satellite {IoT} networks,'' \emph{IEEE Transactions on Aerospace and Electronic Systems}, vol.~57, no.~6, pp. 4269--4279, 2021.

\bibitem{wang2020distributed}
J.~Wang, B.~Zhang, L.~Jia, B.~Zhao, and D.~Guo, ``A distributed collaborative game-theoretic approach in cognitive satellite communication networks,'' \emph{IEEE Access}, vol.~8, pp. 129\,446--129\,460, 2020.

\bibitem{gao2023sum}
Z.~Gao, A.~Liu, X.~Xu, X.~Liang, C.~Han, and X.~Liu, ``Sum data minimization in leo satellite-uav integrated multi-tier computing networks: A game-theoretic multiple access approach,'' \emph{IEEE Transactions on Communications}, vol.~72, no.~3, pp. 1701--1715, 2023.

\bibitem{gao2021files}
Z.~Gao, A.~Liu, C.~Han, and X.~Liang, ``Files delivery and share optimization in leo satellite-terrestrial integrated networks: A noma based coalition formation game approach,'' \emph{IEEE Transactions on Vehicular Technology}, vol.~71, no.~1, pp. 831--843, 2021.

\bibitem{kim2020cognitive}
S.~Kim, ``Cognitive satellite spectrum management scheme based on the cooperative solidarity values,'' \emph{IEEE Access}, vol.~8, pp. 113\,837--113\,846, 2020.

\bibitem{li2024resource}
Z.~Li, C.~Jiang, J.~Sun, and J.~Lu, ``Resource collaboration between satellite and wide-area mobile base stations in integrated satellite-terrestrial network,'' \emph{IEEE Transactions on Mobile Computing}, 2024.

\bibitem{kim2021bargaining}
S.~Kim, ``Bargaining-based spectrum allocation algorithm for high-speed railway communications,'' \emph{IEEE Access}, vol.~9, pp. 71\,651--71\,659, 2021.

\bibitem{li2021agent}
F.~Li, K.-Y. Lam, Z.~Sheng, W.~Lu, X.~Liu, and L.~Wang, ``Agent-based spectrum management scheme in satellite communication systems,'' \emph{IEEE Transactions on Vehicular Technology}, vol.~70, no.~3, pp. 2877--2881, 2021.

\bibitem{hu2020multi}
X.~Hu, X.~Liao, Z.~Liu, S.~Liu, X.~Ding, M.~Helaoui, W.~Wang, and F.~M. Ghannouchi, ``Multi-agent deep reinforcement learning-based flexible satellite payload for mobile terminals,'' \emph{IEEE Transactions on Vehicular Technology}, vol.~69, no.~9, pp. 9849--9865, 2020.

\bibitem{wan2021study}
W.~Wan, Y.~Peng, and N.~Medhin, ``Study of optimal power control by cooperative game for satellite communication subsystems,'' in \emph{6th International Conference on Communication and Electronics Systems}, 2021, pp. 737--746.

\bibitem{wang2021game}
J.~Wang, B.~Zhang, B.~Zhao, G.~Ding, and D.~Guo, ``A game-theoretical learning approach for spectrum trading in cognitive satellite-terrestrial networks,'' \emph{IEEE Communications Letters}, vol.~25, no.~9, pp. 3065--3069, 2021.

\bibitem{zhang2020auction}
X.~Zhang, D.~Guo, K.~An, G.~Zheng, S.~Chatzinotas, and B.~Zhang, ``Auction-based multichannel cooperative spectrum sharing in hybrid satellite-terrestrial {IoT} networks,'' \emph{IEEE Internet of Things Journal}, vol.~8, no.~8, pp. 7009--7023, 2020.

\bibitem{cheng2023satellite}
L.~Cheng and B.~A. Huberman, ``Satellite resource allocation via dynamic auctions and lsh-based predictions,'' in \emph{IEEE 97th Vehicular Technology Conference (VTC2023-Spring)}, 2023, pp. 1--5.

\bibitem{zhang2020vickrey}
X.~Zhang, K.~An, B.~Zhang, Z.~Chen, Y.~Yan, and D.~Guo, ``Vickrey auction-based secondary relay selection in cognitive hybrid satellite-terrestrial overlay networks with non-orthogonal multiple access,'' \emph{IEEE Wireless Communications Letters}, vol.~9, no.~5, pp. 628--632, 2020.

\bibitem{yang2023lightweight}
N.~Yang, D.~Guo, Y.~Jiao, G.~Ding, and T.~Qu, ``Lightweight blockchain-based secure spectrum sharing in space--air--ground-integrated {IoT} network,'' \emph{IEEE Internet of Things Journal}, vol.~10, no.~23, pp. 20\,511--20\,527, 2023.

\bibitem{chen2020correlated}
Z.~Chen, B.~Zhao, K.~An, G.~Ding, X.~Zhang, J.~Xu, and D.~Guo, ``Correlated equilibrium based distributed power control in cognitive satellite-terrestrial networks,'' \emph{IEEE Communications Letters}, vol.~25, no.~3, pp. 945--949, 2020.

\bibitem{liu2020spectrum}
D.~Liu, X.~Yang, N.~Xiao, and S.~Zhu, ``A spectrum allocation algorithm for satellite-terrestrial communication based on game theory,'' in \emph{2020 5th International Conference on Information Science, Computer Technology and Transportation (ISCTT)}.\hskip 1em plus 0.5em minus 0.4em\relax IEEE, 2020, pp. 350--354.

\bibitem{chamberlain2024spectrum}
J.~Chamberlain, J.~T. Johnson, and D.~Starobinski, ``Spectrum sharing between earth exploration satellite and commercial services: An economic feasibility analysis,'' in \emph{IEEE International Symposium on Dynamic Spectrum Access Networks}, 2024, pp. 197--206.

\bibitem{li2024game}
Y.~Li, B.~Wang, Y.~Xu, and H.~Xu, ``Game theory based maritime area detection for cloud-edge collaboration satellite network,'' \emph{Frontiers in Physics}, vol.~12, p. 1387709, 2024.

\bibitem{wang2020novel}
J.~Wang, D.~Guo, and B.~Zhang, ``A novel spectrum sharing scheme for cognitive satellite networks: A game-theoretic approach,'' in \emph{IEEE WCSP}, 2020, pp. 712--717.

\bibitem{chamberlain2024facilitating}
J.~Chamberlain, D.~Starobinski, and J.~T. Johnson, ``Facilitating spectrum sharing with passive satellite incumbents,'' \emph{IEEE Journal on Selected Areas in Communications}, vol.~42, no.~12, pp. 3719--3733, 2024.

\bibitem{yuan2023game}
X.~Yuan, Y.~Li, F.~Qi, W.~Xie, P.~Li, and W.~Li, ``Game theoretic spectrum sharing for energy-efficient 6g ubiquitous {IoT} networks,'' in \emph{2023 International Conference on Information Processing and Network Provisioning (ICIPNP)}.\hskip 1em plus 0.5em minus 0.4em\relax IEEE, 2023, pp. 469--473.

\bibitem{li2019spectral}
F.~Li, K.-Y. Lam, H.-H. Chen, and N.~Zhao, ``Spectral efficiency enhancement in satellite mobile communications: A game-theoretical approach,'' \emph{IEEE Wireless Communications}, vol.~27, no.~1, pp. 200--205, 2019.

\bibitem{li2019spectrum}
F.~Li, K.-Y. Lam, M.~Jia, K.~Zhao, X.~Li, and L.~Wang, ``Spectrum optimization for satellite communication systems with heterogeneous user preferences,'' \emph{IEEE Systems Journal}, vol.~14, no.~2, pp. 2187--2191, 2019.

\bibitem{han2023anti}
R.~Han, M.~Liu, J.~Wang, L.~Bai, and J.~Liu, ``Anti-jamming strategy for satellite internet of things: Beam switching and optimization,'' \emph{IEEE Internet of Things Journal}, vol.~10, no.~23, pp. 20\,254--20\,263, 2023.

\bibitem{zhang2021game}
H.~Zhang, Q.~Li, Y.~Zhang, and X.~Li, ``Game theory based power allocation method for inter-satellite links in leo/meo two-layered satellite networks,'' in \emph{2021 IEEE/CIC International Conference on Communications in China (ICCC)}.\hskip 1em plus 0.5em minus 0.4em\relax IEEE, 2021, pp. 398--403.

\bibitem{wan2023differential}
W.~Wan, J.~M. Cioffi, Y.~Peng, and B.~S. Howard, ``Differential game analysis of energy efficiency for satellite communication subsystems,'' in \emph{Fifth International Conference on Advances in Computational Tools for Engineering Applications}, 2023, pp. 217--222.

\bibitem{zhang2023rate}
S.~Zhang, S.~Zhang, W.~Yuan, and T.~Q. Quek, ``Rate-splitting multiple access-based satellite--vehicular communication system: A noncooperative game theoretical approach,'' \emph{IEEE Open Journal of the Communications Society}, vol.~4, pp. 430--441, 2023.

\bibitem{wang2020admission}
R.~Wang, W.~Kang, G.~Liu, R.~Ma, and B.~Li, ``Admission control and power allocation for noma-based satellite multi-beam network,'' \emph{IEEE Access}, vol.~8, pp. 33\,631--33\,643, 2020.

\bibitem{liu2022spectrum}
Z.~Liu, W.~Lv, and X.~Ren, ``Spectrum allocation scheme based on stable matching in hierarchical cognitive satellite network,'' \emph{IEEE Access}, vol.~10, pp. 134\,549--134\,556, 2022.

\bibitem{tun2024joint}
Y.~K. Tun, G.~D{\'a}n, Y.~M. Park, and C.~S. Hong, ``Joint uav deployment and resource allocation in thz-assisted mec-enabled integrated space-air-ground networks,'' \emph{IEEE Transactions on Mobile Computing}, vol.~24, no.~5, pp. 3794--3808, 2024.

\bibitem{qin2021joint}
P.~Qin, Y.~Zhu, X.~Zhao, X.~Feng, J.~Liu, and Z.~Zhou, ``Joint 3d-location planning and resource allocation for xaps-enabled c-noma in 6g heterogeneous internet of things,'' \emph{IEEE Transactions on Vehicular Technology}, vol.~70, no.~10, pp. 10\,594--10\,609, 2021.

\bibitem{jia2020joint}
Z.~Jia, M.~Sheng, J.~Li, D.~Zhou, and Z.~Han, ``Joint hap access and {LEO} satellite backhaul in 6g: Matching game-based approaches,'' \emph{IEEE Journal on Selected Areas in Communications}, vol.~39, no.~4, pp. 1147--1159, 2020.

\bibitem{li2025game}
W.~Li, L.~Jia, Y.~Chen, Q.~Chen, J.~Yan, and N.~Qi, ``A game-theoretic approach for satellites beam scheduling and power control in a mega hybrid constellation spectrum sharing scenario,'' \emph{IEEE Internet of Things Journal}, 2025.

\bibitem{xiang2024edge}
B.~Xiang, B.~Zhong, A.~Wang, W.~Mao, and L.~Liu, ``Edge computing collaborative offloading strategy for space-air-ground integrated networks,'' \emph{Concurrency and Computation: Practice and Experience}, vol.~36, no.~21, p. e8214, 2024.

\bibitem{chai2024computation}
Z.-Y. Chai, H.-S. Kang, Y.-L. Li, Y.-J. Zhao, and H.~Huang, ``Computation offloading for integrated satellite-terrestrial internet of vehicles in 6g edge network: A cooperative stackelberg game,'' \emph{IEEE Transactions on Intelligent Transportation Systems}, vol.~25, no.~8, pp. 10\,389--10\,404, 2024.

\bibitem{kim2024hierarchical}
S.~Kim, ``Hierarchical aerial offload computing algorithm based on the stackelberg-evolutionary game model,'' \emph{Computer Networks}, vol. 245, p. 110348, 2024.

\bibitem{wang2024two}
Z.~Wang, B.~Lin, Q.~Ye, and H.~Peng, ``Two-tier task offloading for satellite-assisted marine networks: A hybrid stackelberg-bargaining game approach,'' \emph{IEEE Internet of Things Journal}, vol.~12, no.~9, pp. 13\,047--13\,060, 2024.

\bibitem{gao2023game}
Y.~Gao, Z.~Ji, K.~Zhao, T.~De~Cola, and W.~Li, ``Game-based computation offloading and power allocation for leo constellation networks in distributed and dynamic environment,'' \emph{IEEE Internet of Things Journal}, vol.~11, no.~4, pp. 7040--7058, 2023.

\bibitem{zhang2024dogs}
J.~Zhang, J.~Zhang, F.~Shen, F.~Yan, and Z.~Bu, ``Dogs: Dynamic task offloading in space-air-ground integrated networks with game-theoretic stochastic learning,'' \emph{IEEE Internet of Things Journal}, vol.~12, no.~2, pp. 1655--1672, 2024.

\bibitem{sun2021game}
C.~Sun, X.~Wang, H.~Qiu, and Q.~Zhou, ``Game theoretic self-organization in multi-satellite distributed task allocation,'' \emph{Aerospace Science and Technology}, vol. 112, p. 106650, 2021.

\bibitem{qiao2024orbit}
Y.~Qiao, J.~Luo, and S.~Teng, ``An on-orbit data balancing online algorithm for leo satellite cluster: A repeated stochastic game approach,'' in \emph{International Computing and Combinatorics Conference}.\hskip 1em plus 0.5em minus 0.4em\relax Springer, 2024, pp. 66--78.

\bibitem{li2024joint}
P.~Li, Y.~Wang, Z.~Wang, T.~Wang, and J.~Cheng, ``Joint task offloading and resource allocation strategy for hybrid mec-enabled leo satellite networks: A hierarchical game approach,'' \emph{IEEE Transactions on Communications}, vol.~73, no.~5, pp. 3150--3166, 2024.

\bibitem{wang2024dynamic}
H.~Wang and J.~An, ``Dynamic game based task offloading and resource pricing in leo-multi-access edge computing,'' \emph{Computing}, vol. 106, no.~2, pp. 579--606, 2024.

\bibitem{yu2024computation}
C.~Yu, G.~Xie, and Y.~Liu, ``Computation offloading and pricing mechanism based on stackelberg game,'' in \emph{9th International Conference on Intelligent Computing and Signal Processing}, 2024, pp. 163--166.

\bibitem{fang2022matching}
H.~Fang, Y.~Jia, Y.~Wang, Y.~Zhao, Y.~Gao, and X.~Yang, ``Matching game based task offloading and resource allocation algorithm for satellite edge computing networks,'' in \emph{International Symposium on Networks, Computers and Communications}, 2022, pp. 1--5.

\bibitem{cheng2024energy}
L.~Cheng, G.~Feng, Y.~Sun, S.~Qin, F.~Wang, and T.~Q. Quek, ``Energy-constrained satellite edge computing for satellite-terrestrial integrated networks,'' \emph{IEEE Transactions on Vehicular Technology}, no.~2, pp. 3359--3374, 2024.

\bibitem{zhang2024cost}
X.~Zhang, J.~Liu, Z.~Xiong, Y.~Huang, R.~Zhang, S.~Mao, and Z.~Han, ``Cost-effective hybrid computation offloading in satellite-terrestrial integrated networks,'' \emph{IEEE Internet of Things Journal}, 2024.

\bibitem{yang2024joint}
J.~Yang, Y.~Zhang, Z.~Xiao, and Z.~Han, ``Joint access selection and computation offloading in leo ubiquitous edge computing networks: An alternating drl-based approach,'' \emph{IEEE Transactions on Cognitive Communications and Networking}, 2024.

\bibitem{gao2024edge}
Y.~Gao, J.~Liu, S.~Geng, X.~Zhao, Z.~Chen, and H.~Zhou, ``Edge computing task offloading based on game theory for space-air-ground integrated network,'' in \emph{6th International Conference on Communications, Information System and Computer Engineering}, 2024, pp. 627--631.

\bibitem{fan2024graph}
H.~Fan, C.~Sun, J.~Long, L.~Li, Y.~Huo, and S.~Wang, ``Graph-driven resource allocation strategies in satellite {IoT}: A cooperative game theoretic approach,'' \emph{IEEE Internet of Things Journal}, 2024.

\bibitem{zhao2020novel}
L.~Zhao, K.~Yang, Z.~Tan, X.~Li, S.~Sharma, and Z.~Liu, ``A novel cost optimization strategy for sdn-enabled uav-assisted vehicular computation offloading,'' \emph{IEEE Transactions on Intelligent Transportation Systems}, vol.~22, no.~6, pp. 3664--3674, 2020.

\bibitem{lim2021dynamic}
W.~Y.~B. Lim, J.~S. Ng, Z.~Xiong, D.~Niyato, C.~Miao, and D.~I. Kim, ``Dynamic edge association and resource allocation in self-organizing hierarchical federated learning networks,'' \emph{IEEE Journal on Selected Areas in Communications}, vol.~39, no.~12, pp. 3640--3653, 2021.

\bibitem{li2022joint}
Y.~Li, B.~Yang, H.~Wu, Q.~Han, C.~Chen, and X.~Guan, ``Joint offloading decision and resource allocation for vehicular fog-edge computing networks: A contract-stackelberg approach,'' \emph{IEEE Internet of Things Journal}, vol.~9, no.~17, pp. 15\,969--15\,982, 2022.

\bibitem{chen2025game}
Y.~Chen, Y.~Yang, J.~Hu, Y.~Wu, and J.~Huang, ``A game-theoretical approach for distributed computation offloading in leo satellite-terrestrial edge computing systems,'' \emph{IEEE Transactions on Mobile Computing}, to appear.

\bibitem{gao2023multi}
Y.~Gao, K.~Zhao, T.~De~Cola, W.~Li, Y.~Cui, and P.~Hu, ``Multi-user computation offloading in dynamic environment for leo constellation networks,'' \emph{IEEE Transactions on Vehicular Technology}, vol.~73, no.~3, pp. 4453--4458, 2023.

\bibitem{sastry2002decentralized}
P.~S. Sastry, V.~V. Phansalkar, and M.~Thathachar, ``Decentralized learning of nash equilibria in multi-person stochastic games with incomplete information,'' \emph{IEEE Transactions on systems, man, and cybernetics}, vol.~24, no.~5, pp. 769--777, 2002.

\bibitem{monderer1996potential}
D.~Monderer and L.~S. Shapley, ``Potential games,'' \emph{Games and economic behavior}, vol.~14, no.~1, pp. 124--143, 1996.

\bibitem{wei2025network}
F.~Wei, Y.~Wang, G.~Feng, and S.~Qin, ``Network slicing-enabled computation offloading in satellite-terrestrial edge computing networks: A bi-level game approach,'' \emph{IEEE Internet of Things Journal}, vol.~23, no.~12, pp. 11\,576--11\,587, 2025.

\bibitem{li2022game}
P.~Li, Y.~Wang, and Z.~Wang, ``A game-based joint task offloading and computation resource allocation strategy for hybrid edgy-cloud and cloudy-edge enabled leo satellite networks,'' in \emph{IEEE/CIC International Conference on Communications in China}, 2022, pp. 868--873.

\bibitem{tang2021computation}
Q.~Tang, Z.~Fei, B.~Li, and Z.~Han, ``Computation offloading in leo satellite networks with hybrid cloud and edge computing,'' \emph{IEEE Internet of Things Journal}, vol.~8, no.~11, pp. 9164--9176, 2021.

\bibitem{onori2015pontryagin}
S.~Onori, L.~Serrao, and G.~Rizzoni, ``Pontryagin's minimum principle,'' in \emph{Hybrid Electric Vehicles: Energy Management Strategies}.\hskip 1em plus 0.5em minus 0.4em\relax Springer, 2015, pp. 51--63.

\bibitem{wang2022computation}
B.~Wang, J.~Xie, D.~Huang, and X.~Xie, ``A computation offloading strategy for leo satellite mobile edge computing system,'' in \emph{2022 14th International Conference on Communication Software and Networks (ICCSN)}.\hskip 1em plus 0.5em minus 0.4em\relax IEEE, 2022, pp. 75--80.

\bibitem{tran2018joint}
T.~X. Tran and D.~Pompili, ``Joint task offloading and resource allocation for multi-server mobile-edge computing networks,'' \emph{IEEE Transactions on Vehicular Technology}, vol.~68, no.~1, pp. 856--868, 2018.

\bibitem{li2021double}
Z.~Li, C.~Jiang, and L.~Kuang, ``Double auction mechanism for resource allocation in satellite {MEC},'' \emph{IEEE Transactions on Cognitive Communications and Networking}, vol.~7, no.~4, pp. 1112--1125, 2021.

\bibitem{ergun2002improved}
F.~Ergun, R.~Sinha, and L.~Zhang, ``An improved fptas for restricted shortest path,'' \emph{Information Processing Letters}, vol.~83, no.~5, pp. 287--291, 2002.

\bibitem{wang2021profit}
B.~Wang, X.~Li, D.~Huang, and J.~Xie, ``A profit maximization strategy of mec resource provider in the satellite-terrestrial double edge computing system,'' in \emph{IEEE ICCT}, 2021, pp. 906--912.

\bibitem{zhang2023energy}
X.~Zhang, J.~Liu, R.~Zhang, Y.~Huang, J.~Tong, N.~Xin, L.~Liu, and Z.~Xiong, ``Energy-efficient computation peer offloading in satellite edge computing networks,'' \emph{IEEE Transactions on Mobile Computing}, vol.~23, no.~4, pp. 3077--3091, 2023.

\bibitem{cai2022security}
Y.~Cai, H.~Yao, and Y.~Gong, ``Security configuration and pricing scheme for satellite-terrestrial {IoT}: A stackelberg game,'' in \emph{2022 International Wireless Communications and Mobile Computing (IWCMC)}.\hskip 1em plus 0.5em minus 0.4em\relax IEEE, 2022, pp. 237--242.

\bibitem{liao2022secure}
C.~Liao, K.~Xu, H.~Zhu, X.~Xia, Q.~Su, and N.~Sha, ``Secure transmission in satellite-{UAV} integrated system against eavesdropping and jamming: A two-level stackelberg game model,'' \emph{China Communications}, vol.~19, no.~7, pp. 53--66, 2022.

\bibitem{feng2024covert}
S.~Feng, X.~Lu, S.~Sun, E.~Hossain, G.~Wei, and Z.~Ni, ``Covert communication in large-scale multi-tier {LEO} satellite networks,'' \emph{IEEE Transactions on Mobile Computing}, vol.~23, no.~12, pp. 11\,576--11\,587, 2024.

\bibitem{wen2022stackelberg}
X.~Wen, Y.~Ruan, Y.~Li, and R.~Zhang, ``Stackelberg game based secure transmission strategy for cognitive satellite terrestrial networks,'' in \emph{IEEE Global Communications Conference}, 2022, pp. 01--06.

\bibitem{zhang2024dynamic}
Y.~Zhang, F.~Chu, L.~Jia, M.~Yu, and W.~Cao, ``Dynamic anti-jamming strategy in {SIoT}: A stackelberg-matching game approach,'' \emph{IEEE Transactions on Consumer Electronics}, vol.~71, no.~2, pp. 4660--4669, 2024.

\bibitem{liao2023irs}
C.~Liao, K.~Xu, X.~Xia, G.~Hu, C.~Li, Y.~Wang, W.~Xie, X.~Yang, Y.~Shi, and L.~Wan, ``{IRS}-assisted anti-jamming transmission for an integrated satellite-{UAV}-terrestrial network with imperfect {CSI}: A game-based perspective,'' \emph{IEEE Internet of Things Journal}, vol.~10, no.~23, pp. 20\,484--20\,497, 2023.

\bibitem{yin2024uav}
Z.~Yin, J.~Li, Z.~Wang, Y.~Qian, Y.~Lin, F.~Shu, and W.~Chen, ``{UAV} communication against intelligent jamming: A stackelberg game approach with federated reinforcement learning,'' \emph{IEEE Transactions on Green Communications and Networking}, vol.~8, no.~4, pp. 1796--1808, 2024.

\bibitem{gong2024orbit}
Y.~Gong, X.~Liu, H.~Yao, X.~Cheng, A.~Nallanathan, and G.~K. Karagiannidis, ``In-orbit computation and security authentication for satellite-ground twin networks,'' \emph{IEEE Transactions on Vehicular Technology}, vol.~74, no.~2, pp. 3545--3549, 2024.

\bibitem{gong2024computation}
Y.~Gong, H.~Yao, X.~Liu, M.~Bennis, A.~Nallanathan, and Z.~Han, ``Computation and privacy protection for satellite-ground digital twin networks,'' \emph{IEEE Transactions on Communications}, vol.~72, no.~9, pp. 5532--5546, 2024.

\bibitem{liao2024game}
C.~Liao, K.~Xu, G.~Hu, X.~Xia, C.~Wei, W.~Xie, C.~Li, and Y.~Wang, ``Game theory and multi-agent {DRL} based anti-jamming transmission for integrated air-ground network,'' \emph{IEEE Transactions on Vehicular Technology}, vol.~73, no.~12, pp. 19\,565--19\,581, 2024.

\bibitem{zhang2023value}
B.~Zhang, Y.~Zhang, Y.~Wang, and Z.~Yang, ``Value-optimal priority-aware irregular repetition slotted {ALOHA} in satellite-integrated internet of things via noncooperative game,'' \emph{IEEE Internet of Things Journal}, vol.~11, no.~7, pp. 12\,495--12\,509, 2023.

\bibitem{shen2024privacy}
B.~Shen, K.-Y. Lam, F.~Li, and L.~Wang, ``Privacy-aware spectrum pricing and power control optimization for {LEO} satellite internet-of-things,'' \emph{IEEE Transactions on Wireless Communications}, pp. 1--15, to appear.

\bibitem{abdrabou2024game}
M.~Abdrabou and T.~A. Gulliver, ``Game theoretic spoofing detection for space information networks using physical attributes,'' \emph{IEEE Transactions on Communications}, vol.~72, no.~7, pp. 3947--3956, 2024.

\bibitem{liu2021decentralized}
X.~Liu, A.~Yang, C.~Huang, Y.~Li, T.~Li, and M.~Li, ``Decentralized anonymous authentication with fair billing for space-ground integrated networks,'' \emph{IEEE Transactions on Vehicular Technology}, vol.~70, no.~8, pp. 7764--7777, 2021.

\bibitem{gong2020pursuit}
H.~Gong, S.~Gong, and J.~Li, ``Pursuit--evasion game for satellites based on continuous thrust reachable domain,'' \emph{IEEE Transactions on Aerospace and Electronic Systems}, vol.~56, no.~6, pp. 4626--4637, 2020.

\bibitem{xia2023incentive}
Q.~Xia, Z.~Xu, and Z.~Hou, ``Incentive mechanism based on double auction for federated learning in satellite edge clouds,'' in \emph{International Conference on Mobility, Sensing and Networking}, 2023, pp. 660--668.

\bibitem{shen2020enhanced}
D.~Shen, C.~Sheaff, M.~Guo, E.~Blasch, K.~Pham, and G.~Chen, ``Enhanced {GANs} for satellite behavior discovery,'' in \emph{Sensors and Systems for Space Applications XIII}, vol. 11422.\hskip 1em plus 0.5em minus 0.4em\relax SPIE, 2020, pp. 110--121.

\bibitem{wu2023hybrid}
W.~Wu, J.~Chen, and J.~Liu, ``A hybrid optimisation method for intercepting satellite trajectory based on differential game,'' \emph{The Aeronautical Journal}, vol. 127, no. 1312, pp. 900--922, 2023.

\bibitem{barkatsa2025coordinated}
S.~Barkatsa, M.~Diamanti, P.~Charatsaris, S.~Voikos, E.~E. Tsiropoulou, and S.~Papavassiliou, ``Coordinated jamming and poisoning attack detection and mitigation in wireless federated learning networks,'' \emph{IEEE Open Journal of the Communications Society}, to appear.

\bibitem{10816375}
S.~Yang, C.-X. Wang, Y.~Wang, J.~Huang, Y.~Zhou, and E.-H.~M. Aggoune, ``An efficient pre-processing method for 6g dynamic ray-tracing channel modeling,'' \emph{IEEE Transactions on Vehicular Technology}, vol.~74, no.~5, pp. 6941--6953, 2025.

\bibitem{kim2022cooperative}
S.~Kim, ``Cooperative multi-path routing algorithm for integrated satellite-maritime networks,'' \emph{Mobile Information Systems}, vol. 2022, no.~1, p. 8678007, 2022.

\bibitem{zhou2021incentive}
H.~Zhou, T.~Wu, H.~Zhang, and J.~Wu, ``Incentive-driven deep reinforcement learning for content caching and d2d offloading,'' \emph{IEEE Journal on Selected Areas in Communications}, vol.~39, no.~8, pp. 2445--2460, 2021.

\bibitem{tang2024cooperative}
J.~Tang, J.~Li, X.~Chen, K.~Xue, L.~Zhang, Q.~Sun, and J.~Lu, ``Cooperative caching in satellite-terrestrial integrated networks: A region features aware approach,'' \emph{IEEE Transactions on Vehicular Technology}, 2024.

\bibitem{nguyen2023real}
M.-H.~T. Nguyen, T.~T. Bui, L.~D. Nguyen, E.~Garcia-Palacios, H.-J. Zepernick, H.~Shin, and T.~Q. Duong, ``Real-time optimized clustering and caching for 6g satellite-uav-terrestrial networks,'' \emph{IEEE Transactions on Intelligent Transportation Systems}, vol.~25, no.~3, pp. 3009--3019, 2023.

\bibitem{mishra2024minimizing}
S.~K. Mishra, C.~Goyal, C.~Agrawal, and A.~Pratap, ``Minimizing costs for content service providers in 6g space-air-ground integrated networks,'' in \emph{IEEE SPACE}, 2024, pp. 479--483.

\bibitem{liu2021reliable}
J.~Liu, X.~Zhang, R.~Zhang, T.~Huang, and F.~R. Yu, ``Reliable and low-overhead clustering in leo small satellite networks,'' \emph{IEEE Internet of Things Journal}, vol.~9, no.~16, pp. 14\,844--14\,856, 2021.

\bibitem{ni2023fault}
Y.~Ni, H.~Yang, and B.~Jiang, ``Fault-tolerant coverage control of multiple satellites: a differential graphical game approach,'' \emph{IEEE Transactions on Aerospace and Electronic Systems}, vol.~59, no.~5, pp. 5516--5529, 2023.

\bibitem{amozegari2024co}
F.~Amozegari, A.~Kosari, and M.~Fakoor, ``Co-location of geo satellites using differential game theory,'' in \emph{IEEE ICEE}, 2024, pp. 1--8.

\bibitem{11023214}
S.~Liu, Y.~Feng, S.~Yang, B.~Ma, and C.~Li, ``An efficient ray-tracing framework for urban scenarios powered by heterogeneous graph neural networks,'' \emph{IEEE Antennas and Wireless Propagation Letters}, vol.~24, no.~9, pp. 2884--2888, 2025.

\bibitem{anastasopoulos2012feedback}
M.~P. Anastasopoulos, T.~Taleb, P.~G. Cottis, and M.~S. Obaidat, ``Feedback suppression in multicast satellite networks using game theory,'' \emph{IEEE Systems Journal}, vol.~6, no.~4, pp. 657--666, 2012.

\bibitem{wang2022dynamic}
D.~Wang, W.~Wang, Y.~Kang, and Z.~Han, ``Dynamic data offloading for massive users in ultra-dense leo satellite networks based on stackelberg mean field game,'' in \emph{IEEE INFOCOM WKSHPS}, 2022, pp. 1--6.

\bibitem{chen2023remote}
J.~Chen, X.~Di, R.~Xu, H.~Qi, L.~Cong, K.~Zhang, Z.~Xing, X.~He, W.~Lei, and S.~Zhang, ``A remote sensing data transmission strategy based on the combination of satellite-ground link and geo relay under dynamic topology,'' \emph{Future Generation Computer Systems}, vol. 145, pp. 337--353, 2023.

\bibitem{qin2023service}
X.~Qin, T.~Ma, Z.~Tang, X.~Zhang, H.~Zhou, and L.~Zhao, ``Service-aware resource orchestration in ultra-dense leo satellite-terrestrial integrated {6G}: A service function chain approach,'' \emph{IEEE Transactions on Wireless Communications}, vol.~22, no.~9, pp. 6003--6017, 2023.

\bibitem{10938344}
B.~Ma, J.~Ye, S.~Feng, C.~Li, D.~Niyato, and D.~I. Kim, ``Uav-enabled relaying and edge computing: Hybrid network optimization,'' \emph{IEEE Transactions on Vehicular Technology}, vol.~74, no.~7, pp. 11\,644--11\,649, 2025.

\bibitem{zhang2020vcg}
X.~Zhang, B.~Zhang, D.~Guo, Y.~Jia, and Z.~Chen, ``Vcg auction-based multi-relay selection in hybrid satellite-terrestrial overlay networks,'' in \emph{IEEE WCSP}, 2020, pp. 736--741.

\bibitem{zhang2025toward111}
R.~Zhang, G.~Liu, Y.~Liu, C.~Zhao, J.~Wang, Y.~Xu, D.~Niyato, J.~Kang, Y.~Li, S.~Mao \emph{et~al.}, ``Toward edge general intelligence with agentic ai and agentification: Concepts, technologies, and future directions,'' \emph{arXiv preprint arXiv:2508.18725}, 2025.

\bibitem{11049053}
R.~Zhang, C.~Zhao, H.~Du, D.~Niyato, J.~Wang, S.~Sawadsitang, X.~Shen, and D.~I. Kim, ``Embodied ai-enhanced vehicular networks: An integrated vision language models and reinforcement learning method,'' \emph{IEEE Transactions on Mobile Computing}, pp. 1--16, 2025.

\bibitem{tang2022blockchain}
F.~Tang, C.~Wen, L.~Luo, M.~Zhao, and N.~Kato, ``Blockchain-based trusted traffic offloading in space-air-ground integrated networks (sagin): A federated reinforcement learning approach,'' \emph{IEEE J. Sel. Areas Commun.}, vol.~40, no.~12, pp. 3501--3516, 2022.

\end{thebibliography}

\end{document}